\documentclass[10pt]{article}
\usepackage[T1]{fontenc}
\usepackage[utf8]{inputenc}
\usepackage{amsmath}											
\usepackage{amsfonts}
\usepackage{amssymb}
\usepackage{amsthm}
\usepackage{bbm}
\usepackage{longtable}
\usepackage{booktabs}
\usepackage[normalem]{ulem}
\usepackage{caption}
\usepackage{subcaption}
\usepackage{comment}
\usepackage{esint}
\usepackage[margin=1.0in]{geometry}
\usepackage{graphicx}
\usepackage[scr=dutchcal]{mathalfa}
\usepackage{mathtools}
\usepackage[final]{showlabels}
\usepackage{xcolor}
\graphicspath{ {images/} }
\usepackage[section]{placeins}
\DeclareMathOperator{\diag}{diag}

\DeclareMathOperator{\Ran}{Ran}

\newcommand{\norm}[1]{\left\lVert#1\right\rVert}
\newtheorem{proposition}{Proposition}
\newtheorem{lemma}[proposition]{Lemma}
\newtheorem{theorem}[proposition]{Theorem}

\newtheorem{assumption}[proposition]{Assumption}
\newtheorem{corollary}[proposition]{Corollary}

\theoremstyle{remark}
\newtheorem{remark}[proposition]{Remark}

\theoremstyle{definition}
\newtheorem{definition}[proposition]{Definition}

\numberwithin{proposition}{section}
\numberwithin{equation}{section}

\usepackage{tikz}
\usetikzlibrary{backgrounds}

\newcommand{\eps}{\varepsilon}
\newcommand{\Rm}{\mathbb{R}}

\newcommand{\dint}{\displaystyle\int}

\newcommand{\aver}[1]{\langle #1 \rangle}

\newcommand{\fm}{\mathfrak m}

\newcommand{\ow}{{\rm Op}^w}

\newcommand{\A}{\mathcal{A}}

\newcommand{\kdisp}{\mathfrak{f}} 

\newcommand{\cD}{\mathcal{D}}
\newcommand{\cG}{\mathcal{G}}
\newcommand{\cL}{\mathcal{L}}
\newcommand{\cR}{\mathcal{R}}

\newcommand{\cT}{\mathcal{T}}

\newcommand{\bH}{{\rm {\bf H}}}
\newcommand{\bHe}{\bH^\eps} 

\newcommand{\rH}{\mathbb{H}}

\newcommand{\rate}{\mu} 

\newcommand{\ncv}{N_{{\rm CV}}}

\newcommand{\fA}{\mathfrak{A}}
\newcommand{\fB}{\mathfrak{B}}

\newcommand{\inva}{a^\sharp}
\newcommand{\phin}{\phi_{\rm in}}
\newcommand{\phout}{\phi_{\rm out}}
\newcommand{\uni}{\mathfrak{U}}

\newcommand{\cphin}{\check{\phi}_{\rm in}}
\newcommand{\cphout}{\check{\phi}_{\rm out}}

\newcommand{\ds}{\mathscr{s}}

\newcommand{\Opw}{\mathrm{Op}^w}

\newcommand{\C}{\mathbb{C}}

\newcommand{\ze}{\zeta}
\newcommand{\De}{\Delta}
\newcommand{\sig}{\sigma}
\newcommand{\Ga}{\Gamma}
\newcommand{\asharp}{a^\sharp}
\newcommand{\ashift}{a^\sharp}

\newcommand{\ash}{a^\sharp}

\newcommand{\wt}{\tilde{\omega}_p}

\newcommand{\ashl}{a^\sharp_\ell}

\newcommand{\uu}{\mathbf{u}}

\title{Band reduction}
\title{Effective Hamiltonians and asymptotic band coupling}
\author{Guillaume Bal\thanks{Departments of Statistics and Mathematics and Committee on
Computational and Applied Mathematics, University of Chicago, Chicago, IL 60637;
\texttt{guillaumebal@uchicago.edu}}
\and Paul Cazeaux\thanks{Department of Mathematics, Virginia Tech, Blacksburg, VA 24060;
\texttt{cazeaux@vt.edu}}
\and Daniel Massatt\thanks{Department of Mathematical Sciences, New Jersey Institute of
Technology, Newark, NJ 07103; \texttt{daniel.massatt@njit.edu}}
\and Solomon Quinn\thanks{Center for Computational Mathematics, Flatiron Institute, New York,
NY 10010; \texttt{squinn@flatironinstitute.org}}}

\begin{document}

\maketitle

\begin{abstract}
    By a multiple-scale asymptotic expansion, we derive effective macroscopic models for a large class of tight-binding Schr\"odinger equations with slow spatial variations. For wave packets that are spectrally localized to some point $(K;E) \in \mathbb{R}^d \times \mathbb{R}$ in the unperturbed tight-binding dispersion relation, our effective models produce long-time approximations of the tight-binding dynamics, accurate to any prescribed order in the asymptotic parameter. Our framework allows the point $(K;E)$ to be degenerate, meaning that multiple (but not necessarily all) energy bands of the tight-binding Hamiltonian can attain the value $E$ at the frequency $K$. We construct an effective Hamiltonian on the relevant degenerate subspace, which quantifies the coupling between the degenerate and other bands. Several applications and numerical simulations illustrate the theory.
\end{abstract}



\section{Introduction}\label{sec:intro}

Many models of wave propagation in structured media take the form of a Hamiltonian whose
coefficients vary slowly compared with the scale of the underlying microstructure. Examples include
tight-binding models with slowly modulated hoppings and on-site energies, photonic or plasmonic media with slowly varying material parameters, and lattices subject to long-wavelength distortions. 
In each case one is interested in the
propagation of wave packets whose spatial extent is comparable to the scale of the
modulation, and whose frequency lies near a distinguished point of the unperturbed
dispersion relation (for example, a Dirac point or other type of band crossing).
The goal is to derive a \emph{macroscopic}, or \emph{effective}, model describing the dynamics of such wave packets. 

Effective models of this kind have a long history. We refer to \cite{nenciu1991dynamics, panati2003effective, panati2003space} for the construction of effective Hamiltonians describing an isolated family of Bloch bands in a slowly varying external field. See also \cite{teufel2003adiabatic} for a systematic
account, and \cite{gerard1991mathematical} for a closely related reduction of the spectral
problem by a Grushin, or Schur complement, method.
The effective mass theorems of \cite{allaire2005homogenization} provide a homogenized
description of Schr\"odinger dynamics near a Bloch eigenvalue, while
Dirac-type macroscopic models near conical crossings for honeycomb structures were derived in
\cite{fefferman2012honeycomb, fefferman2014wave, li2026pseudo}. The tight-binding limit of such
structures was analyzed in \cite{fefferman2018strongbinding}. The evolution of wave packets in periodic media with a slow spatial modulation was analyzed in \cite{watson2018wavepackets}.

The present paper implements a general framework for effective Schr\"odinger dynamics of tight-binding models. We begin by constructing an approximate solution \eqref{eq:psi_def_higher_order_all} to the tight-binding Schr\"odinger equation. This approximate solution is a truncated asymptotic expansion, with each term determined by the exact solution to a continuum (Dirac-type) initial value problem. Our main result, Theorem \ref{thm:main_higher_order}, establishes the validity of this approximation, providing an error estimate of $O(\delta^p + \delta^{p+\rate} t)$, up to a factor polynomial in
the macroscopic time $\delta t$.
Here, $p \in \mathbb{N}$ is the number of terms in the expansion, and $0 < \rate \le 1$ a fixed parameter.
Corollary \ref{cor:timescale} translates our approximation error into a
range of validity $t \ll \delta^{-\kappa}$ for any $\kappa < 3/2$. This restriction comes from
secular terms of the expansion. Theorem \ref{prop:rH} establishes the validity of a higher-order effective Hamiltonian in approximating the tight-binding dynamics. On sectors whose leading symbol is elliptic, Corollary \ref{cor:rH_eps} shows that the corresponding error is $O(\delta^{p+\rate}t)$ in $L^2$ for all times, so that the effective Hamiltonian remains valid for $t \ll \delta^{-(p+\rate)}$, well beyond the range of the expansion.

The tight-binding models we consider are described by pseudodifferential operators whose symbols, denoted $a_\delta$, satisfy Assumption \ref{assumption:a_higher_order_all}. This assumption requires $a_\delta$ to have slow spatial variations, with its leading-order behavior described by a constant Hermitian matrix $a^0 = \sum_{\ell=1}^L E_\ell P_\ell$. Here, the $E_\ell$ are the distinct eigenvalues of $a^0$ with $P_\ell$ the corresponding orthogonal projections. We will henceforth refer to each $\Ran P_\ell$ as a \emph{spectral sector of $a^0$}, or just \emph{sector} for short. 
In addition to the convergence result Theorem \ref{thm:main_higher_order},
the three primary contributions of this paper are outlined as follows.

\medskip

First, we treat all spectral sectors of $a^0$ simultaneously and to arbitrary order in
$\delta$. Assumption \ref{assumption:a_higher_order_all} requires only that the symbol
admit an expansion in $\delta$ near $K$ whose coefficients grow at most polynomially in the
rescaled momentum, and that each projected leading symbol $b_{0\ell} = P_\ell b_0 P_\ell$
be either bounded or elliptic of order one. The first case covers flat and asymptotically
flat bands, the second Dirac-type sectors.

\medskip

Second, we clarify how the sectors interact. The construction
\eqref{eq:def_phout}--\eqref{eq:def_phin} of Theorem \ref{thm:main_higher_order} produces, for each $\ell$, an evolution equation
governed by the macroscopic Hamiltonian $\bH_{0\ell}$ of that sector alone. Sectors
communicate only via the algebraic correctors $\phout^{j,\ell}$ (cf. \eqref{eq:def_phout}), through a single
redistribution of the initial data at $T=0$ in \eqref{eq:def_phin}.
Proposition \ref{prop:dressed} and Theorem \ref{thm:equiv} further illustrate the independence of each sector, quantifying the error of a modified construction (Definition \ref{def:single}) in which the different sectors decouple.

In Section \ref{sec:effective}, we construct a higher-order effective Hamiltonian \eqref{eq:def_rH} that encodes the coupling between sectors.
The validity of this effective Hamiltonian is established by Theorem \ref{prop:rH}.

\medskip

Third, we work out four applications in detail. The results are collected in Section
\ref{sec:summary}, and the derivations in Sections \ref{sec:applications}--\ref{sec:plasma} of the supplementary material, which recalls the necessary notation in Section \ref{sec:sm_setting}.
The derivations follow a common pattern. We begin with the spectral
data of $a^0$, then the macroscopic operators (the resolvents $\ashl$, the sector
Hamiltonians $\bH_{0\ell}$, the couplings and the self-energies),
then the expansion generated by the prescribed data, and finally, for the coupled chains and the Kekul\'e model, the band-adapted expansion
and the comparison of the two. 

Section \ref{sec:summary_chains} treats two
coupled SSH chains and an SSH chain coupled to a flat band, in which the mechanisms above
are visible in closed form. Section \ref{sec:summary_kekule} treats the $6\times6$ tight-binding
model of graphene with a Kekul\'e-O modulation. 
Here, a Dirac operator with Kekul\'e mass \eqref{eq:sum_kekule_b} describes the dynamics to leading order. We also obtain higher-order corrections to this operator and explicitly calculate an $O(\delta^2)$ approximation of the tight-binding dynamics.
Section \ref{sec:summary_plasma} treats a $9\times9$ cold magnetized plasma model, a
continuous rather than tight-binding model, whose symbol is unbounded in $\xi$. Section
\ref{sec:pde_class} shows that the theory of Section \ref{sec:tb} still applies after a
single hypothesis is changed. Numerical simulations are presented in Section
\ref{sec:numerics}, and further experiments on the effective Hamiltonian, over long times and up to third order, in Section \ref{sec:sm_numerics} of the supplementary material.

\medskip

We emphasize two key distinctions of the present setting compared with the existing literature.
First, no subset of bands is
singled out, as all spectral sectors of $a^0$ are treated at once.
Even sectors on which the initial condition vanishes carry, in general, an $O(\delta)$
contribution to the tight-binding dynamics, through the redistribution mechanism (M2) of Remark
\ref{rem:mechanisms}, and so must be retained as soon as $p \ge 2$. (In the Kekul\'e model of
Section \ref{sec:summary_kekule}, where $b_0$ does not couple the Dirac sector to the
$\Gamma$-point sectors, this contribution is only of order $\delta^2$.) Second, the initial value problem is treated on the same footing as the evolution equation. We consider general initial data, e.g. that are not adapted to an almost invariant subspace such as the one of \cite{panati2003effective}. 
For initial conditions restricted to a single sector $P_\ell$, we quantify the $O(\delta)$ component of the solution orthogonal to $\Ran P_\ell$; see Theorem \ref{thm:main_higher_order} and Remark \ref{rem:p_equals_one}. We also show with Definition \ref{def:single} and Proposition \ref{prop:dressed} that this component can be removed by an $O(\delta)$ modification of the data.

These distinctions are best seen against the space-adiabatic perturbation theory of Nenciu
\cite{nenciu1991dynamics} and of Panati, Spohn and Teufel
\cite{panati2003effective, panati2003space,teufel2003adiabatic}. There, the central objects are
an almost invariant subspace, a unitary intertwiner onto a reference space, and an effective
Hamiltonian on that space, all constructed to arbitrary (in Nenciu's setting, exponential)
order in the adiabatic parameter. The projection onto the relevant band family depends on the
phase-space variables, through the Bloch bundle in the quasi-momentum and through the external
potential in the position variable, and this dependence gives rise to Berry-type terms (connection and curvature).
We refer to \cite{emmrich1996geometry, littlejohn1991geometric} for a geometric analysis of WKB approximations to systems of semiclassical transport equations.
For related work applying to semiclassical pseudodifferential operators with operator-valued symbols, see \cite{martinez2009twisted, nenciu2004semiclassical,
stiepan2013semiclassical}.


In the present setting the leading symbol $a^0$ is a constant
matrix, so that the projections $P_\ell$ and resolvents $\ashl$ (cf. \eqref{eq:inva_higher_order}) depend on neither $X$ nor
$\zeta$, and every Berry-type term is absent by construction. Our approximation of the
tight-binding dynamics reduces to a finite number of Schr\"odinger equations, one per sector,
whose coefficients are given in closed form by Schur complement formulas. Section
\ref{sec:effective} constructs the analogues of the objects of space-adiabatic theory, with a
frequency cutoff. We do not, however, obtain exponential
accuracy.

\paragraph{Notation.} We write $\varphi_\delta$ for the exact solution of the microscopic
evolution problem, $\psi_\delta$ for the approximate solution we construct, $\phi_0$ for
the macroscopic initial condition, and $\phi_\ell = \sum_j \delta^j \phi^j_\ell$, with
$\phi^j_\ell = \phin^{j,\ell} + \phout^{j,\ell}$, for the macroscopic amplitudes in the
sector $\ell$; the labels ${\rm in}$ and ${\rm out}$ refer to $\Ran P_\ell$ and its
orthogonal complement. Slow variables are denoted by capitals, $T = \delta t$ and
$X = \delta x$, and the rescaled momentum by $\zeta = (\xi - K)/\delta$. Operators acting in
the slow variables are written in boldface, $\bH_j := \ow b_j$ and
$\bH_{0\ell} := \ow b_{0\ell}$.
The effective Hamiltonian \eqref{eq:def_rH} of Section \ref{sec:effective} is denoted by $\rH$.
We use $D_z := \frac1i \partial_z$, $\aver{\cdot}$ for
$\aver{Y} = (1+|Y|^2)^{1/2}$, and $S(\fm)$, $S^m$ for the symbol classes of Definitions
\ref{def:symbol_class} and \ref{def:hormander_class}. We use $H^m$ to denote the standard Sobolev space of order $m$, and $H^\infty := \cap_{m \ge 0} H^m$. Constants denoted $C$ may change from
line to line and depend on the order of the Sobolev spaces involved, on $p$, and on 
Sobolev norms of $\phi_0$, but never on $\delta$ or $t$. A list of notation is given in Section \ref{sec:sm_notation} of the supplementary material.

\section{Band reduction for tight-binding Hamiltonians}\label{sec:tb}
Fix $K \in \mathbb{R}^d$ and $p \in \mathbb{N}$ and $\delta_0 > 0$. 
The setting of this paper is described by the following
\begin{assumption}\label{assumption:a_higher_order_all}
    Let 
    \begin{align}\label{eq:a_delta}
        a_\delta(x,\xi) = a(\delta x,\xi;\delta), \qquad (x, \xi) \in \mathbb{R}^d \times \mathbb{R}^d, \quad 0 < \delta \le \delta_0,
    \end{align}
    where $a$ is an $N\times N$ Hermitian-valued symbol in $S(1)$ uniformly in $\delta$.
    Assume there exist a constant Hermitian matrix $a^0$ and Hermitian-valued symbols
    \begin{align}\label{eq:b_j_symbol_class}
        b_{j} (X,\zeta; \delta) = b_{j} \in S (\aver{\zeta}^{j+1}), \qquad j \in \{0, \dots, p\}
    \end{align}
    such that for some $0 < \rate \le 1$,
    \begin{align}\label{eq:decomposition_higher_order_all}
        a(X,K + \delta \zeta;\delta) = a^0 + \sum_{j=0}^{p-1} \delta^{j+1} b_{j}(X, \zeta; \delta) + \delta^{p+\rate} b_{p}(X, \zeta ;\delta), \qquad (X,\zeta) \in \mathbb{R}^d \times \mathbb{R}^d.
    \end{align}
    Let $\{E_\ell\} \subset \mathbb{R}$ denote the eigenvalues of $a^0$ and $\{P_\ell\}$ the corresponding orthogonal projections, so that
    \begin{align*}
        a^0 = \sum_{\ell=1}^L E_\ell P_\ell
    \end{align*}
    for some $1 \le L \le N$. Define $b_{0\ell} := P_\ell b_0 P_\ell$. For each $\ell$, assume that either
    \begin{enumerate}
        \item $b_{0\ell} \in S(1)$; or
        \item $b_{0\ell} \in S^1$ and there exists $c>0$ such that
        \begin{align}\label{eq:ellipticity}
            |\det ( b_{0\ell} (X,\zeta;\delta) + (I_N-P_\ell) \aver{\zeta})|^{1/N} \ge c \aver{\zeta} - c^{-1}, \qquad (X, \zeta; \delta) \in \mathbb{R}^{2d} \times (0,\delta_0].
        \end{align}
    \end{enumerate}
\end{assumption}
We refer to Appendix \ref{sec:pseudo} for definitions of the symbol classes $S (\fm)$ and $S^m$.
For $1 \le \ell \le L$, define the macroscopic Hamiltonian associated with eigenvalue $E_\ell$ by $\bH_{0\ell} := \ow b_{0\ell}$. For $0 \le j \le p$, set $\bH_j := \ow b_j$.
\begin{remark}[On the two symbol classes]\label{rem:symbol_classes}
Let us briefly explain the need for two distinct classes of symbols, $S (\fm)$ and $S^m$, in Assumption \ref{assumption:a_higher_order_all}. By Lemma \ref{lemma:CV}, the requirement $b_j \in S(\aver{\zeta}^{j+1})$ of \eqref{eq:b_j_symbol_class} implies that $\bH_j : H^{s+j+1} \to H^s$ is bounded uniformly in $\delta$. This is all that is required of the $b_j$ for $j \ge 1$, as will be evident in the proofs of Proposition \ref{lemma:res_higher_order_all} and Lemma \ref{lemma:Sobolev_higher_order_all}. 

Regarding the dichotomy for $b_{0\ell}$, we will need the operators $\bH_{0\ell} = \ow b_{0\ell}$ to be self-adjoint on $L^2$, as these operators will describe the effective Schr\"odinger dynamics for the wave-packets we consider.
As described above, the assumption $b_{0\ell} \in S(1)$ gives boundedness of $\bH_{0\ell}$ on every $H^m$ uniformly in $\delta$. Hence this assumption guarantees that the operator $\bH_{0\ell}$ is a bounded self-adjoint operator on $L^2$. On the other hand, for unbounded Hermitian-valued symbols $b_{0\ell} (X, \zeta; \delta) =b_{0\ell} \in S (\aver \zeta)$, one needs to impose additional constraints to ensure self-adjointness of $\bH_{0\ell}$. This is achieved by the ellipticity condition \eqref{eq:ellipticity} and stronger assumption $b_{0\ell} \in S^1 \subset S (\aver \zeta)$; see Lemma \ref{lemma:sa_H0l}.
\end{remark}

\begin{remark}[Taylor expansion]
    If the symbol $a$ from \eqref{eq:a_delta} is smooth in $\mathbb{R}^{2d} \times [0,\delta_0]$, the decomposition \eqref{eq:decomposition_higher_order_all} is obtained by a Taylor expansion of $(\xi, \delta) \mapsto a (X, \xi; \delta)$ about $(\xi, \delta) = (K,0)$, provided that the leading term $a(X,K;0) = a^0$ does not depend on $X$. In this case, $\rate = 1$ and $b_j$ for $j \le p-1$ is independent of $\delta$.
    The possibility $\rate < 1$ is kept for models in which the expansion in $\delta$ does not come from a Taylor series. One example is the interlayer coupling symbol for twisted bilayer graphene as treated in 
    \cite[Proposition 3.5]{bal2026macroscopic}.
\end{remark}

We define the tight-binding Hamiltonian by 
    \begin{equation}\label{eq:H_delta}
        H_\delta :=\ow a_\delta.
    \end{equation}
For $\phi_0 \in H^\infty (\mathbb{R}^d; \mathbb{C}^N)$, let $\varphi_\delta$ be the solution to the microscopic tight-binding evolution problem
    \begin{equation}\label{eq:microWP_higher_order}
      (D_t+H_\delta)\varphi_\delta =0,\qquad \varphi_\delta(0,x)= e^{iK\cdot x} \delta^{\frac d2} \phi_0(\delta x).
    \end{equation}
Here and below, we use the notation $D_z:=\frac 1i\partial_z$ for any variable $z$.

\subsection{Macroscopic model, localized wave packets, and approximation error} \label{subsec:macroscopic}
Recall the notation $\bH_{0\ell} = \ow b_{0\ell}$ with $b_{0\ell}$ from Assumption \ref{assumption:a_higher_order_all}.
Throughout the paper, $H^m$ denotes the standard Sobolev space of order $m$. Our assumptions on $b_{0\ell}$ ensure the following properties. 
\begin{lemma}\label{lemma:sa_H0l}
    Let Assumption \ref{assumption:a_higher_order_all} hold and fix $1 \le \ell \le L$. Then $\bH_{0\ell}$ is self-adjoint on $L^2 (\mathbb{R}^d; \Ran P_\ell)$, with $\Ran P_\ell$ invariant. If $b_{0\ell} \in S(1)$, then $\bH_{0\ell}$ is bounded on $H^m$ for every $m \ge 0$, uniformly in $\delta$. If instead $b_{0\ell} \in S^1$ satisfies \eqref{eq:ellipticity}, then the domain of $\bH_{0\ell}$ is $H^1 (\mathbb{R}^d; \Ran P_\ell)$ and, for any $m \ge 0$,
\begin{align}\label{eq:equivalent_norms_higher_order}
        c \norm{u}_{H^m} \le \norm{(i+\bH_{0\ell})^m u}_{L^2} \le C \norm{u}_{H^m}, \qquad u \in H^m (\mathbb{R}^d; \Ran P_\ell), \quad 0 < \delta \le \delta_0,
\end{align}
with $c, C$ independent of $\delta$.
\end{lemma}
\begin{proof}
    Since $P_\ell$ is a constant orthogonal projection and $b_{0\ell} = P_\ell b_0 P_\ell$ is Hermitian valued, $\bH_{0\ell}$ is symmetric and leaves $\Ran P_\ell$ invariant. When $b_{0\ell} \in S(1)$, Lemma \ref{lemma:CV} gives boundedness on every $H^m$ uniformly in $\delta$, whence self-adjointness. When $b_{0\ell} \in S^1$, Lemma \ref{lemma:CV} gives $\bH_{0\ell} : H^{m+1}\to H^m$ boundedly, and the ellipticity \eqref{eq:ellipticity} provides a parametrix. Indeed, write $M := b_{0\ell} + (I_N-P_\ell)\aver{\zeta}$, which belongs to $S^1$ and satisfies $|\det M|^{1/N} \ge c \aver{\zeta} - c^{-1}$. Since $M^{-1} = (\det M)^{-1}\mathrm{adj}\,M$ and the entries of $\mathrm{adj}\,M$ are bounded by $C\aver{\zeta}^{N-1}$, we obtain $|M^{-1}| \le C \aver{\zeta}^{N-1}/(c\aver{\zeta}-c^{-1})^N \le C \aver{\zeta}^{-1}$ for $\aver{\zeta}\ge \zeta_*$, with $\zeta_*$ and $C$ depending only on $c$ and on finitely many seminorms of $b_{0\ell}$. Let $\tilde\chi\in C^\infty(\mathbb{R}^d)$ vanish for $\aver{\zeta}\le \zeta_*$ and equal one for $\aver{\zeta}\ge 2\zeta_*$ (the matrix $M$ may be singular for small $\zeta$). Differentiating $\partial_\zeta M^{-1} = -M^{-1}(\partial_\zeta M)M^{-1}$ and iterating shows that $\tilde\chi M^{-1} \in S^{-1}$, again uniformly in $\delta$. A standard iteration in the Weyl calculus (see, e.g., \cite[Chapter 4]{Zworski}) then produces $\mathcal Q_\delta = \ow q_\delta$ with $q_\delta \in S^{-1}$ and $\mathcal Q_\delta (i+\bH_{0\ell}) = I + \mathcal K_\delta$ on $\Ran P_\ell$, where $\mathcal K_\delta : H^m \to H^{m+1}$ is bounded uniformly in $\delta$. Essential self-adjointness on $H^1$ and the lower bound in \eqref{eq:equivalent_norms_higher_order} for $m=1$ follow; the upper bound is the boundedness of $\bH_{0\ell}$, and the general case follows by iteration in $m$.
\end{proof}
Since $H_\delta$ is a bounded self-adjoint operator on $L^2 (\mathbb{R}^d; \mathbb{C}^N)$, the problem \eqref{eq:microWP_higher_order} is well posed.
Moreover, for any $m \ge 0$, Lemma \ref{lemma:commutator_bd} and the proof of Lemma \ref{lemma:u_delta_reg} give the bound
\begin{align}\label{eq:apriori_micro}
    \norm{\varphi_\delta (t,\cdot)}_{H^m} \le C (1 + (\delta t)^m) \norm{\phi_0}_{H^m}, \qquad t \ge 0, \quad 0 < \delta \le \delta_0 .
\end{align}

With $\phi_0$ as in \eqref{eq:microWP_higher_order}, define $\phi^0_\ell$ as the solution to
\begin{align}\label{eq:def_phi_0_ell}
    (D_T+\bH_{0\ell}) \phi^0_\ell(T,X;\delta)=0,\qquad \phi^0_\ell(0,X; \delta)=P_\ell \phi_0(X).
\end{align}
Since $\phi^0_\ell (0, X; \delta) \in \Ran P_\ell$ and
$$D_T (I_N-P_\ell) \phi^0_\ell = (I_N-P_\ell) (D_T + \bH_{0\ell}) \phi^0_\ell = 0,$$
we have $\phi^0_\ell (T, X; \delta) \in \Ran P_\ell$ for all $T \ge 0$, $X \in \mathbb{R}^d$, $0 < \delta \le \delta_0$.
Now, define
\begin{align}\label{eq:inva_higher_order}
    \inva_\ell := (a^0 - E_\ell I_N) \vert_{\Ran (I_N-P_\ell)}^{-1} = \sum_{\ell' \ne \ell} (E_{\ell'} - E_\ell)^{-1} P_{\ell'},
\end{align}
and recall the shorthand $\bH_j = \ow b_j$.
With the $\phi^0_\ell$ defined by \eqref{eq:def_phi_0_ell}, set
\begin{align}\label{eq:phi_expansion_higher_order}
    \phi := \sum_{\ell=1}^L \phi_\ell, \qquad \phi_\ell := \sum_{j=0}^p \delta^j \phi^j_\ell,
\end{align}
where the $\phi^j_\ell = \phin^{j,\ell} + \phout^{j,\ell}$ for each $1 \le \ell \le L$ are defined recursively over $j \ge 1$ as follows. Set
\begin{align}\label{eq:def_phout}
    \phout^{j,\ell} &:= -\inva_\ell \left( D_T \phi^{j-1}_\ell + \sum_{i=0}^{j-1} \bH_i \phi^{j-1-i}_\ell\right), \qquad 1 \le j \le p,
\end{align}
and for all $1 \le j \le p-1$ define $\phin^{j,\ell}$ as the solution to the initial value problem
\begin{align}\label{eq:def_phin}
    (D_T + \bH_{0\ell}) \phin^{j,\ell} &= -P_\ell\left(\bH_0 \phout^{j,\ell} + \sum_{i=1}^j \bH_i \phi^{j-i}_\ell\right), \quad \phin^{j,\ell} (0, \cdot\; ; \delta) = -P_\ell \sum_{\ell'=1}^L \phout^{j,\ell'} (0, \cdot\; ; \delta).
\end{align}
Finally, set 
$\phin^{p,\ell} \equiv 0$.

By construction, $\phin^{j,\ell} = P_\ell \phin^{j,\ell}$ and $\phout^{j,\ell} = (I_N - P_\ell) \phout^{j,\ell}$ for every $j,\ell$.
We now 
consider the ansatz
\begin{align}\label{eq:psi_def_higher_order_all}
  \psi_\delta(t,x) = \sum_{\ell=1}^L e^{i(K\cdot x-E_\ell t)} \delta^{\frac d2} \phi_\ell (\delta t,\delta x;\delta).
\end{align}
The main result of this paper is the following
\begin{theorem}\label{thm:main_higher_order}
    Suppose Assumption \ref{assumption:a_higher_order_all} holds for some $0 < \rate \le 1$, and fix $m \in \mathbb{N}_0$. Then 
    \begin{align}\label{eq:estimateHm_higher_order_all}
        \|(\psi_\delta-\varphi_\delta)(t, \cdot)\|_{H^m(\Rm^d; \mathbb{C}^N)} \leq C\delta^p(1 + \delta^{\rate} t) \left(1 + (\delta t)^{2m+2p}\right)
    \end{align}
    uniformly in $t \ge 0$ and $0 < \delta \le \delta_0$. 
\end{theorem}
\begin{proof}
    By construction, we have $\sum_{\ell=1}^L\phi^0_\ell (0,X;\delta) = \phi_0 (X)$ and
    \begin{align*}
        \sum_{\ell=1}^L \phin^{j,\ell} (0,X; \delta) = -\sum_{\ell=1}^L \phout^{j,\ell} (0,X;\delta), \qquad 1 \le j \le p-1.
    \end{align*}
    Hence $\phi (0,X;\delta) - \phi_0 (X) = \delta^p \sum_{\ell=1}^L \phout^{p,\ell} (0,X;\delta)$, from which it follows that $u_\delta := \psi_\delta - \varphi_\delta$ satisfies
    \begin{align}\label{eq:ic_bd_main}
        \|u_\delta (0, \cdot)\|_{H^m} \le C \delta^p.
    \end{align}
    On the other hand, Proposition \ref{lemma:res_higher_order_all} gives
    \begin{align}\label{eq:res_bd_main}
        \norm{(D_t + H_\delta) u_\delta (t, \cdot)}_{H^m} = \norm{(D_t + H_\delta) \psi_\delta (t, \cdot)}_{H^m} \le C \delta^{p+\rate} (1 + (\delta t)^{m+2p}), \qquad t \ge 0, \quad 0 < \delta \le \delta_0.
    \end{align}
    By Lemma \ref{lemma:u_delta_reg} (whose hypothesis \eqref{eq:commutator_bd_reg} is established by Lemma \ref{lemma:commutator_bd}), the residual bounds \eqref{eq:ic_bd_main}-\eqref{eq:res_bd_main} imply the desired estimate \eqref{eq:estimateHm_higher_order_all}.
\end{proof}
The bound \eqref{eq:estimateHm_higher_order_all} for long times can be strengthened, by slowing the growth of the macroscopic amplitudes under an additional hypothesis (Proposition \ref{prop:main_flat}) and by avoiding the commutator factor $(\delta t)^m$ of Lemma \ref{lemma:u_delta_reg} on a lattice (Proposition \ref{prop:expansion_discrete}).
Recalling the dichotomy for $b_{0\ell}$ in Assumption \ref{assumption:a_higher_order_all}, it is useful to define the sets
\begin{align}\label{eq:cL_def}
    \cL_0 := \{\ell \in \{1, \dots, L\} : b_{0\ell} \in S(1)\}, \qquad \cL_1 := \{1, \dots, L\} \setminus \cL_0.
\end{align}
\begin{proposition}\label{prop:main_flat}
    Suppose Assumption \ref{assumption:a_higher_order_all} holds for some $0 < \rate \le 1$. Assume moreover that $b_{0\ell} \equiv 0$ for all 
    $\ell \in \cL_0$. Fix $m \in \mathbb{N}_0$. Then
    \begin{align*}
        \|(\psi_\delta-\varphi_\delta)(t, \cdot)\|_{H^m(\Rm^d; \mathbb{C}^N)} \leq C\delta^p(1 + \delta^{\rate} t) \left(1 + (\delta t)^{m+p-1}\right)
    \end{align*}
    uniformly in $t \ge 0$ and $0 < \delta \le \delta_0$. 
\end{proposition}
\begin{proof}
    We follow the proof of Theorem \ref{thm:main_higher_order} to establish \eqref{eq:ic_bd_main}. By our assumption that $b_{0\ell} \equiv 0$ for all $\ell \in \cL_0$, we can apply Propositions \ref{lemma:res_higher_order_all} and \ref{prop:flat}, strengthening \eqref{eq:res_bd_main} to 
    \begin{align}\label{eq:res_bd_flat}
        \norm{(D_t + H_\delta) u_\delta (t, \cdot)}_{H^m} = \norm{(D_t + H_\delta) \psi_\delta (t, \cdot)}_{H^m} \le C \delta^{p+\rate} (1 + (\delta t)^{p-1}), \qquad t \ge 0, \quad 0 < \delta \le \delta_0.
    \end{align}
    The result then follows from Lemmas \ref{lemma:commutator_bd} and \ref{lemma:u_delta_reg}.
\end{proof}
Theorem \ref{thm:main_higher_order} and Proposition \ref{prop:main_flat} imply the following
\begin{corollary}\label{cor:timescale}
    Suppose Assumption \ref{assumption:a_higher_order_all} holds for some $0 < \rate \le 1$, and fix $m \in \mathbb{N}_0$ and $\epsilon > 0$. 
    Set $\kappa := 2$ if $b_{0\ell} \equiv 0$ for all $\ell \in \cL_0$, and $\kappa := 3/2$ otherwise.
    Then provided $p$ is sufficiently large,
    \begin{align*}
        \lim_{\delta \to 0}\; \sup \left\{ \norm{\psi_\delta (t, \cdot) - \varphi_\delta (t, \cdot)}_{H^m} : t \in [0, \delta^{-\kappa+\epsilon}] \right\} = 0.
    \end{align*}
\end{corollary}

We now establish similar convergence estimates in the case where $H_\delta$ defines a tight-binding Hamiltonian on a lattice.
\begin{proposition}\label{prop:expansion_discrete}
    Let $\cR \subset \mathbb{R}^d$ be a discrete set such that $|x-y| \ge c > 0$ for all distinct elements $x$ and $y$ of $\cR$.
    Suppose that Assumption \ref{assumption:a_higher_order_all} holds for some $0 < \rate \le 1$, and that $H_\delta$ defines a self-adjoint operator on $\ell^2 (\cR; \mathbb{C}^N)$. Then for any $\mathbb{N}_0 \ni m > d/2$, there exists a constant $C > 0$ such that
    \begin{align*}
        \|(\psi_\delta-\varphi_\delta)(t, \cdot)\|_{\ell^2(\cR; \mathbb{C}^N)} \leq C\delta^p(1 + \delta^{\rate} t) \left(1 + (\delta t)^{m+2p}\right)
    \end{align*}
    uniformly in $t \ge 0$ and $0 < \delta \le \delta_0$.
    If, moreover, $b_{0\ell} \equiv 0$ for all $\ell \in \cL_0$, then
    \begin{align*}
        \|(\psi_\delta-\varphi_\delta)(t, \cdot)\|_{\ell^2(\cR; \mathbb{C}^N)} \leq C\delta^p(1 + \delta^{\rate} t) \left(1 + (\delta t)^{p-1}\right)
    \end{align*}
    uniformly in $t \ge 0$ and $0 < \delta \le \delta_0$.
\end{proposition}
\begin{proof}
The first (resp. second) statement follows immediately from \eqref{eq:ic_bd_main}, \eqref{eq:res_bd_main} (resp. \eqref{eq:res_bd_flat}), and the first part of Lemma \ref{lemma:u_delta_reg_discrete}.
\end{proof}

We conclude this subsection with three remarks.
Recall that we refer to each subspace $\Ran P_\ell$ as a \emph{spectral sector of $a^0$}, or just \emph{sector} for short.
\begin{remark}[First-order simplification]\label{rem:p_equals_one}
    If $p=1$ and $\phi_0 \in H^\infty (\mathbb{R}^d; \Ran P_{\ell_0})$ for some $\ell_0 \in \{1, \dots, L\}$, then the function $\psi_\delta$ in \eqref{eq:psi_def_higher_order_all} simplifies considerably. In this case, $\phi_\ell \equiv 0$ for all $\ell \ne \ell_0$, while $\phi_{\ell_0} = \phi_{\ell_0}^0 + \delta \phi_{\ell_0}^1$, where $\phi^0_{\ell_0}$ is defined as the solution to
    \begin{align*}
        (D_T+\bH_{0\ell_0}) \phi^0_{\ell_0}(T,X;\delta)=0,\qquad \phi^0_{\ell_0}(0,X; \delta)=\phi_0(X),
    \end{align*}
    and
    \begin{align*}
        \phi_{\ell_0}^1 = \phout^{1,\ell_0} = -\inva_{\ell_0} \bH_0 \phi_{\ell_0}^0.
    \end{align*}
    The in-sector correctors $\phin^{1,\ell}$ vanish identically since $p=1$. The ansatz carries an $O(\delta)$ error in the initial condition, consistent with \eqref{eq:estimateHm_higher_order_all}. 
\end{remark}
\begin{remark}[Duhamel's formula]
    The definition \eqref{eq:def_phin} implies that for all $1 \le j \le p-1$,
    \begin{align}\label{eq:duhamel}
        \phin^{j,\ell} (T, \cdot \; ; \delta) = - e^{-iT\bH_{0\ell}} P_\ell \sum_{\ell'=1}^L \phout^{j,\ell'} (0, \cdot \; ; \delta) - i \int_0^T e^{-i(T-T')\bH_{0\ell}} P_\ell \left( \bH_0 \phout^{j,\ell} + \sum_{i=1}^j \bH_i \phi^{j-i}_\ell \right)(T', \cdot \; ; \delta) \, dT'.
    \end{align}
\end{remark}
\begin{remark}[Structure of the inter-band coupling]\label{rem:mechanisms}
    The construction \eqref{eq:phi_expansion_higher_order}--\eqref{eq:def_phin} involves no operator coupling between distinct sectors. For each $\ell$, the function $\phin^{j,\ell}$ solves an initial value problem governed by $\bH_{0\ell}$. Inter-band effects enter through exactly three mechanisms.
    \begin{itemize}
        \item[(M1)] \emph{Algebraic admixture.} The corrector $\phout^{j,\ell}$ is determined pointwise in $T$ by \eqref{eq:def_phout}.
        \item[(M2)]\emph{Initial-data redistribution.} The initial condition $-P_\ell \sum_{\ell'} \phout^{j,\ell'} (0, \cdot \; ; \delta)$ prescribed in \eqref{eq:def_phin} is the only channel through which 
        different sectors interact. The resulting contribution to $\phin^{j,\ell}$ involves the free propagator $e^{-i T \bH_{0\ell}}$ applied to this nonzero data.
        \item[(M3)] \emph{Self-energy.} The source term $-P_\ell (\bH_0 \phout^{j,\ell} + \sum_{i \ge 1} \bH_i \phi^{j-i}_\ell)$ in \eqref{eq:def_phin} 
        depends on the functions $\phout^{k,\ell}$ for $k \le j$.
    \end{itemize}
\end{remark}

\subsection{Inter-band coupling and band-adapted initial data}\label{sec:dressed}
In this section, we describe an alternate ansatz which removes the inter-sector coupling mechanism (M2) of Remark \ref{rem:mechanisms}. A similar construction
was used in \cite{bal2024semiclassical} in the analysis of wave packets propagating along curved interfaces in the semiclassical regime. 

\begin{definition}[Single-sector construction]\label{def:single}
    Fix $p \in \mathbb{N}$, $\ell_0 \in \{1, \dots, L\}$ and $\phi_0 \in H^\infty (\mathbb{R}^d ; \Ran P_{\ell_0})$. Define $\check \phi^0_{\ell_0} = \cphin^{0,\ell_0}$ as the solution of \eqref{eq:def_phi_0_ell} with data $\phi_0$, and set $\check \phi^0_\ell \equiv 0$ for $\ell \neq \ell_0$ as well as $\cphout^{0,\ell} \equiv 0$ for all $\ell$. Set
    \begin{align}\label{eq:phi_expansion_higher_order_check}
    \check \phi := \sum_{\ell=1}^L \check \phi_\ell, \qquad \check\phi_\ell := \sum_{j=0}^p \delta^j \check \phi^j_\ell,
    \end{align}
    where the $\check \phi^j_\ell = \cphin^{j,\ell} + \cphout^{j,\ell}$ for each $1 \le \ell \le L$ are defined recursively over $j \ge 1$ as follows. 
    Set
    \begin{align}\label{eq:def_phout_check}
        \cphout^{j,\ell} &:= -\inva_\ell \left( D_T \check \phi^{j-1}_\ell + \sum_{i=0}^{j-1} \bH_i \check \phi^{j-1-i}_\ell\right), \qquad 1 \le j \le p,
    \end{align}
    and for all $1 \le j \le p-1$ define $\cphin^{j,\ell}$ as the solution to the initial value problem
    \begin{align}\label{eq:in_mod}
        (D_T + \bH_{0\ell}) \cphin^{j,\ell} &= -P_\ell\left(\bH_0 \cphout^{j,\ell} + \sum_{i=1}^j \bH_i \check \phi^{j-i}_\ell\right), \quad \cphin^{j,\ell} (0, \cdot\; ; \delta) \equiv 0.
    \end{align}
    Finally, set $\cphin^{p,\ell} \equiv 0$.
    Now, define
    \begin{align}\label{eq:psi_def_higher_order_all_check}
        \check \psi_\delta(t,x) = \sum_{\ell=1}^L e^{i(K\cdot x-E_\ell t)} \delta^{\frac d2} \check \phi_\ell (\delta t,\delta x;\delta)
    \end{align}
    and
    \begin{align}\label{eq:dressed}
        \phi^\delta_0 := \phi_0 + \sum_{j=1}^p \delta^j \cphout^{j,\ell_0} (0, \cdot \; ; \delta) =: \Pi^\delta_{\ell_0} \phi_0.
    \end{align}
\end{definition}
Note that the only difference between the above construction and \eqref{eq:phi_expansion_higher_order}-\eqref{eq:psi_def_higher_order_all} 
comes from the initial conditions in \eqref{eq:def_phin} and \eqref{eq:in_mod}.

The right-hand side of \eqref{eq:dressed} is well defined. Indeed, $\cphout^{1,\ell_0}(0, \cdot \; ; \delta)$ is determined by $\check\phi^0_{\ell_0} (0, \cdot \; ; \delta) = \phi_0$, and inductively $\cphout^{j,\ell_0}(0, \cdot \; ; \delta)$ is determined by the $\check \phi^i_{\ell_0}(0, \cdot \; ; \delta)$ and $D_T \check \phi^i_{\ell_0} (0, \cdot \; ; \delta)$ for $i < j$, all of which are computed from \eqref{eq:def_phi_0_ell} and \eqref{eq:in_mod}. To first order, using $D_T \check \phi^0_{\ell_0} = - \bH_{0\ell_0} \check \phi^0_{\ell_0}$ together with $\ash_{\ell_0} P_{\ell_0} = 0$, we obtain
\begin{align}\label{eq:Pi_dressed}
    \Pi^\delta_{\ell_0} = P_{\ell_0} - \delta \, \ash_{\ell_0} \bH_0 P_{\ell_0} + O(\delta^2).
\end{align}
The operator $\Pi^\delta_{\ell_0}$ maps $\Ran P_{\ell_0}$ into $\mathbb C^N$ and satisfies
$P_{\ell_0}\Pi^\delta_{\ell_0} = P_{\ell_0}$ on $\Ran P_{\ell_0}$, since the correctors
$\cphout^{j,\ell_0}$ take values in $\Ran(I-P_{\ell_0})$.

\begin{proposition}[Band-adapted data]\label{prop:dressed}
    Let Assumption \ref{assumption:a_higher_order_all} hold, and let $\check \phi, \check \phi_\ell, \check \psi_\delta, \phi^\delta_0$ be as in Definition \ref{def:single}. Then
    \begin{enumerate}
        \item $\check \phi_\ell \equiv 0$ for every $\ell \neq \ell_0$ (only the target sector is excited, at every order $j \le p$).
        \item $\check \phi (0, \cdot \; ; \delta) = \phi^\delta_0$ and $\norm{\phi^\delta_0 - \phi_0}_{H^s} \le C_s \delta$ for every $s \ge 0$.
        \item For every $1 \le j \le p-1$,
        \begin{align}\label{eq:duhamel_check}
            \cphin^{j,\ell_0} (T, \cdot \; ; \delta) = - i \int_0^T e^{-i(T-T')\bH_{0\ell_0}} P_{\ell_0} \left( \bH_0 \cphout^{j,\ell_0} + \sum_{i=1}^j \bH_i \check \phi^{j-i}_{\ell_0} \right)(T', \cdot \; ; \delta) \, dT'.
        \end{align}
        If $\bH_{0\ell_0} = 0$, then $\cphin^{j,\ell_0}$ is a finite sum of iterated time integrals of products of the operators 
        $\bH_i$ (for $1 \le i \le j$) applied to $\check \phi^0_{\ell_0} \equiv P_{\ell_0}\phi_0$.
        \item If $\varphi_\delta$ solves \eqref{eq:microWP_higher_order} with initial condition $e^{iK\cdot x} \delta^{\frac d2} \phi^\delta_0 (\delta x)$ in place of $e^{iK \cdot x} \delta^{\frac d2} \phi_0 (\delta x)$, 
        then
        \begin{align}\label{eq:improved}
            \norm{(\check \psi_\delta - \varphi_\delta)(t,\cdot)}_{H^m (\mathbb{R}^d)} \le C \delta^{p+\rate} t \left( 1 + (\delta t)^{2m+2p}\right), \qquad t \ge 0, \quad 0 < \delta \le \delta_0.
        \end{align}
        That is, the term $C \delta^p$ of \eqref{eq:estimateHm_higher_order_all} is absent.
    \end{enumerate}
\end{proposition}
\begin{proof}
    (1) Fix $\ell \neq \ell_0$. We show that $\check \phi^j_\ell \equiv 0$ by induction on $j$. For $j = 0$ this holds by construction. Assume $\check \phi^i_\ell \equiv 0$ for all $i \le j-1$. Then \eqref{eq:def_phout} gives $\cphout^{j,\ell} = -\ashl (D_T \check \phi^{j-1}_\ell + \sum_{i=0}^{j-1} \bH_i \check \phi^{j-1-i}_\ell) = 0$, whence the right-hand side of \eqref{eq:in_mod} vanishes. Since the data vanishes as well and $\bH_{0\ell}$ is self-adjoint, $\cphin^{j,\ell} \equiv 0$ by uniqueness. Thus $\check \phi^j_\ell = \cphin^{j,\ell} + \cphout^{j,\ell} \equiv 0$.

    (2) By \eqref{eq:in_mod} and part 1 of this proposition,
    \begin{align*}
        \check \phi (0) = \check \phi^0_{\ell_0}(0) + \sum_{j \ge 1} \delta^j \left( \cphin^{j,\ell_0}(0) + \cphout^{j,\ell_0}(0)\right) = \check \phi_0 + \sum_{j\ge1} \delta^j \cphout^{j,\ell_0}(0) = \phi^\delta_0 .
    \end{align*}
    The bound on $\phi^\delta_0 - \phi_0$ follows from Lemma \ref{lemma:Sobolev_higher_order_all} (which naturally extends to the setting of Definition \ref{def:single}) evaluated at $T=0$.

    (3) The first part is immediate from Duhamel's formula \eqref{eq:duhamel} applied to vanishing initial data.
    When $\bH_{0\ell_0}=0$, the propagator $e^{-iT\bH_{0\ell_0}}$ is the identity and $\check \phi^0_{\ell_0} \equiv P_{\ell_0} \phi_0$ is constant in $T$. By induction, the formula \eqref{eq:duhamel_check} thus expresses $\cphin^{j,\ell_0}$ as a finite sum of iterated integrals of products of the $\bH_i$ acting on $\check\phi^0_{\ell_0}$. 

    (4) The proof of Theorem \ref{thm:main_higher_order} applies verbatim, except that $\norm{u_\delta (0,\cdot)}_{H^m} = 0$ by item 2 of this proposition, so that the term $(1 + (\delta t)^m) \norm{u_\delta(0,\cdot)}_{H^m}$ of Lemma \ref{lemma:u_delta_reg} is absent.
\end{proof}

\subsubsection*{Arbitrary accuracy of the band-adapted construction}

For each sector $\ell$ and each initial condition $\phi_0 \in H^\infty (\mathbb{R}^d ; \Ran P_\ell)$, Definition \ref{def:single} produces a wave packet
\begin{align}\label{eq:Psi_single}
    \Psi_\delta [\ell, \phi_0]
\end{align}
of the form \eqref{eq:psi_def_higher_order_all_check}, which depends linearly on $\phi_0$.
By Proposition \ref{prop:dressed}, this wave packet excites no sector other than $\ell$, satisfies the 
error estimate \eqref{eq:improved}, and has the initial value
\begin{align}\label{eq:Psi_data}
    \Psi_\delta[\ell,\phi_0] (0,x) = e^{iK\cdot x}\delta^{\frac d2}\bigl( \Pi^\delta_\ell \phi_0 \bigr)(\delta x). 
\end{align}
The discrepancy $\Pi^\delta_\ell \phi_0 - \phi_0 = O(\delta)$ in the initial condition can be treated as new data, to which the same construction applies. One can iterate this procedure to obtain an approximation to the initial condition of arbitrary accuracy.
We now make this precise.

\begin{definition}[Dressing map]\label{def:dressing}
    With $\Pi^\delta_\ell$ as in \eqref{eq:dressed}, set
    \begin{align}\label{eq:Tdelta}
        \mathcal T_\delta := \sum_{\ell=1}^L \Pi^\delta_\ell P_\ell = I_N + \delta \mathcal A_\delta,
        \qquad
        \mathcal A_\delta := \frac1\delta \sum_{\ell=1}^L \bigl( \Pi^\delta_\ell - P_\ell\bigr)P_\ell.
    \end{align}
    For $q \in \mathbb{N}$, let $\mathcal T^{(q)}_\delta := \sum_{k=0}^{q-1}(-\delta \mathcal A_\delta)^k$, so that
    \begin{align}\label{eq:Tq}
        \mathcal T_\delta \,\mathcal T^{(q)}_\delta = I_N - \bigl( -\delta\mathcal A_\delta\bigr)^q .
    \end{align}
\end{definition}

By \eqref{eq:Pi_dressed}, $\mathcal A_\delta = -\sum_\ell \ashl \bH_0 P_\ell + O(\delta)$. Each $\Pi^\delta_\ell - P_\ell$ is a finite sum of compositions of the operators $\bH_i$, $\bH_{0\ell}$ and the constant matrices $\ashl$. By Lemma \ref{lem:Pi_S}, each $\Pi^\delta_\ell$ loses at most $p$ derivatives, so that
\begin{align}\label{eq:A_bounded}
    \norm{\mathcal A_\delta}_{H^{s+p}\to H^s} \le C_s, \qquad 0 < \delta \le \delta_0,
\end{align}
uniformly in $\delta$. The loss of $p$ derivatives is harmless for the smooth data we consider, but it is the reason $\mathcal T_\delta$ is inverted by a truncated Neumann series rather than by a bounded inverse on a fixed Sobolev space.

\begin{theorem}[Band-adapted expansion of arbitrary accuracy]\label{thm:equiv}
    Suppose Assumption \ref{assumption:a_higher_order_all} holds, fix $q \in \mathbb{N}$, let $\phi_0 \in H^\infty (\mathbb{R}^d ; \mathbb{C}^N)$, and set
    \begin{align}\label{eq:psi_ad}
        \psi^{\rm ad}_\delta := \sum_{\ell=1}^L \Psi_\delta \bigl[ \ell, \ P_\ell \mathcal T^{(q)}_\delta \phi_0\bigr].
    \end{align}
    Then, with $m \in \mathbb{N}_0$ and $\varphi_\delta$ the solution of \eqref{eq:microWP_higher_order},
    \begin{align}
        \psi^{\rm ad}_\delta (0,x) &= e^{iK\cdot x}\delta^{\frac d2}\Bigl( \phi_0 - (-\delta \mathcal A_\delta)^q \phi_0\Bigr)(\delta x),
        \label{eq:ad_data}\\
        \norm{(D_t + H_\delta)\psi^{\rm ad}_\delta (t,\cdot)}_{H^m} &\le C \delta^{p+\rate}\bigl( 1 + (\delta t)^{m+2p}\bigr),
        \label{eq:ad_res}\\
        \norm{(\psi^{\rm ad}_\delta - \varphi_\delta)(t,\cdot)}_{H^m (\mathbb{R}^d)} &\le C \bigl( \delta^q + \delta^{p+\rate}t\bigr)\bigl( 1 + (\delta t)^{2m+2p}\bigr),
        \qquad t \ge 0, \quad 0<\delta\le\delta_0 .
        \label{eq:ad_error}
    \end{align}
\end{theorem}
In particular $q=p$ reproduces the accuracy \eqref{eq:estimateHm_higher_order_all} of Theorem \ref{thm:main_higher_order}, while $q>p$ renders the initial-data error negligible compared with the residual of the equation. Moreover each sector $\ell$ evolves under $\bH_{0\ell}$ alone, with no redistribution of data at any order. The inter-band coupling is implemented only by the algebraic operator $\mathcal T^{(q)}_\delta$ acting on the data, and by the correctors $\phout^{j,\ell}$.
\begin{proof}
    Set $A := -\delta \mathcal A_\delta$ and $g_k := A^{k-1}\phi_0$ for $1 \le k \le q$. By the linearity of $\chi \mapsto \Psi_\delta[\ell,\chi]$,
    \begin{align}\label{eq:iteration}
        \psi^{\rm ad}_\delta = \sum_{k=1}^q \ \sum_{\ell=1}^L \Psi_\delta[\ell, P_\ell g_k].
    \end{align}
    By \eqref{eq:Psi_data} and \eqref{eq:Tdelta}, the data for $\sum_\ell \Psi_\delta[\ell,P_\ell g_k]$ is $\mathcal T_\delta g_k = (I_N-A)A^{k-1}\phi_0$. Summing the telescoping series,
    \begin{align*}
        \sum_{k=1}^q (I_N-A)A^{k-1}\phi_0 = (I_N-A)\sum_{k=0}^{q-1}A^k \phi_0 = (I_N - A^q)\phi_0,
    \end{align*}
    we obtain \eqref{eq:ad_data}. 
    The bound \eqref{eq:ad_res} holds because each of the $qL$ terms in \eqref{eq:psi_ad} satisfies the estimate of Proposition \ref{lemma:res_higher_order_all}, with a constant controlled by finitely many Sobolev norms of $g_k$. 
    By \eqref{eq:A_bounded} and $\phi_0 \in H^\infty$, all Sobolev norms of $g_k$ are bounded uniformly in $\delta$. 
    Finally \eqref{eq:ad_error} follows from Lemma \ref{lemma:u_delta_reg}, applicable by Lemma \ref{lemma:commutator_bd}, exactly as in the proof of Theorem \ref{thm:main_higher_order}.
\end{proof}
Note that \eqref{eq:iteration} is the iteration described above Definition \ref{def:dressing}. The function $\sum_\ell \Psi_\delta[\ell, P_\ell g_1]$ is the band-adapted ansatz for the data $\phi_0$. Its initial-data error is $-Ag_1 = g_2$, which is fed back in as a new data, and so on.

\begin{remark}[Comparing the two constructions]
    Note that if $q = p$ in \eqref{eq:psi_ad}, 
    Theorems \ref{thm:main_higher_order} and \ref{thm:equiv} both produce an $O(\delta^p + \delta^{p+\rate}t)$ approximation of the tight-binding solution $\varphi_\delta$ from \eqref{eq:microWP_higher_order}, up to times $t \lesssim \delta^{-1}$. Thus the functions $\psi_\delta$ and $\psi_\delta^{\rm ad}$ agree to $O(\delta^{p+\rate-1})$ on this timescale. The constructions of these functions differ only in the initial conditions imposed on the $\phin^{j,\ell}$ and $\cphin^{j,\ell}$; recall \eqref{eq:def_phin} and \eqref{eq:in_mod}. Indeed, provided the symbols $\{b_j: 0 \le j \le p-1\}$ in \eqref{eq:decomposition_higher_order_all} are all independent of $\delta$, 
    each function $\phin^{j,\ell}$ satisfies a $\delta$-independent (but generally nonzero) initial condition. On the other hand, in the band-adapted expansion \eqref{eq:psi_ad} the leading terms $\cphin^{0,\ell}$ are launched from the $\delta$-dependent data $P_\ell\mathcal T^{(q)}_\delta\phi_0$, while the $\cphin^{j,\ell}$ for $j \ge 1$ identically vanish at $T = 0$.
\end{remark}

\subsection{Effective Hamiltonian}\label{sec:effective}
The previous subsections established the accuracy of two related multiple-scale asymptotic expansions of the solution to the tight-binding Schr\"odinger equation \eqref{eq:microWP_higher_order}. In this subsection, we approximate the tight-binding dynamics by a \emph{reduced} macroscopic Schr\"odinger model, whose Hamiltonian is defined on only one sector. The high-order accuracy of our reduced model is established by Theorem \ref{prop:rH} and Corollary \ref{cor:rH_eps}. 

Let $\chi \in C^\infty_c (\mathbb{R}^d)$ be real-valued and such that $\chi$ is identically one in a neighborhood of the origin. We now introduce a second asymptotic parameter, $0 < \eps \ll 1$, which will control the frequency cutoffs and is chosen such that
\begin{align}\label{eq:def_eta}
    \eta := \delta /\eps^2 \ll 1.
\end{align}
The cutoff function is $\chi_\eps (\zeta) := \chi (\eps \zeta)$. Define $b_j^\eps (X, \zeta) := \chi_\eps (\zeta) b_j (X, \zeta) \chi_\eps (\zeta)$ with $b_j$ as in Assumption \ref{assumption:a_higher_order_all}, and define $\bH_j^\eps := \ow b_j^\eps$. It follows that $\bH_j^\eps$ is bounded on any $H^m$ with 
\begin{align}\label{eq:bH_bd}
    \norm{\bH_j^\eps}_{H^m \to H^m} \le C \eps^{-(j+1)}
\end{align}
uniformly in $(\delta, \eps)$.
For the following construction, the sector $\ell$ is fixed and dropped from the notation. Let $$S : \Ran P \to \Ran (I-P)$$ be a bounded operator depending on $\delta$ and $\eps$, 
such that 
\begin{align}\label{eq:norm_S}
    \norm{S}_{H^m \to H^m} \le C \eps \eta.
\end{align}
This operator will be constructed explicitly below, in Definition \ref{def:S}. Note that $\eps\eta = \delta/\eps$. Throughout, we assume $0<\delta\le c_0\,\eps$ and $\eps\le1$, where $c_0\in(0,1)$ is chosen, for the finitely many Sobolev indices $m$ under consideration, such that \eqref{eq:norm_S} gives $\norm{S}_{H^m\to H^m}\le\frac12$ and $\norm{S^*}_{H^m\to H^m}\le\frac12$ (the bound \eqref{eq:norm_S} holds for $S^*$ as well, since $S^*$ is built from the adjoints of the same operators). Define
\begin{align*}
    \cG := P + S^* S.
\end{align*}
It is evident that $\cG \succeq 1$ on $\Ran P$, meaning that $\cG^{-1} :\Ran P \to \Ran P$ is well defined and positive.
Square roots of these operators are therefore well defined. Set
\begin{align}\label{eq:bH}
    \bHe := \sum_{j=0}^{p-1} \delta^j \bH_j^\eps, 
\end{align}
with the dependence of $\bHe$ on $\delta$ implied. With $\bH_{0\ell} := P \bH_0 P$ as before and $\bH_{0\ell}^\eps := P \bH_0^\eps P$, define
\begin{align}\label{eq:def_rH}
    \rH := \bH_{0\ell} + \cG^{-1/2} \left(
   (P+S^*) \bHe (P+S)+ \delta^{-1}S^* (a^0 - E) S \right) \cG^{-1/2} - \bH_{0\ell}^\eps.
\end{align}
This means that $\rH$ is formally given by
\begin{align*}
    \cG^{-1/2} \left(
   (P+S^*) \bHe (P+S)+ \delta^{-1}S^* (a^0 - E) S \right) \cG^{-1/2},
\end{align*}
where the leading-order term $P \bH_0^\eps P = \bH_{0\ell}^\eps$ in the expansion of $\cG^{-1/2}(P+S^*) \bHe (P+S)\cG^{-1/2}$ is replaced by the un-truncated operator $\bH_{0\ell}$.
The effective Hamiltonian $\rH$ is evidently symmetric. 
Recalling \eqref{eq:def_eta}, \eqref{eq:bH_bd}, and \eqref{eq:norm_S}, the perturbation $\rH - \bH_{0\ell}$ is bounded on any $H^m$ with norm
\begin{align}\label{eq:rH_norm_bd}
    \norm{\rH - \bH_{0\ell}}_{H^m \to H^m} \le C \eta.
\end{align}
Moreover, $$\cG^{-1/2} = P - \frac{1}{2} S^* S + O((\eps \eta)^4) = P + O((\eps \eta)^2).$$ On $L^2$ this holds for $S$ of any size, by the spectral theorem, since $\cG\succeq1$ and $|(1+x)^{-1/2}-1+\frac x2|\le\frac38x^2$ for $x\ge0$. On $H^m$ it follows from the binomial series, which converges since $\norm{S^*S}_{H^m\to H^m}\le\frac14$. Hence the leading-order term in the expansion of $\rH - \bH_{0\ell}$ is
\begin{align*}
    \delta P \bH_{1}^\eps P + S^* \bH_{0}^\eps P + P\bH_{0}^\eps S + \delta^{-1} S^* (a^0 - E) S, 
\end{align*}
with the operator norm of each term bounded by $C\eta$.
Every term of \eqref{eq:def_rH} is composed on both sides with $P$ or with $\cG^{-1/2}$, so that $\rH$ maps functions with values in $\Ran P$ to functions with values in $\Ran P$. We regard $\rH$ as an operator on $L^2 (\mathbb{R}^d; \Ran P)$; in particular $[\rH, P] = 0$. By \eqref{eq:rH_norm_bd}, Lemma \ref{lemma:sa_H0l} and the Kato--Rellich theorem, $\rH$ is self-adjoint with domain of definition
\begin{align*}
    \cD (\rH) = \begin{cases}
    L^2 (\mathbb{R}^d; \Ran P), & \ell \in \cL_0,\\
        H^1 (\mathbb{R}^d; \Ran P), & \ell \in \cL_1,
    \end{cases}
\end{align*}
where we recall the definitions of $\cL_0$ and $\cL_1$ in \eqref{eq:cL_def}.
In the case that $\ell \in \cL_1$, for any $\eta_0 > 0$ there exist constants $0 < c < C$ such that
\begin{align}\label{eq:equivalent_norms_rH}
    c \norm{u}_{H^m} \le \norm{(i+\rH)^m u}_{L^2} \le C \norm{u}_{H^m}, \qquad u \in H^m (\mathbb{R}^d; \Ran P),
\end{align}
uniformly in $(\delta, \eps)$ with $\eta \le \eta_0$, the constants depending on $\eta_0$. Indeed, the upper bound follows from \eqref{eq:rH_norm_bd} and \eqref{eq:equivalent_norms_higher_order}. For the lower bound we argue by induction on $m$, the case $m=0$ being trivial. With $V := \rH - \bH_{0\ell}$, \eqref{eq:equivalent_norms_higher_order} and \eqref{eq:rH_norm_bd} give
\begin{align*}
    \norm{u}_{H^m} \le C\norm{(i+\bH_{0\ell})u}_{H^{m-1}} \le C\bigl(\norm{(i+\rH)u}_{H^{m-1}} + C\eta_0\norm{u}_{H^{m-1}}\bigr).
\end{align*}
By the induction hypothesis, applied to $(i+\rH)u$ and to $u$, and since $\norm{(i+\rH)^{-1}}_{L^2\to L^2}\le1$, both terms on the right-hand side are bounded by $C\norm{(i+\rH)^m u}_{L^2}$.
For $\phi_0 \in H^\infty(\mathbb{R}^d;\Ran P)$, we can therefore define
\begin{align}\label{eq:psi_def_higher_order_rH}
    \psi (t,x) := e^{i(K\cdot x-E t)}\delta^{d/2}
        \Bigl((P+S) \cG^{-1/2}\, e^{-i\delta t\,\rH}\,\mathcal G^{1/2}\phi_0\Bigr)(\delta x).
\end{align}
\begin{definition}\label{def:S}
Let us now precisely define $S$. We construct this operator by
\begin{align*}
    S = \sum_{j=1}^p \delta^j S_j,
\end{align*}
where each $S_j$ depends on $\eps$ and is bounded uniformly in $\delta$.
Setting
\begin{align}\label{eq:def_B}
    B = a^0 - E + \delta \bHe - \delta S \cG^{-1} (P+S^*) \bHe - S \cG^{-1} S^* (a^0 - E)+\delta^{p+\rate} \bH_p,
\end{align}
we define the $S_j$ recursively such that all terms of $O(\delta^j)$, $j \le p$, in
\begin{align}\label{eq:riccati_S}
    (I-P)B(P+S)
\end{align}
vanish (the term $\delta^{p+\rate}\bH_p$ is not eliminated and is treated as a remainder). At order $\delta^j$ this is an equation of the form $(a^0-E)S_j = F_j$, where $F_j$ takes values in $\Ran(I-P)$ and depends only on $S_1,\dots,S_{j-1}$, so that $S_j = a^\sharp F_j$. Since $(I-P-S^*) = (I-S^*)(I-P)$, the same terms then also vanish in $(I-P-S^*)B(P+S)$. Note that this requires a finite number of terms of the expansion
\begin{align*}
    \cG^{-1} \sim P\sum_{j=0}^{\infty} (-S^* S)^j.
\end{align*}
The leading-order terms in $S$ are
\begin{align}\label{eq:leading_order_S}
        S_1 = -a^\sharp \bH_0^\eps P, \qquad S_2 = -a^\sharp (\bH_1^\eps P + \bH_0^\eps S_1 - S_1 P \bH_0^\eps P).
    \end{align}
It is evident that $\norm{S_j}_{H^m \to H^m} \le C\eps^{-j}$ for every $m$, so that $\norm{S}_{H^m \to H^m} \le C \eps \eta$ as desired. We emphasize that all operators in the above expansions are bounded, and so the $S_j$ are all well defined. The reason for defining $S$ in this way will become apparent in the proof of Proposition \ref{prop:res_rH}.
We denote by $S^0 = \sum_{j=1}^p \delta^j S^0_j$ the operator obtained by the same construction with every $\bH_i^\eps$ replaced by $\bH_i$. The $S^0_j$ do not depend on $\eps$, and are in general unbounded with $S^0_j : H^{s+j} \to H^s$ bounded uniformly in $\delta$.
\end{definition}
With $H_\delta$ as in \eqref{eq:H_delta}, let $\varphi$ be the ($\delta$-dependent) solution to the following initial value problem,
    \begin{equation}\label{eq:microWP_higher_order_rH}
      (D_t+H_\delta)\varphi =0,\qquad \varphi(0,x)= e^{iK \cdot x} \delta^{d/2} (P+S) \phi_0 (\delta x).
    \end{equation}
Note the discrepancy in initial conditions between \eqref{eq:microWP_higher_order} and \eqref{eq:microWP_higher_order_rH}.
\begin{theorem}\label{prop:rH}
    Let $\phi_0 \in H^\infty (\mathbb{R}^d; \Ran P)$ and $m,n \in \mathbb{N}$. Then for any $\eta_0 > 0$, there exists $C = C(m,n,\eta_0)$ such that, for all $0 < \delta \le c_0\eps$, $\eps \le 1$, with $\eta \le \eta_0$ and all $t \ge 0$,
    \begin{align*}
        \norm{\psi (t, \cdot) - \varphi (t, \cdot)}_{H^m} \le
        \begin{cases}
        C (\delta^{p+\rate} + \delta \eps^n)\, t\, (1+(\delta t)^{m}), & \ell \in \cL_1,\\
        C (\delta^{p+\rate} + \delta \eps^n)\, t\, (1+(\delta t)^{2m+n+p}), & \ell \in \cL_0.
        \end{cases}
    \end{align*}
\end{theorem}
\begin{proof}
    Since $(\psi - \varphi) \vert_{T=0} \equiv 0$,
    the desired bounds follow from Proposition \ref{prop:res_rH} and Lemma \ref{lemma:u_delta_reg}, with $R_\delta = C(\delta^{p+\rate}+\delta\eps^n)$ and
    with $M = 0$ if $\ell \in \cL_1$ and $M = m+n+p$ if $\ell \in \cL_0$.
\end{proof}
\begin{remark}\label{rem:all_sectors}
    The above result establishes the validity of the macroscopic approximation for the initial condition
    \begin{align*}
        \varphi (0, x) = e^{iK \cdot x} \delta^{d/2} (P+S) \phi_0 (\delta x).
    \end{align*}
    Since our construction applies for every sector, by linearity we obtain an approximation of the tight-binding dynamics with the initial condition
    \begin{align*}
        \varphi (0, x) = e^{iK \cdot x} \delta^{d/2} \Bigl(I+\sum_\ell S^\ell P_\ell\Bigr) \phi_0 (\delta x),
    \end{align*}
    where $S^\ell$ denotes the operator $S$ associated with sector $\ell$, since $\sum_\ell (P_\ell + S^\ell)P_\ell = I + \sum_\ell S^\ell P_\ell$. Each $S^\ell$ is bounded and of $O(\eps \eta)$, so that the operator $I+\sum_\ell S^\ell P_\ell$ is invertible on $H^m$ by a Neumann series, after decreasing $c_0$ if necessary. Therefore, replacing $\phi_0$ by $(I+\sum_\ell S^\ell P_\ell)^{-1} \phi_0$ and applying Theorem \ref{prop:rH} to each sector, we recover a high-order approximation of the tight-binding dynamics for the original initial condition
    \begin{align*}
        \varphi (0, x) = e^{iK \cdot x} \delta^{d/2} \phi_0 (\delta x)
    \end{align*}
    from \eqref{eq:microWP_higher_order}.
\end{remark}
\begin{remark}\label{rem:rH_expansion}
Note that by \eqref{eq:leading_order_S}, the leading-order terms of $\rH$ can be calculated explicitly from \eqref{eq:def_rH}. For fixed $\eps$, we have the formal expansion $\rH \sim \sum_{j=0}^\infty \delta^j \rH_j$, where, with $\Re X := \frac12(X+X^*)$,
\begin{align}\label{eq:rH_leading_order}
\begin{split}
    \rH_0 &= P\bH_{0}P,\\
    \rH_1 &= P \bH_{1}^\eps P - P \bH^\eps_0 a^\sharp \bH^\eps_0 P,\\
    \rH_2 &= P \bH_2^\eps P - 2 \Re (P \bH_1^\eps a^\sharp \bH_0^\eps P) + P \bH_0^\eps a^\sharp \bH_0^\eps a^\sharp \bH_0^\eps P - \Re (P \bH_0^\eps P\bH_0^\eps (a^\sharp)^2 \bH_0^\eps P).
\end{split}
\end{align}
More generally, for $j \ge 1$ we have $\rH_j = P \bH_j^\eps P + \Sigma^{(j-1)}_\ell$, where we will refer to the operators $\Sigma^{(j-1)}_\ell$ as ``self-energies''.

Theorem \ref{prop:rH} still applies if $\rH$ in \eqref{eq:psi_def_higher_order_rH} is replaced by $\tilde \rH := \sum_{j=0}^{p-1}\delta^j \rH_j$. Indeed, $\tilde\rH$ maps $L^2(\mathbb{R}^d;\Ran P)$ into itself and $\tilde\rH - \bH_{0\ell}$ satisfies \eqref{eq:rH_norm_bd}, so that Lemma \ref{lemma:Sobolev_rH} still applies. In the proof of Proposition \ref{prop:res_rH}, the replacement only adds to $\rho_0$ the term $\delta (P+S)\cG^{-1/2}(\rH - \tilde\rH) e^{-iT\tilde\rH}\cG^{1/2}\phi_0$, which is bounded in $H^m$ by $C\delta^{p+1}\norm{e^{-iT\tilde\rH}\cG^{1/2}\phi_0}_{H^{m+p+1}}$. Indeed, a term of order $\delta^k$, $k\ge p$, in the expansion of $\rH$ is a composition of cut-off operators $\bH^\eps_i$ of total order at most $k+1$, so that by \eqref{eq:bH_bd} it maps $H^{m+p+1}$ to $H^m$ with norm at most $C\eps^{p-k}$ (recall that $\delta^{k+1}\eps^{p-k}\le\delta^{p+1}$ since $\delta<\eps$).

Moreover, a frequency cutoff applied to a bounded operator can be removed (at the price of a residual bounded by $C\delta\eps^n$ times a Sobolev norm of $\phi$, of the type already present in Proposition \ref{prop:res_rH}). For example, if $P b_0 (I-P) \in S(1)$, then all instances of $P \bH_0^\eps a^\sharp$ in \eqref{eq:leading_order_S} and \eqref{eq:rH_leading_order} can be replaced by $P \bH_0 a^\sharp$. If the operators $\tilde \rH$ and $S$ can be chosen without any frequency cutoffs (meaning that they are independent of $\eps$, with $S$ and $\rH_j$ bounded for all $1 \le j \le p-1$), then the estimates of Theorem \ref{prop:rH} become
\begin{align*}
        \norm{\psi (t, \cdot) - \varphi (t, \cdot)}_{H^m} \le
        \begin{cases}
        C \delta^{p+\rate} t(1+(\delta t)^{m}), & \ell \in \cL_1,\\
        C \delta^{p+\rate} t(1+(\delta t)^{2m+p+1}), & \ell \in \cL_0.
        \end{cases}
\end{align*}
Note that if $p \ge 2$ and $(I-P) b_0 (I-P)$ is unbounded, then $S_2^0$ of Definition \ref{def:S} contains the unbounded term $a^\sharp \bH_0 a^\sharp \bH_0 P$. For instance, for the coupled SSH chains of Section \ref{sec:summary_chains}, $S^0_2 = (2E)^{-2}(\bH_{02}\beta^* - \beta^*\bH_{01})$ in the sector $\ell = 1$. 
\end{remark}
We now state a consequence of Theorem \ref{prop:rH}, in which the parameters $\eps$ and $\eta$ depend explicitly on $\delta$.
\begin{corollary}\label{cor:rH_eps}
    Let $\phi_0 \in H^\infty (\mathbb{R}^d; \Ran P)$, $m \in \mathbb{N}_0$ and $0 < \theta < 1/2$. Choose $\eps := \delta^\theta$ so that $\eta = \delta^{1-2\theta}$, and fix an integer $n \ge (p+\rate-1)/\theta$. Then there exist $\delta_1>0$ and $C > 0$, depending also on $\theta$ and $n$, such that, for all $0 < \delta \le \delta_1$ and $t \ge 0$,
    \begin{align*}
        \norm{\psi (t, \cdot) - \varphi (t, \cdot)}_{H^m} \le
        \begin{cases}
        C \delta^{p+\rate}\, t\,(1+(\delta t)^{m}), & \ell \in \cL_1,\\
        C \delta^{p+\rate}\, t\,(1+(\delta t)^{2m+n+p}), & \ell \in \cL_0.
        \end{cases}
    \end{align*}
    The same bounds hold if $\rH$ in \eqref{eq:psi_def_higher_order_rH} is replaced by $\tilde\rH$ of Remark \ref{rem:rH_expansion}. If, moreover, $\varphi$ is replaced by the solution $\varphi^0$ of \eqref{eq:microWP_higher_order_rH} with the uncut initial condition $e^{iK\cdot x}\delta^{d/2}((P+S^0)\phi_0)(\delta x)$, the bounds hold with the additional term $C\delta^{p+\rate}(1+(\delta t)^m)$. The same is true if, in addition, $S$ is replaced by $S^0$ in the prefactor $(P+S)$ of \eqref{eq:psi_def_higher_order_rH}, with the additional term $C\delta^{p+\rate}$ if $\ell\in\cL_1$ and $C\delta^{p+\rate}(1+(\delta t)^{m+n+p})$ if $\ell\in\cL_0$.
\end{corollary}
\begin{proof}
    With this choice of parameters, $\delta\eps^n = \delta^{1+n\theta} \le \delta^{p+\rate}$, $\eta \le 1$ and $\delta/\eps = \delta^{1-\theta} \le c_0$ for $$\delta\le\delta_1 := \min(1,\delta_0,c_0^{1/(1-\theta)}),$$ so that the first two statements follow from Theorem \ref{prop:rH} and Remark \ref{rem:rH_expansion}. For the third statement, observe that $S_j - S^0_j$ is a finite sum of compositions in which at least one factor is $\bH_i^\eps - \bH_i$, so that $$\norm{(S-S^0)\phi_0}_{H^m} \le C\delta\eps^n \norm{\phi_0}_{H^{m+n+p}} \le C \delta^{p+\rate}.$$ Note that the first inequality follows from the same arguments used to obtain \eqref{eq:cutoff_bd} below. The desired bound then follows from \eqref{eq:apriori_micro} applied to $\varphi - \varphi^0$. For the prefactor, the same argument gives $$\norm{(S-S^0)w}_{H^m}\le C\delta\eps^n\norm{w}_{H^{m+n+p}},$$ with $w = \cG^{-1/2}e^{-i\delta t\rH}\cG^{1/2}\phi_0$ and $\norm{w}_{H^{m+n+p}}$ bounded by Lemma \ref{lemma:Sobolev_rH}.
\end{proof}
By the following lemma, $(P+S^0)\phi_0$ is precisely the band-adapted initial condition $\Pi^\delta_\ell\phi_0$ of Definition \ref{def:single}, so that Corollary \ref{cor:rH_eps} applies to band-adapted data.

\begin{lemma}\label{lem:Pi_S}
    For every $1 \le \ell \le L$ and $p \in \mathbb{N}$, the operator $\Pi^\delta_\ell$ of \eqref{eq:dressed} satisfies
    \begin{align*}
        \Pi^\delta_\ell = P_\ell + S^0,
    \end{align*}
    with $S^0$ the uncut operator of Definition \ref{def:S}.
\end{lemma}
\begin{proof}
    We treat the explicit powers of $\delta$ in \eqref{eq:decomposition_higher_order_all} as a formal parameter $\lambda$, keeping the symbols $b_j$ fixed, and write $\equiv$ for equality modulo $\lambda^{p+1}$. Let $Q := I-P$, $A := a^0 - E + \sum_{j=0}^{p-1}\lambda^{j+1}\bH_j$, 
    and $\cG^0 := P + (S^0)^*S^0$.

    \emph{Step 1.} The defining property \eqref{eq:riccati_S} of $S^0$ in Definition \ref{def:S} reads $(I-P) B^0 (P+S^0) \equiv 0$, where
    \begin{align*}
        B^0 = A - \lambda S^0 (\cG^0)^{-1} (P+(S^0)^*) \sum_{j=0}^{p-1}\lambda^j\bH_j - S^0 (\cG^0)^{-1} (S^0)^* (a^0 - E)
    \end{align*}
    is the operator $B$ of \eqref{eq:def_B} with $S$ replaced by $S^0$, the cut-off operators $\bH^\eps_j$ replaced by $\bH_j$, the remainder $\delta^{p+\rate}\bH_p$ omitted, and $\delta$ replaced by $\lambda$. Since $(a^0-E)P = 0$,
    it follows that $$R_{\rm new} := QX - S^0(\cG^0)^{-1}(P+(S^0)^*)X \equiv 0,$$ with $X := A(P+S^0)$. Writing $R := QX - S^0PX$, one checks that $(\cG^0)^{-1}(P+(S^0)^*)X - PX = (\cG^0)^{-1}(S^0)^*R$, so that $R_{\rm new} = (I - S^0(\cG^0)^{-1}(S^0)^*)R$. Since $S^0 = O(\lambda)$, the operator $I - S^0(\cG^0)^{-1}(S^0)^*$ is invertible as a formal series, and therefore
    \begin{align}\label{eq:riccati_formal}
        QA(P+S^0) \equiv S^0PA(P+S^0).
    \end{align}

    \emph{Step 2.} Let $u := \sum_{j=0}^{p-1}\lambda^j \cphin^{j,\ell}$ and $v := \sum_{j=1}^p\lambda^j\cphout^{j,\ell}$, so that $\check\phi_\ell = u+v$, with $u = Pu$ and $v = Qv$. By \eqref{eq:def_phout_check} and \eqref{eq:in_mod}, the coefficients of $\lambda^j$, $j \le p$, of $(A+\lambda D_T)(u+v)$ vanish. Projecting with $P$ and $Q$,
    \begin{align*}
        \lambda D_T u \equiv -PA(u+v), \qquad QA(u+v) + \lambda D_T v \equiv 0.
    \end{align*}

    \emph{Step 3.} Let $h := v - S^0u = Qh$. Using \eqref{eq:riccati_formal} applied to $u$, and $\lambda D_T S^0 u = S^0\lambda D_T u \equiv -S^0PA(u+v)$, we obtain
    \begin{align*}
        0 \equiv QA(u+v) + \lambda D_T v \equiv S^0PA(u+S^0u) + QAd - S^0PA(u+v) + \lambda D_Td = QAd - S^0PAd + \lambda D_Td.
    \end{align*}
    Here $QAd = (a^0-E)h + O(\lambda)h$, while $S^0PAd$ and $\lambda D_T h$ carry a factor $\lambda$. Applying $\ashl$, the coefficient of $\lambda^j$ in $h$ is thus determined by its coefficients of lower order. Since $h = O(\lambda)$, induction gives $h \equiv 0$.

    \emph{Step 4.} Since $\cphin^{j,\ell}(0,\cdot) = 0$ for $j \ge 1$, we have $u(0,\cdot) = \phi_0$, so that $\Pi^\delta_\ell\phi_0 = \phi_0 + v(0,\cdot) \equiv (P+S^0)\phi_0$. Both sides are polynomials of degree at most $p$ in $\lambda$, hence they are equal.
\end{proof}

We now improve the long-time estimates of Theorem \ref{prop:rH} and Corollary \ref{cor:rH_eps}, by slowing the growth of $e^{-iT\rH}\phi_0$ when $\rH_0$ (cf. Remark \ref{rem:rH_expansion}) vanishes (Proposition \ref{prop:rH_flat}) and by avoiding the commutator factor $(\delta t)^m$ of Lemma \ref{lemma:u_delta_reg} on a lattice (Proposition \ref{prop:effective_discrete}).
\begin{proposition}\label{prop:rH_flat}
    Take $\phi_0, m, n, \eta_0, c_0$ as in Theorem \ref{prop:rH}, and let $\ell \in \cL_0$. 
    Suppose 
    there exists a function $\nu = \nu (\delta, \eps, \eta)$ such that, for every $k \in \mathbb{N}_0$,
    \begin{align}\label{eq:nu_bound}
        \norm{\rH}_{L^2 \to L^2} + \norm{[\rH, \aver{D_X}]}_{H^k \to H^k} \le C_k\, \nu
    \end{align}
    uniformly in $0 < \delta \le c_0 \eps$, $\eps \le 1$, and $\eta \le \eta_0$. Then
    \begin{align*}
        \norm{\psi (t, \cdot) - \varphi (t, \cdot)}_{H^m} \le
        C (\delta^{p+\rate} + \delta \eps^n)\, t\, (1+ (\delta t)^m)(1+(\delta \nu t)^{m+n+p}), \qquad t \ge 0, 
    \end{align*}
    uniformly in all parameters.
\end{proposition}
\begin{proof}
    Since $(\psi - \varphi) \vert_{T=0} \equiv 0$, the desired bounds follow from Proposition \ref{prop:res_rH_flat} and Lemma \ref{lemma:u_delta_reg_flat}, with $R_\delta = C(\delta^{p+\rate}+\delta\eps^n)$ and $M = m+n+p$.
\end{proof}
\begin{corollary}\label{cor:rH_eps_flat}
    Under the hypotheses and definitions of Corollary \ref{cor:rH_eps}, let $\ell \in \cL_0$ and assume there exists a function $\nu = \nu (\delta)$ such that \eqref{eq:nu_bound} holds uniformly in $\delta > 0$. Then
    \begin{align*}
        \norm{\psi (t, \cdot) - \varphi (t, \cdot)}_{H^m} \le C \delta^{p+\rate} t (1 + (\delta t)^m)(1 + (\delta \nu t)^{m+n+p}), \qquad t \ge 0, \quad 0 < \delta \le \delta_1.
    \end{align*}
\end{corollary}
\begin{proof}
    Fix $0 < \theta < 1/2$, and take $\eps := \delta^\theta$ and $n \ge (p+\rate - 1)/\theta$ in Proposition \ref{prop:rH_flat}.
\end{proof}
We conclude this section by adapting Corollaries \ref{cor:rH_eps} and \ref{cor:rH_eps_flat} to the discrete setting. In the interest of brevity, we do not state the analogues of Theorem \ref{prop:rH} and Proposition \ref{prop:rH_flat}.
\begin{proposition}\label{prop:effective_discrete}
Let $\cR \subset \mathbb{R}^d$ be a discrete set such that $|x-y| \ge c > 0$ for all distinct points $x$ and $y$ in $\cR$.
    Under the hypotheses and definitions of Corollary \ref{cor:rH_eps} with $m>d/2$, assume that $H_\delta$ defines a self-adjoint operator on $\ell^2 (\cR; \mathbb{C}^N)$. Then
    \begin{align*}
        \norm{\psi (t, \cdot) - \varphi (t, \cdot)}_{\ell^2 (\cR; \mathbb{C}^N)} \le
        \begin{cases}
        C \delta^{p+\rate}\, t, & \ell \in \cL_1,\\
        C \delta^{p+\rate}\, t\,(1+(\delta t)^{m+n+p}), & \ell \in \cL_0.
        \end{cases}
    \end{align*}
    If, in addition, $\ell \in \cL_0$ and there exists a function $\nu = \nu (\delta)$ such that \eqref{eq:nu_bound} holds uniformly in $\delta > 0$, then
    \begin{align*}
        \norm{\psi (t, \cdot) - \varphi (t, \cdot)}_{\ell^2 (\cR; \mathbb{C}^N)} \le C \delta^{p+\rate} t (1 + (\delta \nu t)^{m+n+p}), \qquad t \ge 0, \quad 0 < \delta \le \delta_1.
    \end{align*}
\end{proposition}
\begin{proof}
    Proposition \ref{prop:res_rH} gives the estimate
    \begin{align*}
        \norm{(D_t + H_\delta)(\psi (t, \cdot) - \varphi (t, \cdot))}_{H^m} \le
        \begin{cases}
        C (\delta^{p+\rate} + \delta \eps^n), & \ell \in \cL_1,\\
        C (\delta^{p+\rate} + \delta \eps^n) (1+(\delta t)^{m+n+p}), & \ell \in \cL_0.
        \end{cases}
    \end{align*}
    Recalling that $\psi - \varphi$ vanishes at $t = 0$, Lemma \ref{lemma:u_delta_reg_discrete} implies
    \begin{align*}
        \norm{\psi (t, \cdot) - \varphi (t, \cdot)}_{\ell^2} \le \begin{cases}
        C (\delta^{p+\rate} + \delta \eps^n)t, & \ell \in \cL_1,\\
        C (\delta^{p+\rate} + \delta \eps^n) t(1+(\delta t)^{m+n+p}), & \ell \in \cL_0.
        \end{cases}
    \end{align*}
    With $0 < \theta < 1/2$ fixed, our first statement then follows from choosing $\eps := \delta^\theta$ and $n \ge (p+\rate - 1)/\theta$.

    The proof of our second statement is similar.
\end{proof}


\section{Applications}\label{sec:summary}

We now illustrate the theory of Section \ref{sec:tb} on four models: two coupled SSH chains,
an SSH chain coupled to a flat band, the $6\times 6$ tight-binding model of graphene with a
Kekul\'e-O modulation, and a $9\times9$ model of a magnetized cold plasma. The first two are
the simplest settings in which the mechanisms (M1)--(M3) of Remark \ref{rem:mechanisms} are
visible in closed form. The third combines a Dirac sector with two asymptotically flat
sectors. The fourth is a continuous rather than tight-binding model, whose symbol is
unbounded in $\xi$. In each case we record the constant matrix $a^0$ and its spectral
decomposition, the class ($\cL_0$ or $\cL_1$, cf. \eqref{eq:cL_def}) of each sector, the
sector Hamiltonians $\bH_{0\ell}$, the first nontrivial terms of the effective Hamiltonians
\eqref{eq:def_rH}, and the approximation of the microscopic dynamics that follows from
Theorems \ref{thm:main_higher_order} and \ref{prop:rH}. All formulas needed to state these
results are given below. Their derivations, which are direct but lengthy computations, are given in
Sections \ref{sec:applications}, \ref{sec:kekule} and \ref{sec:plasma} of the supplementary
material.

\subsection{Coupled one-dimensional chains}\label{sec:summary_chains}

\paragraph{Two coupled SSH chains.}
Consider two one-dimensional lattices of lattice constant $v>0$, each described by an
SSH-type symbol
\begin{align}\label{eq:sum_ssh_ajj}
    a_{jj}(X,\xi;\delta) = (-1)^{j-1}E I_2 - 2\alpha_j(X)\sin(v\xi/2)\,\sigma_2 + \delta M_j(X),
    \qquad j\in\{1,2\},
\end{align}
with $\sigma_2$ the second Pauli matrix. Here $E>0$, the
$\alpha_j \in C^\infty_b(\mathbb{R})$ satisfy $\alpha_j\ge c>0$, and the
$M_j\in C^\infty_b(\mathbb{R};\C^{2\times2})$ are Hermitian valued. The two chains are
coupled through the off-diagonal block $\delta\beta(X)$, with
$\beta\in C^\infty_b(\mathbb{R};\C^{2\times2})$, so that the full symbol is
\begin{align}\label{eq:sum_ssh_a}
    a(X,\xi;\delta) = \begin{bmatrix} a_{11}(X,\xi;\delta) & \delta\beta(X)\\ \delta\beta^*(X) & a_{22}(X,\xi;\delta)\end{bmatrix} \in \C^{4\times4}.
\end{align}
Expanding about $K=0$ gives $\rate = 1$ and
\begin{align}\label{eq:sum_ssh_a0}
    a^0 = \begin{bmatrix} E I_2 & 0\\ 0 & -E I_2\end{bmatrix}, \qquad
    b_0(X,\zeta) = \begin{bmatrix} b_{01}(X,\zeta) & \beta(X)\\ \beta^*(X) & b_{02}(X,\zeta)\end{bmatrix},
    \qquad b_{0\ell}(X,\zeta) = -v\alpha_\ell(X)\zeta\sigma_2 + M_\ell(X).
\end{align}
Thus $E_1 = E$ and $E_2 = -E$, and both sectors belong to $\cL_1$. Each $\bH_{0\ell} = \ow b_{0\ell}$ is a one-dimensional Dirac operator with Fermi velocity
$v\alpha_\ell$ and mass $M_\ell$. Since the hopping term in \eqref{eq:sum_ssh_ajj} is odd in
$\xi$, one has $b_1 = 0$. 
We therefore find that the effective Hamiltonian \eqref{eq:def_rH} for each sector satisfies
\begin{align}\label{eq:sum_ssh_Heff}
    \rH^{(1)} = \bH_{01} + \frac{\delta}{2E}\,\beta\beta^* + O(\delta^2), \qquad
    \rH^{(2)} = \bH_{02} - \frac{\delta}{2E}\,\beta^*\beta + O(\delta^2).
\end{align}
Recalling Remark \ref{rem:rH_expansion}, we emphasize that no frequency cutoff is required in these first-order terms since the operator $\ow (\beta)$ is bounded.
At second order, since $\ashift_1 = -\frac{1}{2E}P_2$ and $\ashift_2 = \frac{1}{2E}P_1$, Remark
\ref{rem:rH_expansion} gives
\begin{align}\label{eq:sum_ssh_H2}
    \rH^{(\ell)}_2 = P_\ell\bH^\eps_2P_\ell + \frac{1}{4E^2}\Bigl(\beta_\ell\,\bH^\eps_{0\ell'}\,\beta_\ell^* - \Re\bigl(\beta_\ell\beta_\ell^*\,\bH^\eps_{0\ell}\bigr)\Bigr),
    \qquad \beta_1 := \beta, \quad \beta_2 := \beta^*,
\end{align}
where $\ell'\ne\ell$ denotes the other chain and $P_\ell\bH_2^\eps P_\ell$ is the cubic correction to the dispersion of the chain $\ell$.
By Theorem \ref{prop:rH} and Corollary \ref{cor:rH_eps} with $p=2$ and $\rate = 1$, the truncations at order $\delta$ of the above operators $\rH^{(\ell)}$ approximate the tight-binding dynamics started from the band-adapted data $\Pi^\delta_\ell\phi_0$ (Lemma \ref{lem:Pi_S}) with an error $O(\delta^3 (1+t)(1+(\delta t)^m))$ in $H^m$, for all $t \ge 0$.

For the prescribed data $\phi_0 = (\tilde\phi_{01},\tilde\phi_{02})\in H^\infty$ in
\eqref{eq:microWP_higher_order}, the construction of Section \ref{subsec:macroscopic} with
$p=1$ gives
\begin{align}\label{eq:sum_ssh_psi}
    \psi_\delta(t,x) = \delta^{1/2}\left( e^{-iEt}\begin{bmatrix} \tilde\phi^0_1 \\ \frac{\delta}{2E}\beta^*\tilde\phi^0_1\end{bmatrix}
    + e^{iEt}\begin{bmatrix} -\frac{\delta}{2E}\beta\tilde\phi^0_2 \\ \tilde\phi^0_2\end{bmatrix}\right)(\delta t,\delta x),
    \qquad (D_T + \bH_{0\ell})\tilde\phi^0_\ell = 0, \quad \tilde\phi^0_\ell\vert_{T=0} = \tilde\phi_{0\ell},
\end{align}
and 
Proposition \ref{prop:main_flat} yields
$$\norm{(\psi_\delta - \varphi_\delta)(t,\cdot)}_{H^m} \le C(\delta + \delta^2 t)(1+(\delta t)^{m}).$$
Each packet thus acquires an $O(\delta)$ component on the other chain.
At $p = 2$, the redistribution rule \eqref{eq:def_phin} assigns
the data $\phin^{1,1}(0,\cdot) = \frac{1}{2E}\beta\tilde\phi_{02}$ and
$\phin^{1,2}(0,\cdot) = -\frac{1}{2E}\beta^*\tilde\phi_{01}$ to the in-sector correctors.
Part of the data prescribed on one chain is thereby transferred to a wave
packet propagating on the other, with its own group velocity and fast phase. Alternatively, the band-adapted expansion of Theorem \ref{thm:equiv}
launches each chain from the dressed data
\begin{align}\label{eq:sum_ssh_dressed}
    \tilde\phi_{01} + \frac{\delta}{2E}\beta\tilde\phi_{02}, \qquad
    \tilde\phi_{02} - \frac{\delta}{2E}\beta^*\tilde\phi_{01},
\end{align}
after which each chain evolves under its own $\bH_{0\ell}$ with no redistribution of data. The $O(\delta)$ terms of \eqref{eq:sum_ssh_dressed} agree with the redistributed data $\delta\phin^{1,\ell} (0, \cdot)$ above, as stipulated by Theorem \ref{thm:equiv}.

\paragraph{An SSH chain coupled to a flat band.}
Replacing the second chain by a single dispersionless orbital per cell, we consider
\begin{align}\label{eq:sum_flat_a}
    a(X,\xi;\delta) = \begin{bmatrix} a_{11}(X,\xi;\delta) & \delta\gamma(X)\\ \delta\gamma^*(X) & 0\end{bmatrix}\in\C^{3\times3},
    \qquad \gamma\in C^\infty_b(\mathbb{R};\C^2),
\end{align}
with $a_{11}$ as in \eqref{eq:sum_ssh_ajj}. Now $a^0 = \diag(EI_2, 0)$, so that $E_1 = E$ with
$1\in\cL_1$ and $\bH_{01}$ as above, while $E_2 = 0$ with $b_{02} = 0$, so that $2 \in \cL_0$.
Flat sectors of this kind fall outside the scope of the effective mass theorems of
\cite{allaire2005homogenization}, which require a nondegenerate critical point of the band
function, and of the Peierls substitution of \cite{nenciu1991dynamics, panati2003effective},
which requires an isolated band with nonvanishing dispersion. 
Here
$b_1 = 0$, $P_2b_2P_2 = 0$ and $\ashift_2 = \frac1E P_1$, so that every term of Remark
\ref{rem:rH_expansion} containing $\bH_{02} = 0$ vanishes. The effective Hamiltonian on the flat sector is
\begin{align}\label{eq:sum_flat_Heff}
    \rH^{(2)} = -\frac{\delta}{E}\,|\gamma|^2 + \frac{\delta^2}{E^2}\,\gamma^*\,\bH^\eps_{01}\,\gamma + O(\delta^3),
\end{align}
where $\bH^\eps_{01} = P_1\bH^\eps_0P_1$ is the frequency-truncated SSH Hamiltonian of Section \ref{sec:effective}.
The flat band thus acquires a real frequency shift $-|\gamma|^2/E$ at order $\delta$, and at order $\delta^2$ a dispersion inherited from the SSH band. For $p = 2$, only the term $-\frac{\delta}{E}|\gamma|^2$ enters, and Theorem \ref{prop:rH} (case $\ell \in \cL_0$) with Corollary \ref{cor:rH_eps} gives an $O(\delta^3 t)$ approximation of the tight-binding dynamics for times $t \lesssim \delta^{-1}$. The term of order $\delta^2$ enters for $p = 3$.

For the initial condition $\phi_0 = (0,\phi_{00})$ supported in the flat sector, the
construction of Section \ref{subsec:macroscopic} with $p=2$ gives
$\phi_2 = \phi_2^{\rm in} + \phi_2^{\rm out}$, where
\begin{align*}
    \phi_2^{\rm in} &:= \phin^{0,2} + \delta \phin^{1,2} = \phi_{00} + \delta\,\frac{iT}{E}|\gamma|^2\phi_{00}, \\
    \phi_2^{\rm out} &:= \delta \phout^{1,2} + \delta^2 \phout^{2,2} = -\frac{\delta}{E} \phi_{00}\gamma +\frac{\delta^2}{E^2} \left(
        \bH_{01} [\phi_{00} \gamma] - iT |\gamma|^2  \phi_{00}\gamma \right).
\end{align*}
Note that for $T$ bounded, we have $\phi_2^{\rm in} = e^{-i\delta T\Sigma_2}\phi_{00} + O(\delta^2)$, with $\Sigma_2 = -\frac1E|\gamma|^2$.

\subsection{Graphene with Kekul\'e-O modulation}\label{sec:summary_kekule}

A Kekul\'e-O modulation of the nearest-neighbor hoppings of graphene triples the unit cell
and folds the two valleys $K$ and $K'$ onto the $\Gamma$ point; see
\cite{andrade2025topical} for the derivation of the corresponding tight-binding model and
\cite{yang2022topological} for its use in topological lasing cavities. We take the lattice
constant and the hopping amplitude equal to one, with nearest-neighbor vectors
$a_1 = (0,1)^T$, $a_2 = R_{2\pi/3}a_1$, $a_3 = R_{-2\pi/3}a_1$, and Kekul\'e wave vector
$G = \frac{4\pi}{3\sqrt3}(1,0)^T$. In the basis $(\fA_1,\fA_2,\fA_3,\fB_1,\fB_2,\fB_3)$ of
sublattice $\times$ folded-valley degrees of freedom, the symbol is the $6\times6$ Hermitian matrix
\begin{align}\label{eq:sum_kekule_symbol}
    a(X,\xi;\delta) = \begin{pmatrix} 0 & A\\ A^* & 0\end{pmatrix}, \qquad
    A = \begin{pmatrix} \kdisp_0 & 0 & 0 \\ 0 & \kdisp_{-1} & 0 \\ 0 & 0 & \kdisp_1 \end{pmatrix}
    + \delta\begin{pmatrix}
      0 & \De \kdisp_1 & \De^* \kdisp_{-1}\\
      \De^* \kdisp_1 & 0 & \De \kdisp_0\\
      \De \kdisp_{-1} & \De^* \kdisp_0 & 0
    \end{pmatrix},
\end{align}
built from the folded dispersion functions $\kdisp_n(\xi) = \kdisp(\xi+nG)$, $n\in\{-1,0,1\}$,
with $\kdisp(\xi) = \sum_{j=1}^3 e^{i\xi\cdot a_j}$, and the complex Kekul\'e order parameter
$\De(X) = \De \in C^\infty_b(\mathbb{R}^2;\C)$. We expand about the folded Dirac point $\xi = 0$,
writing $A(X,\delta\ze) = \sum_{j\ge0}\delta^jA_j(X,\ze)$ with $\ze = \ze_1+i\ze_2$. Then
$\rate = 1$ and $b_j = \begin{psmallmatrix} 0 & A_{j+1}\\ A_{j+1}^* & 0\end{psmallmatrix}$, where
\begin{align}\label{eq:sum_kekule_A01}
\begin{split}
    A_0 &= \begin{pmatrix} 3&0&0\\0&0&0\\0&0&0 \end{pmatrix}, \qquad
    A_1(X,\ze) = \begin{pmatrix} 0 & 0 & 0 \\ 0 & \tfrac{3}{2}\ze & 3\De(X) \\ 0 & 3\De^*(X) & -\tfrac{3}{2}\bar\ze \end{pmatrix},\\ 
    A_2(X,\zeta) &= \begin{pmatrix}
    -\tfrac{3}{4}|\ze|^2 & -\tfrac{3}{2}\De(X)\bar\ze & \tfrac{3}{2}\De^*(X)\ze \\
    -\tfrac{3}{2}\De^*(X)\bar\ze & \tfrac{3}{8}\bar\ze^2 & 0 \\
    \tfrac{3}{2}\De(X)\ze & 0 & \tfrac{3}{8}\ze^2
  \end{pmatrix}, \\
  A_3(X,\zeta) &= \begin{pmatrix}
    \tfrac{1}{16}(\ze^3 - \bar\ze^3) & \tfrac{3}{8}\De(X)\ze^2 & \tfrac{3}{8}\De^*(X)\bar\ze^2 \\
    \tfrac{3}{8}\De^*(X)\ze^2 & -\tfrac{3}{16}\ze|\ze|^2 & -\tfrac{3}{4}\De(X)|\ze|^2 \\
    \tfrac{3}{8}\De(X)\bar\ze^2 & -\tfrac{3}{4}\De^*(X)|\ze|^2 & \tfrac{3}{16}\bar\ze|\ze|^2
  \end{pmatrix}.
  \end{split}
\end{align}

\paragraph{Spectral data.} The matrix
$a^0 = \begin{psmallmatrix} 0 & A_0\\ A_0 & 0\end{psmallmatrix}$, with $A_0 = \diag(3,0,0)$,
has the eigenvalues
\begin{align}\label{eq:sum_kekule_spec}
    E_+ = 3 \ (\text{rank } 1), \qquad E_0 = 0\ (\text{rank } 4), \qquad E_- = -3\ (\text{rank } 1),
\end{align}
with $P_\pm = \uu_\pm\uu_\pm^*$, $\uu_\pm = \frac{1}{\sqrt2}(e_1,\pm e_1)^T$, and
$P_0 = \diag(0,1,1,0,1,1)$. The sector corresponding to $E_0$ carries the two valleys on both sublattices,
while $E_\pm$ come from the folded $\Gamma$ point. Since $A_1$ has a vanishing first row and
column, $P_\pm b_0 = b_0 P_\pm = 0$. Hence $\pm\in\cL_0$ with
$b_{0,\pm} = 0$, and the inter-sector coupling at order one vanishes identically.

\paragraph{The Dirac sector.} In the basis $(\fA_2,\fA_3,\fB_2,\fB_3)$ of $\Ran P_0$, the
effective Hamiltonian is
\begin{align}\label{eq:sum_kekule_H0}
    \rH^{(0)} = \ow b_{0,0} + \sum_{j=1}^{2}\delta^j\,\ow \bigl(\chi_\eps^2\, b_{j,0}\bigr) + O(\delta^3), \qquad b_{j,0} := P_0 b_j P_0,
\end{align}
where, with $\ze = \ze_1 + i\ze_2$ and $\sig_0 = I_2$,
\begin{align}\label{eq:sum_kekule_b}
\begin{split}
    b_{0,0}(X,\ze) &= \tfrac32\bigl(\ze_1\,\sig_1\otimes\sig_3 - \ze_2\,\sig_2\otimes\sig_0\bigr)
    + 3\bigl(\Re\De(X)\,\sig_1\otimes\sig_1 - \Im\De(X)\,\sig_1\otimes\sig_2\bigr),\\
    b_{1,0}(\ze) &= \tfrac38 \bigl(\Re(\ze^2)\,\sig_1\otimes\sig_0 + \Im(\ze^2)\,\sig_2\otimes\sig_3\bigr),\\
    b_{2,0}(X,\ze) &= -\tfrac{3}{16}|\ze|^2\bigl(\ze_1\,\sig_1\otimes\sig_3 - \ze_2\,\sig_2\otimes\sig_0\bigr)
    - \tfrac34|\ze|^2\bigl(\Re\De(X)\,\sig_1\otimes\sig_1 - \Im\De(X)\,\sig_1\otimes\sig_2\bigr).
\end{split}
\end{align}
The leading term $\ow b_{0,0}$ is a
Dirac operator with Kekul\'e mass. Since
$b_{0,0}^2 = (\tfrac94|\ze|^2 + 9|\De|^2)I_4$, the symbol $b_{0,0}$ is elliptic in the sense of
\eqref{eq:ellipticity}, so that $0\in\cL_1$. Note that $b_{1,0} \in S^2$ and $b_{2,0} \in S^3$.
No self-energy contributes through order $\delta^2$, since $b_0\ashift_0 = \ashift_0 b_0 = 0$.

\paragraph{The $\Gamma$-point sectors.} The rank-one sectors $E_\pm$ are asymptotically flat,
with
\begin{align}\label{eq:sum_kekule_Hpm}
    \rH^{(\pm)} = \delta\,\ow\bigl(\mp\tfrac34 \chi_\eps (\zeta)^2 |\ze|^2\bigr) + O(\delta^3),
\end{align}
that is, quadratic dispersions with effective masses $m^*_\pm = \mp\frac23$.

\paragraph{Inter-sector coupling.} The coupling between the Dirac sector and the
$\Gamma$-point sectors is carried by the covectors $\Ga_\pm := \uu_\pm^* b_1 P_0$, which in the
basis $(\fA_2,\fA_3,\fB_2,\fB_3)$ of $\Ran P_0$ read
\begin{align}\label{eq:sum_kekule_Gamma}
    \Ga_\pm(X,\ze) = \frac{3}{2\sqrt2}\Bigl(\mp\De(X)\,\ze,\ \pm\De^*(X)\,\bar\ze,\ -\De(X)\,\bar\ze,\ \De^*(X)\,\ze\Bigr).
\end{align}
These functions are linear in $\ze$ and proportional to $\De$, with $|\Ga_\pm|^2 = \frac92|\De|^2|\ze|^2$.
The coupling therefore requires both a nonzero Kekul\'e modulation and a nonzero macroscopic
momentum. Because the leading coupling is carried by $b_1$ rather than $b_0$, the operators
$S_1$ of \eqref{eq:leading_order_S} vanish for all three sectors, and the coupling first appears in
the effective Hamiltonians at order $\delta^3$, through the self-energies
$\Sigma^{(2)}_\ell := -P_\ell\bH_1\ashl\bH_1P_\ell$,
\begin{align}\label{eq:sum_kekule_Sigma}
    \Sigma^{(2)}_0 = \tfrac13\Bigl(\ow(\Ga_-)^*\ow(\Ga_-)-\ow(\Ga_+)^*\ow(\Ga_+)\Bigr), \qquad
    \Sigma^{(2)}_\pm = \pm \tfrac13\, \ow (\Ga_\pm)\, \ow (\Ga_\pm)^*.
\end{align}
In the wave function this coupling appears earlier, at order
$\delta^2$, through the algebraic corrector
\begin{align}\label{eq:sum_kekule_out2}
    \phout^{2,0} = -\ashift_0\bH_1\phi^0_0 = -\tfrac13\,\uu_+\,\ow(\Ga_+)\phi^0_0 + \tfrac13\,\uu_-\,\ow(\Ga_-)\phi^0_0,
    \qquad \ashift_0 = \tfrac13(P_+-P_-).
\end{align}
For expansions with $p\ge3$, the coupling covectors $\Gamma_\pm$ appear also in the data
\begin{align}\label{eq:sum_kekule_seed}
    \phin^{2,\pm}(0,\cdot) = -P_\pm\phout^{2,0}(0,\cdot) = \pm\tfrac13\,\ow(\Ga_\pm)\,\phi_0
\end{align}
transferred to the $\Gamma$-point sectors by the redistribution rule \eqref{eq:def_phin}. The latter are nearly dispersionless packets carrying the fast phases
$e^{\mp3it}$, which separate from the Dirac packet.

\paragraph{Approximation of the tight-binding dynamics.} For data
$\phi_0\in H^\infty(\mathbb{R}^2;\Ran P_0)$ in the Dirac sector and $p = 2$, there is no
redistribution of data at order $\delta$, since $(I_6-P_0)b_0P_0 = 0$. Thus the construction of
Section \ref{subsec:macroscopic} reduces to
\begin{align}\label{eq:sum_kekule_psi}
\begin{split}
    \psi_\delta(t,x) &= \delta\bigl(\phi^0_0 + \delta\phin^{1,0} + \delta^2\phout^{2,0}\bigr)(\delta t,\delta x),\\
    (D_T + \ow b_{0,0})\phi^0_0 &= 0, \quad \phi^0_0(0,\cdot) = \phi_0, \qquad
    (D_T + \ow b_{0,0})\phin^{1,0} = -\ow (b_{1,0})\,\phi^0_0, \quad \phin^{1,0}(0,\cdot) = 0,
\end{split}
\end{align}
with $\phout^{2,0}$ given by \eqref{eq:sum_kekule_out2}.
By Theorem \ref{thm:main_higher_order}, this provides an $O(\delta^2)$ approximation of the
tight-binding dynamics for $t\lesssim\delta^{-1}$. Since $0\in\cL_1$ and $S_1 = -\ashift_0\bH_0P_0 = 0$, the band-adapted data is
$\Pi^\delta_0\phi_0 = \phi_0+\delta^2S^0_2\phi_0$ with $S^0_2\phi_0 = \phout^{2,0}(0,\cdot)$ by Lemma \ref{lem:Pi_S}. For this initial condition, Theorem \ref{prop:rH} and Corollary \ref{cor:rH_eps} with $p=2$ and $\tilde\rH = \ow b_{0,0}+\delta\ow (\chi_\eps^2 b_{1,0})$
give an error $O(\delta^3(1+t)(1+(\delta t)^m))$ in $H^m$, for all $t \ge 0$.
Data that excite the sectors $\pm\in\cL_0$ are covered by the case $\ell\in\cL_0$ of Theorem \ref{prop:rH} (and Proposition \ref{prop:rH_flat} with $\nu= \eta$ there).

\paragraph{The Kekul\'e vortex.}\label{sec:summary_kekule_vortex}
The case of principal interest for applications is a vortex profile
$\De(X) = \De_0(R)e^{i\vartheta}$ in polar coordinates $X = R(\cos\vartheta,\sin\vartheta)$, with
$\De_0(0) = 0$ and $\De_0$ bounded away from zero at infinity, so that $\De$ has winding number
one. The leading term $\ow b_{0,0}$ of \eqref{eq:sum_kekule_H0} is then a Dirac operator with a
mass vortex (the Kekul\'e mass coupling the two valleys) whose spectrum contains a topologically protected zero mode localized near the
vortex core by the index computation of Jackiw and Rossi \cite{jackiw1981zero}. 
As shown in \cite{hou2007electron}, the zero mode depends only on $R$ and is given explicitly, up to normalization and in the basis $(\fA_2,\fA_3,\fB_2,\fB_3)$ of $\Ran P_0$, by
\begin{align}\label{eq:zero_mode}
    \phi_0 = \begin{bmatrix}
        \vec{0} \\ \phi_0^\fB
    \end{bmatrix} \in \mathbb{C}^4, \qquad \phi_0^\fB = \begin{bmatrix}
        \phi_0^{\fB_2} \\ -i \phi_0^{\fB_2}
    \end{bmatrix}, \qquad \phi_0^{\fB_2}(R) = e^{-2 \int_{0}^R \De_0 (R') dR'}.
\end{align}
Since the zero mode is stationary, it is a natural test case for the
approximation \eqref{eq:sum_kekule_psi}; see Section \ref{sec:numerics_kekule}. A rigorous spectral comparison between the
tight-binding and macroscopic Hamiltonians is left for future work.

\subsection{Magnetized cold plasmas}\label{sec:summary_plasma}

The last model is not of tight-binding type. Its symbol is a first-order differential symbol,
unbounded in $\xi$, so that Assumption \ref{assumption:a_higher_order_all} does not apply as
stated. We therefore begin by identifying the hypothesis that has to be modified.

\subsubsection{Continuous models: symbols unbounded in $\xi$}\label{sec:pde_class}

The requirement that $a \in S(1)$ be Hermitian-valued (Assumption \ref{assumption:a_higher_order_all}) implies by Lemma \ref{lemma:CV} that $H_\delta$ is a bounded self-adjoint operator on $L^2 (\mathbb{R}^d; \mathbb{C}^N)$. (Recall that $H_\delta$ is defined in terms of $a$ by \eqref{eq:a_delta} and \eqref{eq:H_delta}.) The self-adjointness of $H_\delta$ is sufficient to ensure that the tight-binding evolution problem \eqref{eq:microWP_higher_order} is well posed. The only other result that requires the uniform boundedness of derivatives of $a$ is Lemma \ref{lemma:commutator_bd}. (This lemma allows Lemma \ref{lemma:u_delta_reg} to be applied in the proof of Theorem \ref{thm:main_higher_order}.)
This suggests the following variant of Assumption \ref{assumption:a_higher_order_all}, which covers continuous models.
\begin{assumption}\label{assumption:pde}
    Let $a_\delta$ be as in \eqref{eq:a_delta}, where $a = a(X,\xi;\delta)$ is an $N \times N$ Hermitian-valued symbol in $S(\aver{\xi}^{m_0})$ uniformly in $\delta$, for some $m_0 \ge 0$. Assume that
    \begin{align}\label{eq:slow_part_bounded}
        \partial^\beta_X a \in S(1) \quad \text{uniformly in } \delta, \qquad \text{for every } \beta \in \mathbb{N}^d_0 \text{ with } |\beta| \ge 1,
    \end{align}
    and that $H_\delta = \ow a_\delta$ is self-adjoint on $L^2 (\mathbb{R}^d ; \mathbb{C}^N)$, with domain satisfying $H^{m_0}(\mathbb{R}^d; \mathbb{C}^N) \subset \mathcal D(H_\delta)$. Assume moreover that \eqref{eq:b_j_symbol_class}, \eqref{eq:decomposition_higher_order_all}, and the dichotomy for the $b_{0\ell}$ of Assumption \ref{assumption:a_higher_order_all} all hold.
\end{assumption}

Condition \eqref{eq:slow_part_bounded} says, roughly, that the part of the symbol that grows in $\xi$
has coefficients independent of the slow variable. It holds in particular when
$a = a_{\rm free}(\xi) + a_{\rm slow}(X,\xi;\delta)$ with $a_{\rm slow} \in S(1)$ (the converse is not true in general). This is
satisfied by the cold plasma symbol \eqref{eq:sum_plasma_symbol} below, whose only
$x$-dependence is through the bounded block $\omega_p (\delta x)$, and more generally by
Maxwell- and Dirac-type operators with slowly varying material coefficients.

\begin{proposition}\label{prop:pde_class}
    Under Assumption \ref{assumption:pde}, the results Lemma \ref{lemma:commutator_bd}, Proposition \ref{lemma:res_higher_order_all}, Lemma \ref{lemma:Sobolev_higher_order_all}, Theorem \ref{thm:main_higher_order}, Proposition \ref{prop:main_flat}, Proposition \ref{prop:dressed}, Theorem \ref{prop:rH}, Corollary \ref{cor:rH_eps}, Lemma \ref{lem:Pi_S}, Proposition \ref{prop:rH_flat}, and Corollary \ref{cor:rH_eps_flat} all hold without modification, as does the bound \eqref{eq:apriori_micro}.
\end{proposition}
\begin{proof}
    Lemma \ref{lemma:commutator_bd} was proved under the hypothesis \eqref{eq:slow_part_bounded}, so that $\norm{[\nabla_x, H_\delta]}_{H^m \to H^m} \le C \delta$ and Lemma \ref{lemma:u_delta_reg} applies. The problem \eqref{eq:microWP_higher_order} is well posed by the assumed self-adjointness of $H_\delta$, and \eqref{eq:apriori_micro} follows from Lemma \ref{lemma:u_delta_reg} applied to $u_\delta = \varphi_\delta$ with vanishing right-hand side. All remaining ingredients of the proofs of 
    results listed in this proposition are unchanged. Indeed, the identity \eqref{eq:HD} is valid for any symbol for which the Weyl quantization is defined, the decomposition of $\tilde H_\delta$ follows from \eqref{eq:decomposition_higher_order_all}, and the boundedness of $a^0$ on $H^s$ and of $\bH_j : H^{s+j+1}\to H^s$ follows from Lemma \ref{lemma:CV} and \eqref{eq:b_j_symbol_class}. In the proof of Theorem \ref{prop:rH}, the unboundedness of $H_\delta$ enters only through the well-posedness of \eqref{eq:microWP_higher_order_rH} and through Lemma \ref{lemma:u_delta_reg}. The frequency truncations, the estimates of $\bH_j - \bH_j^\eps$, and Lemma \ref{lemma:Sobolev_rH} involve only the symbols $b_j \in S(\aver{\zeta}^{j+1})$ and the sector Hamiltonians $\bH_{0\ell}$, which are the same as under Assumption \ref{assumption:a_higher_order_all}.
\end{proof}

\begin{remark}\label{rem:pde_class_gain}
    For the cold plasma model of Section \ref{sec:summary_plasma_model} one has $m_0 = 1$, and the estimate of Lemma \ref{lemma:commutator_bd} in fact holds with $C\delta^2$, since the $x$-dependent block of \eqref{eq:sum_plasma_affine} carries an explicit factor $\delta$. The decomposition \eqref{eq:decomposition_higher_order_all} holds there with $b_j = 0$ for $j \ge 1$ and any $p$, so that Theorem \ref{thm:main_higher_order} applies for every $p$. Thus Corollary \ref{cor:timescale} gives validity of the expansion up to $t \ll \delta^{-\kappa}$ for any $\kappa < 3/2$. Moreover, $\bH_p = 0$, so that for the sectors $\pm\omega_- \in \cL_1$, Theorem \ref{prop:rH} and Corollary \ref{cor:rH_eps} hold for every $p$ with $\delta^{p+\rate}$ replaced by $\delta^{p+1}$. For band-adapted data the effective dynamics are therefore accurate to $O(\delta^{p+1}(1+t))$ in $L^2$, for all $t \ge 0$ and every $p$.
\end{remark}

\subsubsection{The cold plasma model}\label{sec:summary_plasma_model}

A cold magnetized plasma, in two spatial dimensions and with a background magnetic field along
$\hat e_z$, is described by the electron fluid velocity $\mathbf v$, the electric field
$\mathbf E$ and the magnetic field $\mathbf B$. Setting
$\varphi = (\mathbf v,\mathbf E,\mathbf B)^T\in\C^9$, the linearized equations
\begin{align}\label{eq:sum_plasma_eqs}
  \partial_t\mathbf{v} = \Omega\hat{e}_z\times\mathbf{v} - \omega_p(\delta x)\mathbf{E},
  \qquad
  \partial_t\mathbf{E} = \omega_p(\delta x)\mathbf{v} + \nabla\times\mathbf{B},
  \qquad
  \partial_t\mathbf{B} = -\nabla\times\mathbf{E}
\end{align}
take the form $(D_t + H_\delta)\varphi = 0$, where $\Omega>0$ is the cyclotron frequency and
$\omega_p>0$ the slowly varying plasma frequency. (The more common form of these equations is
obtained by the relabelling $(\Omega,\mathbf E,\mathbf B)\mapsto(-\Omega,-\mathbf E,-\mathbf B)$.)
With an out-of-plane wavenumber $k_z\ne0$ fixed, $H_\delta = \ow a_\delta$ has the
$9\times9$ Hermitian symbol $a_\delta(x,\xi) = a(\delta x,\xi)$,
\begin{align}\label{eq:sum_plasma_symbol}
  a(X,\xi) = \begin{pmatrix}
    i\Omega\hat{e}_z\times & -i\omega_p(X) & 0 \\
    i\omega_p(X) & 0 & -\xi\times \\
    0 & \xi\times & 0
  \end{pmatrix},
  \qquad \xi=(\xi_1,\xi_2,k_z)^T,
\end{align}
where $\xi\times$ denotes the matrix of $v\mapsto\xi\times v$. Let $\omega_-\in(0,\Omega)$ be the
unique solution of
\begin{align}\label{eq:sum_def_omega_minus}
    k_z^2 = \omega_-^2\Bigl(1-\frac{\omega_-}{\omega_--\Omega}\Bigr) = \frac{\Omega\,\omega_-^2}{\Omega-\omega_-}.
\end{align}
This is the value of the plasma frequency at which the plasma (Langmuir) mode crosses the lower
left-hand polarized wave. We assume that
$\omega_p(X) = \omega_- + \delta\wt(X)$ with $\wt\in C^\infty_b(\mathbb{R}^2)$, and expand
about zero in-plane momentum. The symbol is then exactly affine in $\delta$,
\begin{align}\label{eq:sum_plasma_affine}
    a_\delta(x,\xi) = a^0 + \delta\, b_0\bigl(\delta x,\tfrac{\xi-K}{\delta}\bigr), \qquad
    b_0(X,\ze) = \begin{pmatrix} 0 & -i\wt(X) & 0\\ i\wt(X) & 0 & -\ze\times\\ 0 & \ze\times & 0\end{pmatrix},
\end{align}
where $K = (0,0,k_z)^T$ and $a^0$ is the value of \eqref{eq:sum_plasma_symbol} at $\omega_p = \omega_-$ and $\xi = K$. Thus $b_j = 0$ for every $j\ge1$. Here $x\in\mathbb{R}^2$ and $\zeta = (\zeta_1,\zeta_2)$ are in-plane, so that, in the notation of Section \ref{sec:tb}, $d = 2$ and the in-plane reference momentum is $0$, while $k_z$ is a fixed parameter of the symbol. Accordingly, no factor $e^{iK\cdot x}$ appears below. Assumption
\ref{assumption:pde} holds with $m_0 = 1$ and $\rate = 1$, for every $p$.

\paragraph{Spectral data.} The spectrum of $a^0$ is
\begin{align}\label{eq:sum_plasma_spec}
    \sigma(a^0) = \bigl\{\pm\omega_-\ (\text{mult.\ }2),\ \pm\omega_R,\ \pm\omega_{L+},\ 0\bigr\},
    \qquad 0<\omega_-<\omega_R<\omega_{L+},
\end{align}
so that $L = 7$. Here $\omega_R>0$ and $\omega_{L+}>\Omega$ are the right-hand and upper
left-hand polarized frequencies; that is, the solutions $\lambda$ of
\begin{align*}
    \frac{k_z^2}{\lambda^2} = 1 - \frac{\omega_-^2}{\lambda(\lambda+\Omega)} \quad (\lambda = \omega_R),
    \qquad
    \frac{k_z^2}{\lambda^2} = 1 - \frac{\omega_-^2}{\lambda(\lambda-\Omega)} \quad (\lambda = \omega_{L+}),
\end{align*}
the second equation having $\omega_-$ as its other positive solution. The two degenerate sectors
$\pm\omega_-$ are of class $\cL_1$. The five simple sectors are of class $\cL_0$, with
\begin{align}\label{eq:sum_plasma_flat}
    b_{0,R} = f_R\,\wt, \qquad b_{0,L+} = f_{L+}\,\wt, \qquad b_{0,0} = 0, \qquad b_{0,-\ell} = -\overline{b_{0,\ell}},
\end{align}
where
\begin{align*}
    f_R = \frac{2\omega_-}{N_R(\Omega+\omega_R)}, \qquad f_{L+} = \frac{2\omega_-}{N_{L+}(\omega_{L+}-\Omega)}, \qquad
    N_j = \frac{\omega_-^2}{(\Omega\pm\omega_j)^2} + 1 + \frac{k_z^2}{\omega_j^2},
\end{align*}
with the sign $+$ for $j = R$ and $-$ for $j \in\{-, L+\}$, writing $\omega_j$ for the corresponding frequency. The last identity in \eqref{eq:sum_plasma_flat} follows from the
antiunitary symmetry $\Gamma_P\,\overline{a^0}\,\Gamma_P = -a^0$, with $\Gamma_P = \diag(I_3,I_3,-I_3)$. 

\paragraph{The topological sector.} The eigenspace of $\omega_-$ is spanned by the orthonormal vectors
\begin{align}\label{eq:sum_plasma_evecs}
  \Psi_1 = \frac{1}{\sqrt{2}}\begin{pmatrix}\hat{e}_z\\i\hat{e}_z\\0\end{pmatrix},
  \qquad
  \Psi_2 = \frac{1}{\sqrt{N_-}}\begin{pmatrix}
    i\frac{\omega_-}{\Omega-\omega_-}\hat{e}_+\\\hat{e}_+\\-i\frac{k_z}{\omega_-}\hat{e}_+
  \end{pmatrix},
  \qquad
  \hat{e}_+ = \frac{1}{\sqrt{2}}\begin{pmatrix}1\\ i\\0\end{pmatrix}.
\end{align}
In this basis, and with $\ze = \ze_1 + i\ze_2$,
\begin{align}\label{eq:sum_plasma_Dirac}
    b_{0,\omega_-}(X,\ze) = \bigl(\tfrac12 - \beta\bigr)\wt(X)\, I_2
    - \tfrac{\ze_2}{\alpha}\sigma_1 - \tfrac{\ze_1}{\alpha}\sigma_2
    + \bigl(\tfrac12 + \beta\bigr)\wt(X)\,\sigma_3,
\end{align}
where $\alpha = 2\omega_-\sqrt{N_-}/k_z$ and $\beta = \omega_-/((\Omega-\omega_-)N_-)$. Thus
$\bH_{0,\omega_-} = \ow b_{0,\omega_-}$ is a Dirac operator with Fermi velocity $\alpha^{-1}$,
mass $(\frac12 + \beta)\wt(X)$ and potential
$(\frac12 - \beta)\wt(X)$. A change of sign of $\wt$, that is of
$\omega_p - \omega_-$, across an interface is the topological phase transition analyzed in
\cite{frazier2025topological, fu2022dispersion}. Since $b_j = 0$ for $j\ge1$,
\begin{align}\label{eq:sum_plasma_Heff}
    \rH^{(\omega_-)} = \bH_{0,\omega_-} + \delta\,\Sigma_{\omega_-} + O(\delta^2), \qquad
    \Sigma_{\omega_-} = \sum_{\ell\ne\omega_-}\frac{\mathbf M_{\omega_-,\ell}\mathbf M_{\ell,\omega_-}}{E_{\omega_-} - E_\ell},
\end{align}
with the couplings $\mathbf M_{\ell\ell'} := P_\ell\bH_0P_{\ell'}$. Since these couplings are unbounded, they enter $\rH^{(\omega_-)}$ with the frequency cutoff of Section \ref{sec:effective}, that is, with $\bH_0$ replaced by $\bH_0^\eps$ (Remark \ref{rem:rH_expansion}). Every one of the six other
sectors contributes. The contributions of $-\omega_-$ and of the zero mode have the symbols
\begin{align}\label{eq:sum_plasma_sigma}
    \frac{k_z^2 |\ze|^2}{8N_-\omega_-^3}\, I_2
    \qquad\text{and}\qquad
    \frac{|\ze|^2}{2N_-\omega_-}\begin{pmatrix}0&0\\0&1\end{pmatrix},
\end{align}
which are quadratic in $\ze$, while the $\wt$-dependent couplings involve only the sectors
$-\omega_R$ and $\omega_{L+}$; all couplings are listed in Section \ref{sec:plasma_ops}. In particular $\Sigma_{\omega_-}$ is a
second-order operator, so that the leading correction to the first-order Dirac operator
$\bH_{0,\omega_-}$ is of second order, with a coefficient of size $\delta$.

\paragraph{Approximation of the microscopic dynamics.} For an initial condition
$\phi_0\in H^\infty(\mathbb{R}^2;\Ran P_{\omega_-})$ in the topological sector and $p = 1$, set
\begin{align}\label{eq:sum_plasma_psi}
    \psi_\delta(t,x) := \delta\, e^{-i\omega_- t}\bigl(\phi^0_{\omega_-} + \delta\,\phout^{1,\omega_-}\bigr)(\delta t,\delta x),
    \qquad
    \phout^{1,\omega_-} = \sum_{\ell\ne\omega_-}\frac{P_\ell\,\bH_0\,\phi^0_{\omega_-}}{\omega_- - E_\ell},
\end{align}
where $\phi^0_{\omega_-}$ solves the Dirac equation $(D_T + \bH_{0,\omega_-})\phi^0_{\omega_-} = 0$
with $\phi^0_{\omega_-}(0,\cdot) = \phi_0$. Proposition
\ref{prop:pde_class} and Theorem \ref{thm:main_higher_order} then give
\begin{align*}
    \norm{(\psi_\delta - \varphi_\delta)(t,\cdot)}_{H^m} \le C_m(\delta + \delta^2 t)\bigl(1 + (\delta t)^{2m+2}\bigr),
    \qquad t\ge0,\quad 0<\delta\le1.
\end{align*}
Since $b_j = 0$ for $j\ge1$, the expansion may be continued to arbitrary order, and by Remark
\ref{rem:pde_class_gain} the effective dynamics remain valid up to $t\ll\delta^{-\kappa}$ for
any $\kappa<3/2$.

\section{Numerical simulations}\label{sec:numerics}

To illustrate the results of Sections \ref{sec:tb} and \ref{sec:summary}, we compare the tight-binding dynamics \eqref{eq:microWP_higher_order} with three macroscopic approximations: 
\begin{itemize}
    \item the construction of Theorem \ref{thm:main_higher_order} with the prescribed data, with $p = 1, 2$ (and $p = 3$ for the Kekul\'e-O model);
    \item the band-adapted construction of Definition
\ref{def:single} with $p = 2$, started from the dressed data $\Pi^\delta_\ell\phi_0$ of
\eqref{eq:dressed};
    \item the reduced model obtained by truncating the effective Hamiltonian
\eqref{eq:def_rH} after its first correction. In this case, the amplitude $w$ in the sector $\ell$ of the data solves
\begin{align}\label{eq:num_reduced}
    (D_T + \bH_{0\ell} + \delta\rH_1)\,w = 0, \qquad w(0,\cdot) = \phi_0,
\end{align}
with $\rH_1 = P_\ell\bH_1P_\ell - P_\ell\bH_0\ashl\bH_0P_\ell$ the first-order coefficient of Remark \ref{rem:rH_expansion}. 
Recalling \eqref{eq:psi_def_higher_order_rH} and the fact that $\cG^{\pm1/2} = P_\ell + O(\delta^2)$, we approximate the tight-binding dynamics by the function
\begin{align*}
    \psi^\sharp (t,x) = e^{i(K\cdot x-E t)}\delta^{d/2} \bigl( (P_\ell + \delta S_1^0 + \delta^2 S_2^0)w \bigr) (\delta t, \delta x),
\end{align*}
with $S_1^0$, $S_2^0$ from Definition \ref{def:S}.
\end{itemize}

At $t = 0$, our approximation $\psi^\sharp$ coincides with the dressed data.
Moreover, up to $O(\delta^2)$ and frequency-truncation errors, the function $\psi^\sharp$ agrees with $\psi$ of Theorem \ref{prop:rH} and Corollary \ref{cor:rH_eps}, for all times $t \lesssim \delta^{-1}$.
For sectors of class $\cL_1$, the $L^2$-error associated with $\psi^\sharp$
is therefore of order
$\delta^2 + \delta^3 t$ for all $t \ge 0$; for sectors of class $\cL_0$, the term $\delta^3t$ carries an additional factor polynomial in $\delta t$. This applies to the coupled chains of Section \ref{sec:summary_chains}, the Kekul\'e-O modulated graphene of Section \ref{sec:summary_kekule}, and, by Proposition \ref{prop:pde_class}, the cold plasma model of Section \ref{sec:summary_plasma}.

All operators are used here without frequency cutoff (on the pseudo-spectral grid of Section \ref{sec:numerics_methods}, the discretization itself truncates frequencies at the grid scale). 
Further experiments on the effective Hamiltonian, up to third order and over much longer times, are reported in Section \ref{sec:sm_numerics} of the supplementary material. The code reproducing all the figures of this section and of Section \ref{sec:sm_numerics} is available online \cite{cazeaux2026code}.

\subsection{Numerical methods}\label{sec:numerics_methods}
All computations are conducted with periodic boundary conditions. 
The tight-binding Hamiltonians are assembled as sparse matrices on periodic lattices, the slowly varying coefficients being evaluated at bond midpoints, and the microscopic evolution is computed by Krylov approximations of the exponential semigroup. 
The macroscopic equations are discretized on the corresponding periodic domains by a pseudo-spectral method: all functions are approximated by truncated Fourier series, so that $D_X$ is represented by a diagonal matrix, while multiplication operators are applied on a uniform grid. 
The correctors are computed from the recursions \eqref{eq:def_phout}--\eqref{eq:def_phin} and \eqref{eq:leading_order_S}, with $P_\ell$ and $\ashl$ the spectral projections and reduced resolvents of $a^0$; the closed forms of Section \ref{sec:summary} serve only as checks. 
The macroscopic solution, which does not depend on $\delta$, is then interpolated onto the lattice. For the continuum model of cold plasma, whose symbol is affine in $\xi$, the microscopic problem is solved on the macroscopic grid in the variables $(T, X)$. Errors are measured in the relative $\ell^2$ norm ($L^2$ for the plasma).

Except for the zero mode of Section \ref{sec:numerics_kekule} and the interface mode of Section \ref{sec:numerics_plasma}, the data are Gaussian wave packets
\begin{align*}
    \phi_0 = c\,\mathcal P(D_X)\,(g\,v), \qquad g(X) = e^{-|X-X_0|^2/(2\sigma^2)}\,e^{i\zeta_0\cdot X},
\end{align*}
with center $X_0$, width $\sigma$ and momentum $\zeta_0$, all in the slow variables. 
Here $\mathcal P(\zeta)$ is a spectral projection associated with the matrix $b_{0,\ell}(X_0,\zeta)$, and $v$ is a fixed vector. If $b_{0,\ell}$ is scalar-valued, then $\mathcal{P}$ is the identity. Otherwise, $\mathcal{P}$ is chosen such that $b_{0,\ell} (X_0, \zeta)$ is positive on $\Ran \mathcal{P} (\zeta)$. Since $b_{0,\ell}$ is a Dirac-type symbol in all such cases, 
this ensures that the direction of wave-packet propagation agrees with the direction of $ \zeta_0$. The constant $c$ normalizes $\phi_0$ in $L^2$.


\subsection{Coupled chains: rates of convergence}\label{sec:numerics_chains}

\paragraph{Two coupled SSH chains.}
We take $E = 1$, $v = 2$ and $\alpha_1 = \alpha_2 = 1$ in \eqref{eq:sum_ssh_ajj}, and
\begin{align*}
    M_1(X) = \bar m\,\sigma_3 + V_0\,\mathrm{sech}^2(X/2)\,\sigma_0, \qquad M_2 = -\bar m\,\sigma_3, \qquad
    \beta(X) = \beta_0\bigl(1 + \mathrm{sech}^2(X/2)\bigr) I_2,
\end{align*}
with $\bar m = 1/2$, $V_0 = 2$ and $\beta_0 = 1$. Recalling that $\beta$ is the coupling matrix of \eqref{eq:sum_ssh_a}, the above definitions implement a smooth barrier on the first chain, across which the coupling is twice as strong.
The simulation domain is the periodic interval $[-32, 32]$. 

The initial condition is a
Gaussian wave packet on the positive branch of the first chain, of width $\sigma=2$ and macroscopic momentum
$\zeta_0 = 1$, centered at $X_0 = -10$. 
Its energy is close to the height of the barrier, which
splits it into a reflected and a transmitted wave packet. Through the coupling, the self-energy
$\frac{1}{2E}\beta\beta^*$ of \eqref{eq:sum_ssh_Heff} acts on the first chain as an additional
potential which is strongest near the barrier.

\begin{figure}[tb]
  \centering
  \includegraphics[width=5.125in]{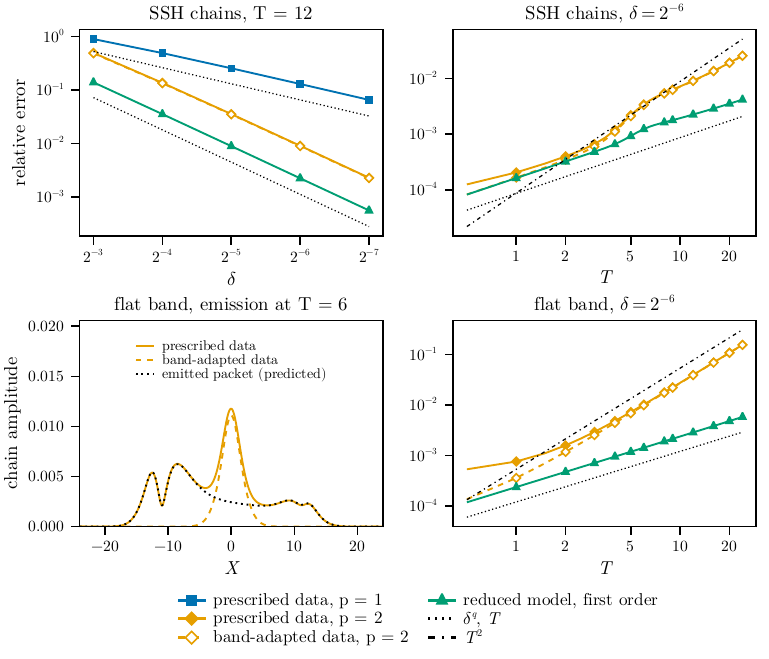}
  \caption{Coupled chains. Top: two SSH chains, relative $\ell^2$ error at $T = 12$ as a function of
  $\delta$ (left), and as a function of $T$ for $\delta = 2^{-6}$ (right); the two expansions with
  $p = 2$ nearly coincide. Bottom left: SSH chain coupled to a flat band, flat-band data; amplitude of
  the SSH component of the tight-binding solution at $T = 6$, $\delta = 2^{-6}$, for the prescribed
  and band-adapted data, and amplitude of the predicted emitted packet. Bottom right: same model,
  relative $\ell^2$ error as a function of $T$ for $\delta = 2^{-6}$. Filled markers: prescribed
  data, $p = 1, 2$. Open markers: band-adapted data, $p = 2$. Triangles: reduced model
  \eqref{eq:num_reduced}. Dotted lines: $\delta^q$ or $T$ for $q=1,2$. Dash-dotted lines: $T^2$.}
  \label{fig:numerics_chains}
\end{figure}

Figure~\ref{fig:numerics_chains} (top left) confirms the rates $\delta$ and $\delta^2$ of Theorem~\ref{thm:main_higher_order} for $p = 1, 2$, for both prescribed and band-adapted data. 
The reduced model \eqref{eq:num_reduced} converges at the same rate $\delta^2$, with a smaller error prefactor. 
As a function of $T$ (top right), the error of both expansions with $p = 2$ grows faster than linearly. 
The error for the reduced model on the other hand grows only linearly in $T$, consistent with Proposition \ref{prop:effective_discrete}. 

\paragraph{An SSH chain coupled to a flat band.}
We keep the first chain without the barrier ($M_1 = \bar m\,\sigma_3$) and take, in
\eqref{eq:sum_flat_a},
\begin{align*}
    \gamma(X) = \gamma_0\bigl(1 + \mathrm{sech}^2(X/2)\bigr)\begin{bmatrix} 1 \\ i/2\end{bmatrix},
    \qquad \gamma_0 = 0.6.
\end{align*}
The data is a Gaussian with $\sigma=2$, $\zeta_0 = 0$, $X_0 = 0$ on the flat band. 
With the prescribed data, $O(\delta)$ wave packets are emitted into the SSH band at $T = 0$ by the mechanism (M2) of Remark \ref{rem:mechanisms} and travel away along its two branches (Figure~\ref{fig:numerics_chains}, bottom left), matching the predicted packet $\delta E^{-1}e^{-iT\bH_{01}}[\gamma\phi_{00}]$. 
They are absent for the band-adapted data. 
In the flat sector, where $\bH_{02} = 0$, the dynamics at this order is determined by the phase $e^{i\delta T|\gamma|^2/E}$,
which the expansions contain only to first order in $\delta T$. 
Their error therefore grows quadratically in $T$, as predicted by the second part of Proposition \ref{prop:expansion_discrete}. 
The reduced model includes the phase; its error grows only linearly, and remains smaller by more than an order of magnitude at the final time (bottom right). 
The flat sector is of class $\cL_0$, for which the second part of Proposition \ref{prop:effective_discrete} (with $\nu = \delta$ there) gives
an error of order $\delta^3t$ for all times $t \lesssim \delta^{-2}$.

\subsection{The Kekul\'e vortex}\label{sec:numerics_kekule}
We take the vortex of Section \ref{sec:summary_kekule_vortex} with
$\De_0(R) = R\,(1 + R^2)^{-1/2}e^{-(R/9)^8}$, the last factor making $\De$ periodic on the torus
$[-12\sqrt3, 12\sqrt3)\times[-15, 15)$. Since the total winding number of $\De$ on the torus vanishes, this factor necessarily closes the Kekul\'e gap near the boundary of the domain, at distance about $9$ from the core.

\begin{figure}[tb]
  \centering
  \includegraphics[width=5.125in]{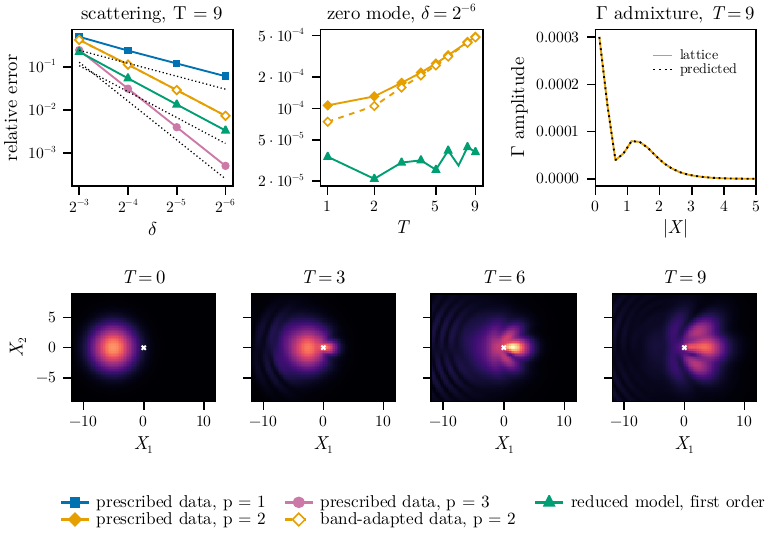}
  \caption{Kekul\'e vortex. 
  Top left: scattering wave packet, relative $\ell^2$ error at $T = 9$ as a  function of $\delta$. 
  Top middle: zero-mode data, relative $\ell^2$ error as a function of $T$ for  $\delta = 2^{-6}$. 
  Top right: radial profile of the $\Gamma$-sector amplitude at $T = 9$, $\delta = 2^{-5}$, for the band-adapted zero-mode data: tight-binding solution and predicted  corrector $\delta^2\phout^{2,0}$. 
  Bottom: amplitude of the Dirac-sector component of the tight-binding solution for the scattering wave packet at $T = 0, 3, 6, 9$, $\delta = 2^{-5}$; the cross marks the core of the vortex. 
  Markers as in Figure~\ref{fig:numerics_chains}. Circles: prescribed data, $p = 3$. Dotted lines: $\delta^q$ for $q=1,2,3$.}
  \label{fig:numerics_kekule}
\end{figure}

\paragraph{Zero mode.}
We take as data the zero mode $\phi_0$ of $\ow b_{0,0}$ localized at the vortex core, given by \eqref{eq:zero_mode}. 
(On the torus, the kernel of $\ow b_{0,0}$ also contains modes supported near the
boundary, where the Kekul\'e gap closes.) 
Since the lattice is bipartite, the sublattice operator
$J := \diag(I_3,-I_3)$ anticommutes with $H_\delta$ and with every $P_0b_jP_0$. The mode $\phi_0$ is
an eigenvector of $J$, so that $\langle\phi_0, b_{j,0}\phi_0\rangle = 0$. It follows that $\phin^{1,0}$ is bounded uniformly in time, as \eqref{eq:def_phin} gives
\begin{align*}
    (D_T + \bH_{00}) \phin^{1,0} = -P_0 \bH_1 \phi_0, \qquad \phin^{1,0} (0, \cdot) \equiv 0.
\end{align*}
Note that the secular term vanishes because the source $-P_0 \bH_1 \phi_0$ is orthogonal to $\phi_0$. Hence for $p=2$, the function $\phi$ of \eqref{eq:phi_expansion_higher_order} remains bounded uniformly in $T$, 
so that the approximation error in Proposition \ref{prop:expansion_discrete} can be improved to $C \delta^2(1+T)$. That is, we have linear (rather than quadratic) growth of the error in $T$, consistent with Figure \ref{fig:numerics_kekule} (top middle). 
The same plot shows that the reduced model error remains small for all times $T \le 9$.

The projection of the tight-binding and approximate solutions onto the $\Gamma$-point bands is illustrated in Figure \ref{fig:numerics_kekule} (top right). Here, the approximate solution is constructed by the expansion \eqref{eq:def_phout}-\eqref{eq:def_phin}, with $p=2$. Thus its projection onto the $\Gamma$-point bands is given by $\delta^2 \phout^{2,0}$ of \eqref{eq:sum_kekule_out2}. 

\paragraph{Scattering off the vortex.}
Next, a Gaussian wave packet on the positive branch of the Dirac sector, of width $\sigma = 2.5$ and momentum $\zeta_0 = (1.25, 0)$, is launched from $X_0 = (-5, 0)$ towards the core, where it is scattered (Figure~\ref{fig:numerics_kekule}, bottom row).
Since this branch is doubly degenerate, we take $v = (1, \tfrac{i}{2}, 0.4, -0.3)$ in the basis $(\fA_2,\fA_3,\fB_2,\fB_3)$ of $\Ran P_0$.
After the interaction, the expansions converge at the rates predicted by Theorems
\ref{thm:main_higher_order} and \ref{thm:equiv}, 
and the
reduced model converges at the rate $\delta^2$, consistent with Theorem \ref{prop:rH}; see Figure \ref{fig:numerics_kekule} (top left).
The reduced model is again more accurate than both expansions with $p = 2$.

\subsection{Cold plasma interface}\label{sec:numerics_plasma}
\begin{figure}[tb]
  \centering
  \includegraphics[width=5.125in]{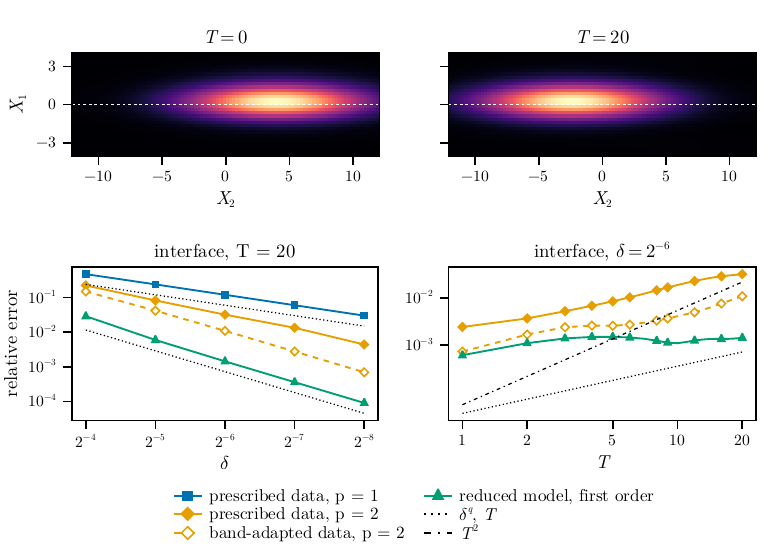}
  \caption{Cold plasma interface. Top: amplitude of the $\omega_-$ component of the solution at
  $T = 0$ and $T = 20$, $\delta = 2^{-6}$, with $X_2$ on the horizontal axis; the dashed line is the interface $X_1 = 0$. Bottom:
  relative $L^2$ error at $T = 20$ as a function of $\delta$ (left), and as a function of $T$ for
  $\delta = 2^{-6}$ (right). Markers as in Figure~\ref{fig:numerics_chains}.}
  \label{fig:numerics_plasma}
\end{figure}
With $k_z = 0.4$, $\Omega = 1$, we take
\begin{align*}
    \wt(X) = \omega_{\max}\tanh\Bigl(\frac{X_{\max}}{\pi W}\sin\frac{\pi X_1}{X_{\max}}\Bigr),
\end{align*}
which is close to $\omega_{\max}\tanh(X_1/W)$ near $X_1 = 0$ and periodic on $[-X_{\max}, X_{\max})^2$, with $\omega_{\max} = 1.5$, $W = 2$ and $X_{\max} = 16$. 
The mass $(\frac12 + \beta)\wt(X)$ of $\bH_{0,\omega_-}$ in \eqref{eq:sum_plasma_Dirac} changes sign with $\wt$ across the interface $X_1 = 0$. This topological phase transition induces a chiral mode localized to the interface; see \cite{frazier2025topological, fu2022dispersion} for more details.

The data is a wave packet carefully constructed on this mode as follows. For each momentum $\zeta_2$ along the interface, we compute the generalized eigenfunction of $\bH_{0,\omega_-}$ localized along the interface $X_1 = 0$ in the bulk gap. We build a superposition of these functions with Gaussian weights $e^{-\sigma^2 (\zeta_2 - \zeta_0)^2/2}$ with $\sigma=5$, $\zeta_0 = 0.6$, centered at $X_{0} = (0,4)$. 
The wave packet propagates along the interface in a single direction, at the group velocity of the interface mode, and remains localized on it (Figure~\ref{fig:numerics_plasma}, top). The band-adapted construction and the reduced model converge at the rate $\delta^2$, the latter with a much smaller error (bottom left). 
With prescribed data, the expansion with $p = 2$ is still in a pre-asymptotic regime at the final time. 
Its error is dominated by the $O(\delta)$ wave packet emitted into the sector $\omega_R \approx 0.44$, which is large because $\omega_R$ is close to $\omega_- \approx 0.33$. Compared with the estimates of Theorems \ref{thm:main_higher_order}, \ref{thm:equiv}, and \ref{prop:rH}, at fixed $\delta$ the errors grow slowly in $T$ (bottom right).


\section*{Acknowledgments.}
The work of GB was supported in part by NSF grant DMS-2606421 and ONR grant N00014-26-1-2017. DM’s research was supported by AFRL grant FA9550-24-1-0177.
PC's work was supported in part by NSF grant DMS-2608677 and Simons Travel Support for Mathematicians 966604. The Flatiron Institute is a division of the Simons Foundation.
\section*{Use of generative AI.}
Generative AI (Claude Opus 5.5) was used to adapt PC's existing code to the models of this paper,
and to run and validate the numerical experiments. It was also used to audit the manuscript for internal consistency and gaps in
its arguments and proofs, and to help with bibliographical data. All such output was checked by the authors,
who are alone accountable for the correctness of statements and proofs.

\appendix

\section{Pseudodifferential operators}\label{sec:pseudo}
We introduce notation on pseudo-differential operators used throughout the text and refer to \cite{Bony, DS, Hormander, Zworski} for additional details. Let $\mathcal{S} (\mathbb{R}^d) \otimes \mathbb{C}^n$ be the Schwartz space of vector-valued functions and 
$\mathcal{S}' (\mathbb{R}^d) \otimes \mathbb{C}^n$ its dual. 
Given a symbol $a(x,\xi) \in \mathcal{S}' (\mathbb{R}^d \times \mathbb{R}^d) \otimes \mathbb{C}^{n \times n}$, we define the Weyl quantization of $a$ as the operator 
\begin{align}\label{eq:weylquanth}
    \ow(a) \psi (x) :=
    \frac{1}{(2\pi)^d} \int_{\mathbb{R}^{2d}}
    e^{i(x-y)\cdot \xi}
    a(\frac{x+y}{2}, \xi) \psi (y) dy d\xi,
    \qquad
    \psi \in \mathcal{S} (\mathbb{R}^d) \otimes \mathbb{C}^n.
\end{align}

For $(x, \xi) = X\in \mathbb{R}^{2d}$, we define $\aver{X} := \sqrt{1 + |X|^2}$. A function $\fm: \mathbb{R}^{2d} \rightarrow [0,\infty)$ is called an {\em order function} if there exist constants $C_0 > 0$, $N_0 > 0$ such that
$\fm(X) \le C_0 \aver{X-Y}^{N_0} \fm(Y)$ for all $X,Y \in \mathbb{R}^{2d}$. 
Note that $\aver{X}^m$ is an order function for any $m \in \mathbb{R}$.

\begin{definition}\label{def:symbol_class}
We say that $a \in S(\fm)$ ({\em symbol class} associated to the order function $\fm$) if for every $\alpha \in \mathbb{N}^{2d}$, there exists $C_\alpha > 0$ such that $|\partial^\alpha a (X)| \le C_\alpha \fm(X)$
for all $X \in \mathbb{R}^{2d}$. 
If $a = a (X; \delta)$ also depends on some parameter $\delta \in \mathbb{R}$, we say that $a \in S (\fm)$ 
if all the constants $C_\alpha$ can be chosen independent of $\delta$. 
\end{definition}

We also use the following H\"ormander classes of symbols defined in \cite{hormander1971fourier} and \cite[Chapter 18]{Hormander}. 
\begin{definition}\label{def:hormander_class}
    For $m \in \mathbb{R}$, let $S^m$ denote the set of $a \in C^\infty (\mathbb{R}^d \times \mathbb{R}^d)$ such that for every $(\alpha, \beta) \in \mathbb{N}^d \times \mathbb{N}^d$,
\begin{align*}
    |\partial^\alpha_\xi \partial^\beta_x a (x,\xi)| \le C_{\alpha, \beta} \aver{\xi}^{m - |\alpha|}, \qquad (x, \xi) \in \mathbb{R}^d \times \mathbb{R}^d.
\end{align*}
If $a = a (x,\xi; \delta)$ also depends on some parameter $\delta \in \mathbb{R}$, we say that $a \in S^m$ if all the constants $C_{\alpha, \beta}$ can be chosen independent of $\delta$.
\end{definition}

We next recall the following Calder\'on-Vaillancourt result \cite{calderon1972class}:
\begin{lemma}\label{lemma:CV}
    If $a = a(x,\xi)$ belongs to $S(\aver{\xi}^m)$, then for any $s \ge 0$, $\ow a : H^{s+m} \to H^s$ is bounded. Moreover, there exists $\ncv> 0$ (independent of $s$) such that
    \begin{align*}
        \norm{\ow (a) f}_{H^s (\mathbb{R}^d)} \le C \max_{|\alpha+\beta| \le \ncv+s} \norm{\aver{\xi}^{-m}\partial^\alpha_x \partial^\beta_\xi a (x,\xi)}_{L^\infty (\mathbb{R}^{2d})} \norm{f}_{H^{s+m} (\mathbb{R}^d)}
    \end{align*}
    uniformly in $a \in S(\aver{\xi}^m)$ and $f \in H^{s+m} (\mathbb{R}^d)$.
\end{lemma}

Finally, we establish a commutator estimate for the Hamiltonians considered in this paper. Note that only the derivatives of the symbol in the slow variable enter, so that the estimate is insensitive to growth of $a$ in $\xi$ as long as the growing part of the symbol has $x$-independent coefficients; this is what allows the continuous models of Section \ref{sec:pde_class} to be treated together with the tight-binding ones.

\begin{lemma}\label{lemma:commutator_bd}
    Suppose $H_\delta = \ow a_\delta$ with $a_\delta$ as in \eqref{eq:a_delta}, and assume that $\partial^\beta_X a \in S(1)$ uniformly in $\delta$ for every multi-index $\beta$ with $|\beta| \ge 1$. Then for all $m \in \mathbb{N}$, the operator
    $$[\nabla_x, H_\delta] : H^{m}(\mathbb{R}^d; \mathbb{C}^N) \to H^m(\mathbb{R}^d; \mathbb{C}^{d \times N})$$
    is bounded, with
    \begin{align}\label{eq:assumption_commutator_bd_reg}
        \norm{[\nabla_x, H_\delta]}_{H^{m} \to H^m} \le C \delta, \qquad 0 < \delta \le 1.
    \end{align}
\end{lemma}
\begin{proof}
    We have $a_\delta (x,\xi) = a (\delta x, \xi; \delta)$, so that the symbol of $[\nabla_x, H_\delta]$ is $\delta (\nabla_X a) (\delta x, \xi; \delta)$; see, e.g. \cite{DS, Zworski} for more details. This symbol is $O(\delta)$ in $S(1)$, and we conclude by Lemma \ref{lemma:CV} that \eqref{eq:assumption_commutator_bd_reg} holds.
\end{proof}

\section{Classical results on the Schr\"odinger equation}
\begin{lemma}[\cite{bal2026macroscopic}, Lemma C.1]\label{lemma:elliptic_reg}
    Let $\phi_0 \in H^\infty (\mathbb{R}^d; \mathbb{C}^N)$ and $s \ge 0$.
    Let $\bH$ be a self-adjoint operator on $L^2 (\mathbb{R}^d; \mathbb{C}^N)$ and $\Lambda$ a bounded operator from $H^s (\mathbb{R}^d; \mathbb{C}^N)$ to $L^2 (\mathbb{R}^d; \mathbb{C}^N)$ satisfying $[\bH, \Lambda] = 0$ and 
    \begin{align}\label{eq:equivalent_norms_appendix}
        C_1 \norm{u}_{H^s} \le \norm{\Lambda u}_{L^2} \le C_2 \norm{u}_{H^s}, \qquad u \in H^s
    \end{align}
    for some $0 < C_1 < C_2$. Then the solution $\phi = \phi (T,X)$ to 
    \begin{align*}
        (D_T + \bH) \phi (T,X) = r(T,X), \qquad \phi (0,X) = \phi_0 (X)
    \end{align*}
    satisfies
    \begin{align}\label{eq:phi_S_bd}
        \norm{\phi (T, \cdot)}_{H^s} \le C_1^{-1} C_2 \left( \norm{\phi_0}_{H^s} + \int_0^T \norm{r (T', \cdot)}_{H^s} dT'\right), \qquad T \ge 0.
    \end{align}
\end{lemma}
\begin{lemma}\label{lemma:bdd_reg}
    Let $\phi_0 \in H^\infty (\mathbb{R}^d; \mathbb{C}^N)$ and $s \in \mathbb{N}_0$. Let $\bH$ be a bounded self-adjoint operator on $L^2 (\mathbb{R}^d; \mathbb{C}^N)$. Suppose there exists a linear operator $\Lambda$ on $L^2 (\mathbb{R}^d; \mathbb{C}^N)$ such that for all $\ds \in \{0, \dots, s-1\}$, the operators
    \begin{align*}
        \Lambda: H^{\ds+1} (\mathbb{R}^d; \mathbb{C}^N) \to H^{\ds} (\mathbb{R}^d; \mathbb{C}^N), \qquad [\bH, \Lambda] : H^{\ds} (\mathbb{R}^d; \mathbb{C}^N) \to H^{\ds} (\mathbb{R}^d; \mathbb{C}^N)
    \end{align*}
    are both bounded, with
    \begin{align*}
        C_1 \norm{u}_{H^{\ds+1}} \le \norm{\Lambda u}_{H^{\ds}} \le C_2 \norm{u}_{H^{\ds+1}}, \qquad u \in H^{\ds}
    \end{align*}
    for some $0 < C_1 < C_2$.
    Then the solution to
    \begin{align*}
        (D_T + \bH) \phi (T,X) = r(T,X), \qquad \phi (0,X) = \phi_0 (X)
    \end{align*}
    satisfies
    \begin{align}\label{eq:bdd_reg}
        \norm{\phi (T, \cdot)}_{H^s} \le C\sum_{\ds=0}^s T^{\ds} \left( \norm{\phi_0}_{H^{s-\ds}} + \int_0^T \norm{r (T', \cdot)}_{H^{s-\ds}} dT'\right), \qquad T \ge 0.
    \end{align}
\end{lemma}
\begin{proof}
    We proceed by induction on $s$. The base case $s=0$ in fact holds with $C=1$, and is a standard result following immediately from the self-adjointness of $\bH$; see, e.g. \cite[Equation (C.3)]{bal2026macroscopic}.

    Now, let $S \in \mathbb{N}$ and suppose that \eqref{eq:bdd_reg} holds for all $0 \le s \le S-1$. Then
    \begin{align*}
        \Lambda^{S} r = \Lambda^{S} (D_T + \bH) \phi = (D_T + \bH)\Lambda^{S} \phi - [\bH,\Lambda^{S}]\phi,
    \end{align*}
    so that the base case $s=0$, applied to $\Lambda^S \phi$, implies
    \begin{align*}
        \norm{\Lambda^{S} \phi (T, \cdot)}_{L^2} \le C \left( \norm{\Lambda^{S} \phi_0}_{L^2} + \int_0^T \norm{\Lambda^{S} r (T', \cdot) + [\bH, \Lambda^{S}] \phi (T', \cdot)}_{L^2} dT' \right).
    \end{align*}
    Now, our assumptions on $\Lambda$ imply that for $g = \phi_0, r (T, \cdot)$,
    \begin{align*}
        \norm{\phi (T, \cdot)}_{H^{S}} \le C\norm{\Lambda^S \phi (T, \cdot)}_{L^2}, \quad \norm{\Lambda^S g}_{L^2} \le C\norm{g}_{H^S}, \quad \norm{[\bH, \Lambda^S] \phi (T, \cdot)}_{L^2} \le C \norm{\phi (T, \cdot)}_{H^{S-1}}.
    \end{align*}
    The last inequality follows from the fact that
    $[\bH, \Lambda^S] = \sum_{i=0}^{S-1} \Lambda^{i}[\bH,\Lambda]\Lambda^{S-1-i}$, where
    each summand maps $H^{S-1}$ to $L^2$ boundedly, with $\Lambda^{S-1-i} : H^{S-1}\to H^{i}$ and
    $\Lambda^i : H^i \to L^2$ by the first hypothesis, and $[\bH,\Lambda]$ bounded on
    $H^{i}$ by the second. Hence
    \begin{align*}
        \norm{\phi (T, \cdot)}_{H^S} \le C \left( \norm{\phi_0}_{H^S} + \int_0^T \norm{r (T', \cdot)}_{H^S} dT' + T \sup_{T' \in [0,T]} \norm{\phi(T', \cdot)}_{H^{S-1}}\right).
    \end{align*}
    By our inductive hypothesis, the norm $\norm{\phi(T', \cdot)}_{H^{S-1}}$ can be bounded by the right-hand side of \eqref{eq:bdd_reg} with $s = S-1$. It follows that \eqref{eq:bdd_reg} in fact holds for $s = S$.
\end{proof}
\begin{corollary}\label{cor:flat}
    Take $\phi_0$, $s$, $\bH$, $\Lambda$ as in Lemma \ref{lemma:bdd_reg}. Define $\phi$ as the solution to
    \begin{align*}
        (D_T + \nu \bH) \phi (T,X; \nu) = r(T,X;\nu), \qquad \phi (0,X; \nu) = \phi_0 (X),
    \end{align*}
    where $r (T, \cdot \; ; \nu) \in H^s$ for all $T \ge 0$ and $\nu > 0$.
    Then 
    \begin{align*}
        \norm{\phi (T, \cdot \; ; \nu)}_{H^s} \le C\sum_{\ds=0}^s (\nu T)^{\ds} \left( \norm{\phi_0}_{H^{s-\ds}} + \int_0^{T} \norm{r (T', \cdot \; ; \nu)}_{H^{s-\ds}} dT'\right), \qquad T \ge 0, \quad \nu > 0.
    \end{align*}
\end{corollary}
\begin{proof}
    Set $\Phi (\cT ,X) := \phi (\cT/\nu, X)$, then
    \begin{align*}
        (D_\cT + \bH )\Phi (\cT,X) = \nu^{-1} r(\cT/\nu, X; \nu),
    \end{align*}
    so that Lemma \ref{lemma:bdd_reg} implies
    \begin{align*}
        \norm{\Phi (\cT, \cdot \; ; \nu)}_{H^s} \le C\sum_{\ds=0}^s \cT^{\ds} \left( \norm{\phi_0}_{H^{s-\ds}} + \nu^{-1} \int_0^\cT \norm{r (\cT'/\nu, \cdot \; ; \nu)}_{H^{s-\ds}} d\cT'\right), \qquad \cT \ge 0, \quad \nu > 0.
    \end{align*}
    The result then follows from setting $\cT := \nu T$ in the above estimate.
\end{proof}
\begin{lemma}\label{lemma:u_delta_reg}
    Fix $m \in \mathbb{N}_0$. Suppose $\{ H_\delta : 0 < \delta \le \delta_0\}$ is a family of 
    symmetric operators on $L^2 (\mathbb{R}^d; \mathbb{C}^N)$ such that for any $0 < \delta \le \delta_0$ and $k \in \{0, 1, \dots, m-1\}$, the operator
    $$[\nabla_x, H_\delta] : H^{k}(\mathbb{R}^d; \mathbb{C}^N) \to H^k(\mathbb{R}^d; \mathbb{C}^{d \times N})$$
    is bounded, with
    \begin{align}\label{eq:commutator_bd_reg}
        \norm{[\nabla_x, H_\delta]}_{H^{k} \to H^k} \le C \delta, \qquad 0 < \delta \le \delta_0.
    \end{align}
    Let $R_\delta \ge 0$, and let $u_\delta = u_\delta (t,x)$ with $t \ge 0$ and $x \in \mathbb{R}^d$ be a family of functions satisfying 
    \begin{align*}
        \norm{(D_t + H_\delta) u_\delta (t, \cdot)}_{H^{m}} \le R_\delta (1 + (\delta t)^M), \qquad t \ge 0, \quad 0 < \delta \le \delta_0,
    \end{align*}
    for some $M \ge 0$. Then for some $C > 0$,
    \begin{align*}
        \norm{u_\delta (t, \cdot)}_{H^m} \le C \left( R_\delta\, t (1 + (\delta t)^{m+M}) + (1+(\delta t)^m)\norm{u_\delta (0, \cdot)}_{H^m} \right), \qquad t \ge 0, \quad 0 < \delta \le \delta_0.
    \end{align*}
\end{lemma}
\begin{proof}
    We have $(D_t + H_\delta) u_\delta = R_\delta r_\delta$,
    where $\norm{r_\delta (t, \cdot)}_{H^k} \le 1 + (\delta t)^M$ uniformly in $t$ and $\delta$. It follows that $ (u_\delta, D_t u_\delta)_{H^k} = R_\delta (u_\delta, r_\delta)_{H^k} - (u_\delta, H_\delta u_\delta)_{H^k}$,
    hence
    \begin{align}\label{eq:dt}
        -i \partial_t \left( \norm{u_\delta (t, \cdot)}_{H^k}^2 \right) = 2 i\Im (u_\delta, D_t u_\delta)_{H^k} = 2 i\Im \left( R_\delta (u_\delta, r_\delta)_{H^k} - (u_\delta, H_\delta u_\delta)_{H^k}\right). 
    \end{align}
    Since $H_\delta$ is symmetric, we know that
    \begin{align}\label{eq:uHu}
        (u_\delta, H_\delta u_\delta)_{H^k}
        &= \sum_{|\alpha| \le k} \int_{\mathbb{R}^d} \overline{\partial^\alpha_x u_\delta (t,x)} H_\delta \partial^\alpha_x u_\delta (t,x) {\rm d} x + 
        \sum_{|\alpha| \le k} \int_{\mathbb{R}^d} \overline{\partial^\alpha_x u_\delta (t,x)} [\partial^\alpha_x, H_\delta] u_\delta (t,x) {\rm d} x,
    \end{align}
    with the first term on the above right-hand side real. By expanding the commutator $[\partial^\alpha_x, H_\delta]$ as a sum of terms involving commutators of $H_\delta$ with only first-order derivatives, our assumption \eqref{eq:commutator_bd_reg} implies that the second term on the right-hand side of \eqref{eq:uHu} is bounded by $C \delta \norm{u_\delta(t, \cdot)}_{H^{k}}\norm{u_\delta(t, \cdot)}_{H^{k-1}}$, and thus
    \begin{align*}
        \partial_t \left( \norm{u_\delta (t, \cdot)}_{H^k}^2 \right) \le 2 R_\delta \norm{u_\delta (t, \cdot)}_{H^k} \norm{r_\delta(t, \cdot)}_{H^k} + C \delta \norm{u_\delta(t, \cdot)}_{H^k}\norm{u_\delta(t, \cdot)}_{H^{k-1}}
    \end{align*}
    by \eqref{eq:dt}.
    Note that the second term on the right-hand side is understood to vanish if $k = 0$.
    It follows that
    \begin{align*}
        \partial_t \norm{u_\delta (t, \cdot)}_{H^k}\le R_\delta \norm{r_\delta(t, \cdot)}_{H^k} + C \delta \norm{u_\delta(t, \cdot)}_{H^{k-1}},
    \end{align*}
    and so
    \begin{align*}
        \norm{u_\delta (t, \cdot)}_{H^k} \le R_\delta \int_0^t \norm{r_\delta(s, \cdot)}_{H^k}d s + C \delta \int_0^t \norm{u_\delta(s, \cdot)}_{H^{k-1}} d s + \norm{u_\delta (0, \cdot)}_{H^k}.
    \end{align*}
    The result then follows from induction on $k$.
\end{proof}
\begin{lemma}\label{lemma:u_delta_reg_flat}
    Under the hypotheses of Lemma \ref{lemma:u_delta_reg}, assume there exists a function $\nu = \nu (\delta)$ such that
    \begin{align*}
        \norm{(D_t + H_\delta) u_\delta (t, \cdot)}_{H^{m}} \le R_\delta (1 + (\delta \nu t)^M), \qquad t \ge 0, \quad 0 < \delta \le \delta_0.
    \end{align*}
    Then there exists a constant $C>0$ such that
    \begin{align*}
        \norm{u_\delta (t, \cdot)}_{H^m} \le C \left( R_\delta\, t (1 + (\delta t)^{m})(1 + (\delta \nu t)^M) + (1+(\delta t)^m)\norm{u_\delta (0, \cdot)}_{H^m} \right), \qquad t \ge 0, \quad 0 < \delta \le \delta_0.
    \end{align*}
\end{lemma}
\begin{proof}
    Following the proof of Lemma \ref{lemma:u_delta_reg}, we find that
    \begin{align*}
        \norm{u_\delta (t, \cdot)}_{H^k} \le R_\delta \int_0^t (1+(\delta \nu s)^M) ds + C \delta \int_0^t \norm{u_\delta(s, \cdot)}_{H^{k-1}} d s + \norm{u_\delta (0, \cdot)}_{H^k},
    \end{align*}
    and conclude by induction on $k$, using that $s \mapsto 1+(\delta\nu s)^M$ is nondecreasing.
\end{proof}
\begin{lemma}\label{lemma:u_delta_reg_discrete}
    Let $\cR \subset \mathbb{R}^d$ be a discrete set such that $|x-y| \ge c > 0$ for all distinct $x,y \in \cR$.
    Under the assumptions of Lemma \ref{lemma:u_delta_reg} with $m > d/2$, suppose $H_\delta$ defines a self-adjoint operator on $\ell^2 (\cR; \mathbb{C}^N)$. 
    Then
    \begin{align*}
        \norm{u_\delta (t, \cdot)}_{\ell^2 (\cR; \mathbb{C}^N)} \le C \left( R_\delta\, t (1 + (\delta t)^{M}) + \norm{u_\delta (0, \cdot)}_{\ell^2 (\cR; \mathbb{C}^N)} \right), \qquad t \ge 0, \quad 0 < \delta \le \delta_0.
    \end{align*}
    If, in addition, the hypotheses of Lemma \ref{lemma:u_delta_reg_flat} are satisfied, then
    \begin{align*}
        \norm{u_\delta (t, \cdot)}_{\ell^2 (\cR; \mathbb{C}^N)} \le C \left( R_\delta\, t (1 + (\delta \nu t)^{M}) + \norm{u_\delta (0, \cdot)}_{\ell^2 (\cR; \mathbb{C}^N)} \right), \qquad t \ge 0, \quad 0 < \delta \le \delta_0.
    \end{align*}
\end{lemma}
\begin{proof}
    The assumptions of Lemma \ref{lemma:u_delta_reg} and self-adjointness of $H_\delta$ on $\ell^2 := \ell^2 (\cR; \mathbb{C}^N)$ imply that
    \begin{align*}
        -i \partial_t \left( \norm{u_\delta (t, \cdot)}_{\ell^2}^2 \right) = 2 i\Im \left( R_\delta (u_\delta, r_\delta)_{\ell^2} - (u_\delta, H_\delta u_\delta)_{\ell^2}\right) = 2 iR_\delta \Im (u_\delta, r_\delta)_{\ell^2},
    \end{align*}
    where $\norm{r_\delta (t, \cdot)}_{H^m} \le 1 + (\delta t)^M$.
    On the other hand, our assumptions on $\cR$ imply the classical inequality
    \begin{align*}
        \norm{u}_{\ell^2} \le C_m \norm{u}_{H^m}, \qquad u \in H^m (\mathbb{R}^d; \mathbb{C}^n), \quad m > d/2.
    \end{align*}
    Applying Cauchy-Schwarz, we thus conclude that
    \begin{align*}
        \partial_t \norm{u_\delta (t, \cdot)}_{\ell^2} \le R_\delta \norm{r_\delta (t, \cdot)}_{\ell^2} 
        \le C R_\delta (1+(\delta t)^M).
    \end{align*}
    Our first statement follows from integrating the above inequality with respect to $t$.

    A parallel argument proves the desired bound under the additional constraint of Lemma \ref{lemma:u_delta_reg_flat}.
\end{proof}

\section{Proof of main result}\label{sec:proof_main}
In this section, we state and prove Proposition \ref{lemma:res_higher_order_all}, which is required to establish Theorem \ref{thm:main_higher_order}. The proof of this proposition in turn uses Lemma \ref{lemma:Sobolev_higher_order_all}, which is also contained below.

For $\ell \in \{1, \dots, L\}$, define
\begin{align*}
    \psi_\delta^\ell (t,x) := e^{i(K\cdot x-E_\ell t)} \delta^{\frac d2} \phi_\ell(\delta t,\delta x;\delta),
\end{align*}
so that \eqref{eq:psi_def_higher_order_all} reads
$
    \psi_\delta = \sum_{\ell=1}^L \psi_\delta^\ell.
$
Recall the definitions of $\cL_0$ and $\cL_1$ in \eqref{eq:cL_def}.
\begin{proposition}\label{lemma:res_higher_order_all}
    Suppose Assumption \ref{assumption:a_higher_order_all} holds, and take $H_\delta$ as in \eqref{eq:H_delta}. 
    Then
    \begin{align*}
        \norm{(D_t + H_\delta) \psi^\ell_\delta (t, \cdot)}_{H^s} \le C_s \delta^{p+\rate}\begin{cases}
            1 + (\delta t)^{s+2p}, & t \ge 0, \quad 0 < \delta \le \delta_0, \quad s \ge 0, \quad \ell \in \cL_0,\\
            1 + (\delta t)^{p-1}, & t \ge 0, \quad 0 < \delta \le \delta_0, \quad s \ge 0, \quad \ell \in \cL_1.
        \end{cases} 
    \end{align*}
\end{proposition}
\begin{proof}
    Fix $\ell \in \{1, \dots, L\}$, and define
    \begin{align}\label{eq:tHdelta}
        \tilde H_\delta := \ow \tilde a_\delta, \qquad \tilde a_\delta (X, \zeta) = a (X, K + \delta \zeta; \delta),
    \end{align}
    an operator in the slow variable $X$.
    The definitions of $H_\delta$ and $\psi_\delta^\ell$ imply that
    \begin{align}\label{eq:HD}
    \begin{split}
        H_\delta \psi_\delta^\ell (t,x) &= \frac{1}{(2\pi)^d} \dint_{\Rm^{2d}} e^{i(x-y)\cdot \xi} a(\delta\frac{x+y}2,\xi;\delta) e^{i(K\cdot y-E_\ell t)} \delta^{\frac d2} \phi_\ell(\delta t,\delta y;\delta) dy d\xi\\
        &=
        \dfrac{e^{i(K\cdot x-E_\ell t)}}{(2\pi)^d}  \dint_{\Rm^{2d}} e^{i(\delta x-\delta y)\cdot \zeta}  a(\frac{\delta x+\delta y}2,K+\delta \zeta;\delta)  \delta^{\frac d2} \phi_\ell(\delta t,\delta y;\delta) d \delta y d\zeta\\
        &=e^{i (K \cdot x - E_\ell t)} \delta^{d/2} \tilde H_\delta \phi_\ell (\delta t, \delta x; \delta)
    \end{split}
    \end{align}
    and
    \begin{align*}
        D_t \psi_\delta^\ell (t,x) = e^{i(K\cdot x-E_\ell t)} \delta^{\frac d2} (\delta D_T\phi_\ell -E_\ell \phi_\ell)(\delta t,\delta x;\delta),
    \end{align*}
    so that with $\uni_\delta$ the operator defined by $\uni_\delta u (t,x) := \delta^{\frac d2} u(\delta t, \delta x)$, we have
    \begin{align}\label{eq:psi_U_all}
        (H_\delta + D_t) \psi_\delta^\ell = e^{i (K \cdot x - E_\ell t)}\uni_\delta (\tilde H_\delta + \delta D_T - E_\ell) \phi_\ell.
    \end{align}
    The decomposition \eqref{eq:decomposition_higher_order_all} implies that
    \begin{align*}
        \tilde H_\delta = a^0 + \sum_{j=0}^{p-1} \delta^{j+1} \bH_j + \delta^{p+\mu} \bH_p,
    \end{align*}
    with $a^0 : H^s \to H^s$ 
    and the $\bH_j :H^{s + j+1} \to H^{s}$ bounded uniformly in $\delta$.
    Defining $\phin^{0,\ell} := \phi^0_\ell$ and $\phout^{0,\ell} \equiv 0$, we then have
    \begin{align*}
        (\tilde H_\delta + \delta D_T - E_\ell)\phi_\ell = \left( a^0 - E_\ell + \delta D_T + \sum_{j=0}^{p-1} \delta^{j+1} \bH_j + \delta^{p+\mu} \bH_p\right) \sum_{i=0}^p \delta^i (\phin^{i,\ell} + \phout^{i,\ell}).
    \end{align*}
    Collecting terms of $O(\delta^k)$ for $k \le p$ 
    and using that $(a^0 - E_\ell) \phin^{i,\ell} \equiv 0$, we find that $$(\tilde H_\delta + \delta D_T - E_\ell) \phi_\ell = \sum_{k=1}^p \delta^k r^k_\ell + \delta^{p+\rate} r^{p+1}_\ell,$$
    where
    \begin{align*}
        r^k_\ell := (D_T + \bH_{0}) \phin^{k-1,\ell} + \sum_{j=1}^{k-1} \bH_j \phin^{k-1-j,\ell} + (a^0 - E_\ell) \phout^{k,\ell} + D_T \phout^{k-1,\ell} + \sum_{j=0}^{k-1} \bH_j \phout^{k-1-j,\ell},
    \end{align*}
    and $r^{p+1}_\ell = r^{p+1}_\ell (T, X; \delta)$ satisfies
    \begin{align*}
        \norm{r^{p+1}_\ell (T, \cdot \; ; \delta)}_{H^s} \le C_s \left( \max_{0 \le j \le p} \norm{\phi^{j}_\ell (T, \cdot\; ; \delta)}_{H^{s+p+1}} + \delta^{1-\rate} \norm{D_T \phi^{p}_\ell (T, \cdot\; ; \delta)}_{H^s} \right), \qquad T \ge 0, \quad 0 < \delta \le \delta_0.
    \end{align*}
    The definitions \eqref{eq:def_phin} and \eqref{eq:def_phout} of $\phin^{j,\ell}$ and $\phout^{j,\ell}$ imply that
    \begin{align*}
        P_\ell r^k_\ell &= (D_T + \bH_{0\ell}) \phin^{k-1,\ell} + P_\ell\sum_{j=1}^{k-1} \bH_j \phin^{k-1-j,\ell} + P_\ell \sum_{j=0}^{k-1} \bH_j \phout^{k-1-j,\ell}\\
        &= (D_T + \bH_{0\ell}) \phin^{k-1,\ell} + P_\ell \left( \sum_{j=1}^{k-1} \bH_j \phi^{k-1-j}_\ell + \bH_0 \phout^{k-1,\ell}\right) = 0
    \end{align*}
    and
    \begin{align*}
        (I_N-P_\ell) r^k_\ell &= (I_N-P_\ell) \left( \sum_{j=0}^{k-1} \bH_j \phin^{k-1-j,\ell} + (a^0 - E_\ell) \phout^{k,\ell} + D_T \phi^{k-1}_\ell + \sum_{j=0}^{k-1} \bH_j \phout^{k-1-j,\ell}\right)\\
        &= (I_N-P_\ell) \left( \sum_{j=0}^{k-1} \bH_j \phi^{k-1-j}_\ell + (a^0 - E_\ell) \phout^{k,\ell} + D_T \phi^{k-1}_\ell \right) = 0,
    \end{align*}
    hence $r^k_\ell \equiv 0$ for all $k \le p$. 
    Recalling \eqref{eq:psi_U_all}, we conclude that
    \begin{align*}
        \norm{(D_t + H_\delta) \psi_\delta^\ell (t, \cdot)}_{H^s} \le C_s \left( \delta^{p+\rate} \max_{0 \le j \le p} \norm{\phi^j_\ell (\delta t, \cdot\; ; \delta)}_{H^{s+p+1}} + \delta^{p+1}\norm{D_T \phi^p_\ell (\delta t, \cdot\; ; \delta)}_{H^s}\right) 
    \end{align*}
    uniformly in $t \ge 0$ and $0 < \delta \le \delta_0$.
    The result then follows from Lemma \ref{lemma:Sobolev_higher_order_all}.
\end{proof}
\begin{lemma}\label{lemma:Sobolev_higher_order_all}
    Suppose Assumption \ref{assumption:a_higher_order_all} holds. 
    For any $k, s \ge 0$ and $1 \le \ell \le L$, there exists $C > 0$ such that the functions $\phi^j_\ell$ defined by \eqref{eq:def_phi_0_ell}, \eqref{eq:def_phout} and \eqref{eq:def_phin} satisfy
    \begin{align*}
        \sum_{j=0}^p \norm{D_T^k \phi^j_\ell (T,\cdot \; ; \delta)}_{H^s} \le C \begin{cases}
            1+T^{s+p-1}, & T \ge 0, \quad 0 < \delta \le \delta_0, \quad \ell \in \cL_0,\\
            1+T^{p-1}, & T \ge 0, \quad 0 < \delta \le \delta_0, \quad \ell \in \cL_1.
        \end{cases}
    \end{align*}
\end{lemma}
For ease of notation below, define $\phin^{0,\ell} := \phi^0_\ell$ and $\phout^{0,\ell} \equiv 0$ so that $\phi^j_\ell = \phin^{j,\ell} + \phout^{j,\ell}$ for all $j$ and $\ell$.
\begin{proof}
    A simple inductive argument using the definitions \eqref{eq:def_phout} and \eqref{eq:def_phin} of $\phout^{j,\ell}$ and $\phin^{j,\ell}$ reveals that the initial conditions $\phin^{j,\ell} (0, \cdot\; ; \delta) = -P_\ell \sum_{\ell'=1}^L \phout^{j,\ell'} (0, \cdot\; ; \delta)$ are all bounded uniformly in $\delta$, in any Sobolev space. We will therefore handle the cases $\ell \in \cL_1$ and $\ell \in \cL_0$ separately, respectively applying Lemmas \ref{lemma:elliptic_reg} and \ref{lemma:bdd_reg}.

    Assume first that $\ell \in \cL_1$.
    Recalling that $\phout^0 \equiv \phin^p \equiv 0$, it suffices to show that for any $0 \le j \le p-1$,
    \begin{align}\label{eq:ind_all}
        \norm{D_T^k \phin^{j,\ell} (T,\cdot \; ; \delta)}_{H^s} + \norm{D_T^k \phout^{j+1,\ell} (T,\cdot \; ; \delta)}_{H^s}\le C_{k,s} (1+T^{j})
    \end{align}
    uniformly in $T \ge 0$ and $0 < \delta \le \delta_0$. 
    We prove \eqref{eq:ind_all} by induction in $j$.
    Assumption \ref{assumption:a_higher_order_all} implies that $b_{0\ell} \in S^1$ is elliptic so that \eqref{eq:equivalent_norms_higher_order} holds. Since $\phi^0_\ell (T,X; \delta) \in \Ran P_\ell$ for all $(T, X; \delta)$, it follows from Lemma \ref{lemma:elliptic_reg} with $\Lambda = (i+\bH_{0\ell})^{s+k+1}$ that $$\norm{\phi^0_\ell (T, \cdot \; ; \delta)}_{H^{s+k+1}} + \norm{D^k_T \phi^0_\ell (T, \cdot \; ; \delta)}_{H^{s}} \le C.$$ Indeed, the function $\chi := D^k_T \phi^0_\ell$ satisfies
    \begin{align*}
        (D_T + \bH_{0\ell}) \chi = 0, \qquad \chi \vert_{T=0} = (-\bH_{0\ell})^k P_\ell \phi_0,
    \end{align*}
    so that Lemma \ref{lemma:elliptic_reg} applies to $\chi$ as well.
    The base case $j=0$ then follows from the estimate
    \begin{align*}
        \norm{D_T^k \phout^{1,\ell} (T,\cdot \; ; \delta)}_{H^s} &= \norm{\inva_\ell \bH_0 \bH_{0\ell}^k \phi^0_\ell (T, X; \delta)}_{H^s} \le C\norm{\phi^0_\ell (T, \cdot \; ; \delta)}_{H^{s+k+1}}. 
    \end{align*}
    
    Suppose now that for some $1 \le J \le p-1$, the bound \eqref{eq:ind_all} is satisfied for all $j \le J-1$ and all $k, s \ge 0$. 
    Then the defining equation \eqref{eq:def_phin} for $\phin^{J,\ell}$ and Lemma \ref{lemma:elliptic_reg} with $\Lambda = (i + \bH_{0\ell})^s$ imply that
    \begin{align}\label{eq:induction_in_1}
        \begin{split}
        &\norm{D^k_T \phin^{J,\ell} (T,\cdot \; ; \delta)}_{H^s}\\
        &\hspace{1cm} \le C \left\{ \norm{D_T^k \phin^{J,\ell} (0, \cdot \; ; \delta)}_{H^s} + T \sup_{T' \in [0,T]} \norm{\bH_0 D^k_T \phout^{J,\ell} (T', \cdot \; ; \delta) + \sum_{i=1}^J \bH_i D^k_T \phi^{J-i}_\ell(T', \cdot \; ; \delta) }_{H^s} \right\}\\
        &\hspace{1cm} \le C (1+T^{J}).
        \end{split}
    \end{align} 
    Here, we used that the $\bH_i$ commute with $D_T$ to obtain the first inequality, while
    the last inequality follows from the uniform boundedness of $\bH_i :H^{s + i + 1} \to H^s$ and our inductive hypothesis.
    Similarly, the definition \eqref{eq:def_phout} of $\phout^{j,\ell}$ and uniform boundedness of $\inva : H^s \to H^s$ imply that
    \begin{align*}
        \norm{D_T^k \phout^{J+1,\ell} (T,\cdot \; ; \delta)}_{H^s} \le C \left( \norm{D_T^{k+1} \phi^{J}_\ell (T,\cdot \; ; \delta)}_{H^s} + \sum_{i=0}^{J} \norm{\bH_i D_T^k \phi^{J-i}_\ell(T,\cdot \; ; \delta)}_{H^s} \right) \le C (1 + T^{J}),
    \end{align*}
    with the last inequality following from \eqref{eq:induction_in_1} and the inductive hypothesis.
    Thus we have verified that \eqref{eq:ind_all} holds for $j=J$, concluding the proof in the case that $\ell \in \cL_1$.

    \medskip
    
    Suppose instead that $\ell \in \cL_0$. It then suffices to show that for any $0 \le j \le p-1$,
    \begin{align}\label{eq:ind_all_0}
        \norm{D_T^k \phin^{j,\ell} (T,\cdot \; ; \delta)}_{H^s} + \norm{D_T^k \phout^{j+1,\ell} (T,\cdot \; ; \delta)}_{H^s}\le C_{k,s} (1+T^{s+j})
    \end{align}
    uniformly in $T \ge 0$ and $0 < \delta < \delta_0$. We again proceed by induction. For the base case, we use the boundedness of $\bH_{0\ell}$ on $H^s$ to write
    \begin{align*}
        \norm{D^k_T \phin^{0,\ell} (T, \cdot \; ; \delta)}_{H^s} = \norm{\bH_{0\ell}^k \phin^{0,\ell}(T, \cdot \; ; \delta)}_{H^s} \le C\norm{\phin^{0,\ell}(T, \cdot \; ; \delta)}_{H^s} 
    \end{align*}
    and
    \begin{align*}
        \norm{D^k_T \phout^{1,\ell} (T, \cdot \; ; \delta)}_{H^s} = \norm{\inva_\ell \bH_{0\ell}^{k+1} \phin^{0,\ell}(T, \cdot \; ; \delta)}_{H^s} \le C\norm{\phin^{0,\ell}(T, \cdot \; ; \delta)}_{H^s}. 
    \end{align*}
    Lemma \ref{lemma:bdd_reg} with $\Lambda = \aver{D_X}$, and the definition \eqref{eq:def_phi_0_ell} of $\phi^0_\ell =\phin^{0,\ell}$ imply that $$\norm{\phin^{0,\ell}(T, \cdot \; ; \delta)}_{H^s} \le C (1 + T^s),$$ establishing \eqref{eq:ind_all_0} when $j = 0$.
    
    Suppose now that for some $1 \le J \le p-1$, the bound \eqref{eq:ind_all_0} is satisfied for all $j \le J-1$ and $k, s \ge 0$. Lemma \ref{lemma:bdd_reg} and the definition \eqref{eq:def_phin} of $\phin^{J,\ell}$ imply that
    \begin{align*}
        &\norm{D^k_T \phin^{J,\ell} (T,\cdot \; ; \delta)}_{H^s}\\
        &\hspace{0.2cm} \le C \sum_{\ds=0}^s T^\ds \left\{ \norm{D_T^k \phin^{J,\ell} (0, \cdot \; ; \delta)}_{H^{s-\ds}} + T \sup_{T' \in [0,T]} \norm{\bH_0 D^k_T \phout^{J,\ell} (T', \cdot \; ; \delta) + \sum_{i=1}^J \bH_i D^k_T \phi^{J-i}_\ell(T', \cdot \; ; \delta) }_{H^{s-\ds}} \right\}\\
        &\hspace{0.2cm} \le C (1+T^{s+J}),
    \end{align*}
    with the last inequality following from the inductive hypothesis and boundedness of the $\bH_i$ on $H^{s - \ds}$. 
    As before, we conclude by the definition \eqref{eq:def_phout} of $\phout^{J,\ell}$ that
    \begin{align*}
        \norm{D_T^k \phout^{J+1,\ell} (T,\cdot \; ; \delta)}_{H^s} \le C \left( \norm{D_T^{k+1} \phi^{J}_\ell (T,\cdot \; ; \delta)}_{H^s} + \sum_{i=0}^{J} \norm{\bH_i D_T^k \phi^{J-i}_\ell(T,\cdot \; ; \delta)}_{H^s} \right) \le C (1 + T^{s+J}),
    \end{align*}
    establishing \eqref{eq:ind_all_0} when $j = J$.
\end{proof}
The above estimates for $\ell \in \cL_0$ can be improved if $\bH_{0\ell} = 0$.
\begin{proposition}\label{prop:flat}
    Suppose Assumption \ref{assumption:a_higher_order_all} holds, take $H_\delta$ as in \eqref{eq:H_delta}, and let $\ell \in \{1, \dots, L\}$ such that $\bH_{0\ell} = 0$. Then
    \begin{align*}
        \norm{(D_t + H_\delta) \psi^\ell_\delta (t, \cdot)}_{H^s} \le C_s \delta^{p+\rate}(
            1 + (\delta t)^{p-1}), \qquad t \ge 0, \quad 0 < \delta \le \delta_0, \quad s \ge 0.
    \end{align*}
\end{proposition}
\begin{proof}
    This result follows immediately from Lemma \ref{lemma:bH_zero} and the proof of Proposition \ref{lemma:res_higher_order_all}.
\end{proof}
\begin{lemma}\label{lemma:bH_zero}
Suppose Assumption \ref{assumption:a_higher_order_all} holds. Let $\ell \in \{1, \dots, L\}$ such that $\bH_{0\ell} = 0$.
    For any $k, s \ge 0$, there exists $C > 0$ such that the functions $\phi^j_\ell$ defined by \eqref{eq:def_phi_0_ell}, \eqref{eq:def_phout} and \eqref{eq:def_phin} satisfy
    \begin{align*}
        \sum_{j=0}^p \norm{D_T^k \phi^j_\ell (T,\cdot \; ; \delta)}_{H^s} \le C 
            (1+T^{p-1}), \qquad T \ge 0, \quad 0 < \delta \le \delta_0.
    \end{align*}    
\end{lemma}
\begin{proof}
    The function $\phin^{0,\ell}$ is independent of $T$, which implies that \eqref{eq:ind_all} holds when $j=0$. 
    The estimate \eqref{eq:ind_all} for $0 \le j \le p-1$ is then verified by induction on $j$ as before.
\end{proof}
\section{Validity of the effective Hamiltonian}
We establish the following proposition, which is used in the proof of Theorem \ref{prop:rH}.
\begin{proposition}\label{prop:res_rH}
    Let $\phi_0 \in H^\infty (\mathbb{R}^d; \Ran P)$ and $m,n \in \mathbb{N}$. Then for any $\eta_0 > 0$, there exists $C = C(m,n,\eta_0)$ such that, for all $0 < \delta \le c_0\eps$, $\eps \le 1$, with $\eta \le \eta_0$ and all $t \ge 0$,
    \begin{align*}
        \norm{(D_t + H_\delta)\psi (t, \cdot)}_{H^m} \le
        \begin{cases}
        C (\delta^{p+\rate} + \delta \eps^n), & \ell \in \cL_1,\\
        C (\delta^{p+\rate} + \delta \eps^n) (1+(\delta t)^{m+n+p}), & \ell \in \cL_0.
        \end{cases}
    \end{align*}
\end{proposition}
\begin{proof}
    Following the proof of Proposition \ref{lemma:res_higher_order_all}, we find that
    \begin{align*}
        r (t,x) := (D_t + H_\delta)\psi (t,x) = e^{i(K\cdot x-E t)} \delta^{\frac d2} \left(a^0 - E + \sum_{j=0}^{p-1} \delta^{j+1} \bH_j + \delta^{p+\rate} \bH_p +\delta D_T \right) [\phi] (\delta t, \delta x),
    \end{align*}
    where 
    \begin{align*}
        \phi (T, \cdot) = (P+S) \cG^{-1/2}\, e^{-iT\,\rH}\,\mathcal G^{1/2}\phi_0.
    \end{align*}
    As in \cite[Proof of Lemma B.2]{bal2026macroscopic}, we have $$r(t,x) = e^{i(K \cdot x - Et)} \delta^{d/2} (\rho_0 (\delta t, \delta x) + \rho_1 (\delta t, \delta x)),$$ where with $\bHe$ the frequency-cut-off macroscopic Hamiltonian defined by \eqref{eq:bH},
    \begin{align}\label{eq:rho}
        \rho_0 (T,X) = \left(a^0 - E + \delta\bHe + \delta^{p+\rate} \bH_p +\delta D_T \right) [\phi] (T,X), \quad \norm{\rho_1 (T,\cdot)}_{H^m} \le C \delta \eps^n \norm{\phi (T, \cdot)}_{H^{m+n+p}}.
    \end{align}
    Indeed, the error associated with each truncation is $\bH_j - \bH_j^\eps = \ow\bigl((1-\chi_\eps^2) b_j\bigr)$. Since $1-\chi^2$ vanishes near the origin and is bounded with all its derivatives, $|\partial_\zeta^\beta(1-\chi^2(\eps\zeta))| \le C_{\beta,n}(\eps|\zeta|)^n\aver{\zeta}^{-|\beta|}$ for all $\beta$ and $n$, uniformly in $0<\eps\le1$. Hence $(1-\chi_\eps^2)b_j \in \eps^n S(\aver{\zeta}^{j+1+n})$ uniformly in $\eps$, and Lemma \ref{lemma:CV} implies that
    \begin{align}\label{eq:cutoff_bd}
        \norm{(\bH_j - \bH_j^\eps) \phi (T, \cdot)}_{H^m} \le C \eps^n \norm{\phi (T, \cdot)}_{H^{m+n+j+1}}.
    \end{align}
    
    It remains to control $\rho_0$. We write
    \begin{align*}
        (P+S^*) D_T \phi &= -\cG^{1/2} \rH e^{-iT \rH} \cG^{1/2} \phi_0\\
        &= -((P+S^*) \bHe (P+S)+ \delta^{-1} S^* (a^0 - E) S) \cG^{-1/2} e^{-iT \rH} \cG^{1/2} \phi_0 -\cG^{1/2} (\bH_{0\ell} -\bH_{0\ell}^\eps) e^{-iT \rH} \cG^{1/2} \phi_0\\
        &= -(P+S^*) \left( \bHe + \delta^{-1} (a^0 - E)\right) \phi - \cG^{1/2} (\bH_{0\ell} -\bH_{0\ell}^\eps) \cG^{-1/2} (P+S^*) \phi,
    \end{align*}
    and again conclude that the truncation error decays super-algebraically in $\eps$,
    \begin{align*}
        \norm{\delta \cG^{1/2} (\bH_{0\ell} -\bH_{0\ell}^\eps) \cG^{-1/2} (P+S^*) \phi (T, \cdot)}_{H^m} \le C \delta \eps^n \norm{\cG^{-1/2} (P+S^*) \phi (T, \cdot)}_{H^{m+n+1}}.
    \end{align*}
    It follows from the uniform boundedness of $S^*$ and $\cG^{-1/2}$ in $(\delta, \eps)$ that
    \begin{align}\label{eq:rho_0_bd}
    \begin{split}
        \norm{(P+S^*) \rho_0 (T, \cdot)}_{H^m} &\le \delta^{p+\rate} \norm{(P+S^*) \bH_p \phi (T, \cdot)}_{H^m} + C \delta \eps^n \norm{\phi (T, \cdot)}_{H^{m+n+1}} \\
        &\le C \left( \delta^{p+\rate} \norm{\phi(T, \cdot)}_{H^{m+p+1}} + C \delta\eps^n \norm{\phi (T, \cdot)}_{H^{m+n+1}}\right).
    \end{split}
    \end{align}
    On the other hand, a calculation reveals that
    \begin{align*}
        (I - P - S^*) \rho_0 = (I - P - S^*) (B-\mathcal E) \phi, 
    \end{align*}
    where
    \begin{align*}
        B &= a^0 - E + \delta \bHe - \delta S \cG^{-1} (P+S^*) \bHe - S \cG^{-1} S^* (a^0 - E)+\delta^{p+\rate} \bH_p,\\
        \mathcal E &= \delta S \cG^{-1/2} (\bH_{0\ell} - \bH_{0\ell}^\eps) \cG^{-1/2} (P+S^*).
    \end{align*}
    As above, we find that
    \begin{align*}
        \norm{(I-P-S^*) \mathcal E \phi (T, \cdot)}_{H^m} \le C \delta\eps^n \norm{\phi (T, \cdot)}_{H^{m+n+1}}.
    \end{align*}
    To control the rest, we recall Definition \ref{def:S}, which constructs $S$ by eliminating from $(I - P) B (P+S)$, and hence from $(I - P - S^*) B (P+S) = (I-S^*)(I-P)B(P+S)$, all $O(\delta^j)$ terms for $j \le p$. The symbols $b_j^\eps$ are bounded in $S(\aver{\zeta}^{j+1})$ uniformly in $\eps$, so that $\norm{\bH_j^\eps}_{H^{s+j+1} \to H^s} \le C$ uniformly in $\eps$, while $\norm{\bH_j^\eps}_{H^{s+k} \to H^s} \le C\eps^{k-j-1}$ for $0\le k\le j+1$ by \eqref{eq:bH_bd}. Every remaining term of order $\delta^k$, $k \ge p+1$, in $(I-P-S^*)B(P+S)$ other than the one containing $\bH_p$ is a composition of the $\bH_i^\eps$ and of constant matrices of total order at most $k$. It therefore maps $H^{m+p+1}$ to $H^m$ with norm at most $C\eps^{\min(0,\,p+1-k)}$, and $\delta^k\eps^{p+1-k} = \delta^{p+1}(\delta/\eps)^{k-p-1} \le \delta^{p+1}$ since $\delta<\eps$. The infinitely many terms generated by $\cG^{-1}$ are summed using $\norm{S}_{H^m\to H^m}\le\frac12$.
    It follows that $(I - P - S^*) B (P+S) = B_0 + \delta^{p+\rate} (I - P - S^*) \bH_p (P+S)$, where
    \begin{align*}
        \norm{B_0}_{H^{m+p+1} \to H^m} \le C \delta^{p+1}, \qquad \norm{\bH_p}_{H^{m+p+1} \to H^{m}} \le C.
    \end{align*}
    By the boundedness of $S^*$ and $\cG^{-1}$, we have therefore shown that
    \begin{align*}
        \norm{(I-P-S^*) \rho_0 (T, \cdot)}_{H^m} &\le C \left( \delta^{p+1} \norm{\cG^{-1} (P+S^*)\phi (T, \cdot)}_{H^{m+p+1}} + \delta^{p+\rate} \norm{\phi (T, \cdot )}_{H^{m+p+1}} + \delta \eps^n \norm{\phi (T, \cdot)}_{H^{m+n+1}}\right)\\
        &\le C \left( \delta^{p+\rate} \norm{\phi (T, \cdot )}_{H^{m+p+1}} + \delta\eps^n \norm{\phi (T, \cdot)}_{H^{m+n+1}}\right).
    \end{align*}
    Combining this with \eqref{eq:rho_0_bd}, we conclude that
    \begin{align*}
        \norm{\rho_0(T, \cdot)}_{H^m} \le C\left( \delta^{p+\rate} \norm{\phi (T, \cdot )}_{H^{m+p+1}} + \delta\eps^n \norm{\phi (T, \cdot)}_{H^{m+n+1}}\right).
    \end{align*}
    Recalling \eqref{eq:rho}, we have now shown that $\rho := \rho_0 + \rho_1$ satisfies
    \begin{align*}
        \norm{\rho(T, \cdot)}_{H^m} \le C\left( \delta^{p+\rate} \norm{\phi (T, \cdot )}_{H^{m+p+1}} + \delta\eps^n \norm{\phi (T, \cdot)}_{H^{m+n+p}}\right).
    \end{align*}
    The result then follows from Lemma \ref{lemma:Sobolev_rH} and the uniform boundedness of $P+S$ and $\cG^{\pm 1/2}$.
\end{proof}
\begin{lemma}\label{lemma:Sobolev_rH}
    Let $\phi_0 \in H^\infty (\mathbb{R}^d; \Ran P)$ and $m \in \mathbb{N}_0$. Take $\rH$ as in \eqref{eq:def_rH}. Then for any $\eta_0 > 0$, there exists a constant $C > 0$ such that, for $0<\delta\le c_0\eps$ and $\eps\le1$, the function
    $\phi := e^{-iT \rH} \phi_0$ satisfies
    \begin{align*}
        \norm{\phi (T, \cdot)}_{H^m} \le
        \begin{cases}
        C, & \ell \in \cL_1, \quad 0 < \eta \le \eta_0,\\
        C \left( 1+T^{m}\right), & \ell \in \cL_0, \quad 0 < \eta \le \eta_0,
        \end{cases}
        \qquad T \ge 0.
    \end{align*}
\end{lemma}
\begin{proof}
    If $\ell \in \cL_0$, then $\rH$ is uniformly bounded in $L^2$ and the result follows from Lemma \ref{lemma:bdd_reg} with $\Lambda = \aver{D_X}$. Note that its hypothesis on $[\rH,\Lambda]$ holds uniformly in $(\delta,\eps)$, meaning that for any $k \ge 0$,
    \begin{align*}
        \norm{[\rH,\Lambda]}_{H^k\to H^k}\le C, \qquad 0 < \delta \le c_0 \eps, \quad \eps \le 1.
    \end{align*}
    Indeed, $[\bH_{0\ell},\Lambda]$ is bounded on every $H^k$ since $b_{0\ell}\in S(1)$, and each term $\delta^j\bH_i^\eps$ entering $\rH-\bH_{0\ell}$ has a commutator with $\Lambda$ bounded on $H^k$ by the same quantity $C\delta^j\eps^{-(i+1)}$ as the term itself, since $\zeta$-derivatives of $b_i^\eps$ do not increase its size. 

    Suppose now that $\ell \in \cL_1$. Since $\rH$ acts on $L^2(\mathbb{R}^d;\Ran P)$ and $\phi_0 \in \Ran P$, we have $\phi(T,\cdot) \in \Ran P$ for all $T$. The operator $\Lambda := (i+\rH)^m$ commutes with $\rH$, is bounded from $H^m (\mathbb{R}^d; \Ran P)$ to $L^2$, and satisfies the norm equivalence \eqref{eq:equivalent_norms_rH}. Lemma \ref{lemma:elliptic_reg} with $r \equiv 0$ therefore gives $\norm{\phi (T, \cdot)}_{H^m} \le C \norm{\phi_0}_{H^m}$ for all $T \ge 0$.
\end{proof}
We now improve Proposition \ref{prop:res_rH} in the case that the leading-order term of $\rH$ vanishes.
\begin{proposition}\label{prop:res_rH_flat}
    Take $\phi_0, m, n, \eta_0, c_0, H_\delta, \psi$ as in Proposition \ref{prop:res_rH}, and let $\ell \in \cL_0$. Suppose there exists a function $\nu = \nu (\delta, \eps, \eta)$ such that \eqref{eq:nu_bound} holds uniformly in $0 < \delta \le c_0 \eps$, $\eps \le 1$, and $\eta \le \eta_0$. Then
    \begin{align*}
        \norm{(D_t + H_\delta) \psi (t, \cdot)}_{H^m} \le C (\delta^{p+\mu} + \delta \eps^n) (1+ (\delta \nu t)^{m+n+p})
    \end{align*}
    uniformly in all parameters.
\end{proposition}
\begin{proof}
    This result follows immediately from Lemma \ref{lemma:Sobolev_rH_flat} and the proof of Proposition \ref{prop:res_rH}.
\end{proof}
\begin{lemma}\label{lemma:Sobolev_rH_flat}
    Take $\phi_0, m, \rH, \eta_0, c_0, \phi$ as in Lemma \ref{lemma:Sobolev_rH}, and let $\ell \in \cL_0$. Suppose there exists a function $\nu = \nu (\delta, \eps, \eta)$ such that \eqref{eq:nu_bound} holds uniformly in $0 < \delta \le c_0 \eps$, $\eps \le 1$, and $\eta \le \eta_0$. Then
    \begin{align*}
        \norm{\phi (T, \cdot)}_{H^m} \le
        C \left( 1+(\nu T)^{m}\right), \qquad 0 < \delta \le c_0 \eps, \quad 0 < \eps \le 1, \quad 0 < \eta \le \eta_0,
        \quad T \ge 0.
    \end{align*}
\end{lemma}
\begin{proof}
The proof follows the $\ell \in \cL_0$ case of Lemma \ref{lemma:Sobolev_rH}, but applies Corollary \ref{cor:flat} with $\bH := \nu^{-1}\rH$ and $\Lambda := \aver{D_X}$ instead of Lemma \ref{lemma:bdd_reg}. By \eqref{eq:nu_bound}, $\bH$ is bounded on $L^2$ and $[\bH,\Lambda]$ is bounded on every $H^k$, uniformly in $(\delta,\eps,\eta)$, so that the hypotheses of Corollary \ref{cor:flat} hold uniformly.
\end{proof}

\bibliographystyle{siam}
\bibliography{refs}

\clearpage
\setcounter{page}{1}
\renewcommand{\thepage}{SM\arabic{page}}
\setcounter{section}{0}
\renewcommand{\thesection}{SM\arabic{section}}
\setcounter{equation}{0}
\renewcommand{\theequation}{SM\arabic{equation}}
\setcounter{proposition}{0}
\setcounter{figure}{0}
\renewcommand{\thefigure}{SM\arabic{figure}}
\setcounter{table}{0}
\renewcommand{\thetable}{SM\arabic{table}}

\begin{center}
    {\LARGE Supplementary material}\\[1.5ex]
    {\Large Effective Hamiltonians and asymptotic band coupling}\\[2ex]
    {\large Guillaume Bal, Paul Cazeaux, Daniel Massatt, and Solomon Quinn}
\end{center}
\vspace{2ex}

\noindent This supplement contains the derivations of the results summarized in Section
\ref{sec:summary} of the main paper: the two coupled one-dimensional chains in Section
\ref{sec:applications}, graphene with a Kekul\'e-O modulation in Section \ref{sec:kekule}, and
the magnetized cold plasma in Section \ref{sec:plasma}. To make it readable on its own, Section
\ref{sec:sm_setting} recalls the setting, the notation and the formulas of the main paper that are
used in the derivations. Section \ref{sec:sm_numerics} reports further numerical experiments on the
effective Hamiltonians of Section \ref{sec:effective}, and Section \ref{sec:sm_notation} lists the notation of both documents. Sections, equations and results whose numbers carry the prefix SM belong
to this supplement; all others refer to the main paper. Citations refer to the bibliography of
the main paper.

\section{Setting and notation}\label{sec:sm_setting}

We recall the constructions of the main paper that are used below, and refer to the main paper
for the precise assumptions and for the proofs.

\paragraph{Microscopic problem.} Let $a = a(X,\xi;\delta)$ be an $N\times N$ Hermitian-valued
symbol, set $a_\delta(x,\xi) := a(\delta x,\xi;\delta)$ and $H_\delta := \ow a_\delta$, with
$\ow$ the Weyl quantization, and consider
\begin{align}\label{eq:sm_micro}
    (D_t + H_\delta)\varphi_\delta = 0, \qquad \varphi_\delta(0,x) = e^{iK\cdot x}\delta^{\frac d2}\phi_0(\delta x),
\end{align}
where $D_z := \frac1i\partial_z$, $K\in\mathbb{R}^d$ is a reference quasi-momentum and
$\phi_0\in H^\infty(\mathbb{R}^d;\C^N)$ a macroscopic profile. Slow variables are denoted by
$T = \delta t$ and $X = \delta x$, and the rescaled momentum by $\zeta = (\xi-K)/\delta$.
Operators acting in the slow variables are written in boldface.

\paragraph{Expansion of the symbol and spectral sectors.} We assume (Assumption
\ref{assumption:a_higher_order_all}) that, for some $p\in\mathbb{N}$ and $0<\rate\le1$,
\begin{align}\label{eq:sm_decomp}
    a(X,K+\delta\zeta;\delta) = a^0 + \sum_{j=0}^{p-1}\delta^{j+1}b_j(X,\zeta;\delta) + \delta^{p+\rate}b_p(X,\zeta;\delta),
    \qquad b_j \in S(\aver{\zeta}^{j+1}),
\end{align}
with $a^0 = \sum_{\ell=1}^L E_\ell P_\ell$ a constant Hermitian matrix, $E_\ell$ its distinct
eigenvalues and $P_\ell$ the corresponding orthogonal projections. Each $\Ran P_\ell$ is called a
(spectral) sector. We write
\begin{align}\label{eq:sm_ops}
    \bH_j := \ow b_j, \qquad b_{0\ell} := P_\ell b_0P_\ell, \qquad \bH_{0\ell} := \ow b_{0\ell}, \qquad
    \ashl := (a^0-E_\ell)\big\vert_{\Ran(I_N-P_\ell)}^{-1} = \sum_{\ell'\ne\ell}\frac{P_{\ell'}}{E_{\ell'}-E_\ell}.
\end{align}
Each sector is either of class $\cL_0$, meaning that $b_{0\ell}\in S(1)$ (bounded, e.g.\ flat
or asymptotically flat bands), or of class $\cL_1$, meaning that $b_{0\ell}\in S^1$ is elliptic
in the sense that
\begin{align}\label{eq:sm_ellipticity}
    |\det(b_{0\ell}(X,\zeta;\delta) + (I_N-P_\ell)\aver{\zeta})|^{1/N} \ge c\aver{\zeta} - c^{-1}
\end{align}
for some $c>0$ (e.g.\ Dirac-type sectors). For continuous models such as the cold plasma, the
requirement $a\in S(1)$ is replaced by Assumption \ref{assumption:pde}; all results recalled
below then hold unchanged (Proposition \ref{prop:pde_class}).

\paragraph{Expansion with prescribed data.} For each sector $\ell$, let $\phi^0_\ell$ solve
\begin{align}\label{eq:sm_phi0}
    (D_T + \bH_{0\ell})\phi^0_\ell = 0, \qquad \phi^0_\ell(0,\cdot) = P_\ell\phi_0,
\end{align}
and set $\phin^{0,\ell} := \phi^0_\ell$, $\phout^{0,\ell} := 0$. For $1\le j\le p$, define
$\phi^j_\ell := \phin^{j,\ell} + \phout^{j,\ell}$ recursively by the algebraic correctors
\begin{align}\label{eq:sm_phout}
    \phout^{j,\ell} := -\ashl\Bigl(D_T\phi^{j-1}_\ell + \sum_{i=0}^{j-1}\bH_i\phi^{j-1-i}_\ell\Bigr) \in \Ran(I_N-P_\ell),
\end{align}
and, for $1\le j\le p-1$, by the solutions $\phin^{j,\ell}\in\Ran P_\ell$ of
\begin{align}\label{eq:sm_phin}
    (D_T + \bH_{0\ell})\phin^{j,\ell} = -P_\ell\Bigl(\bH_0\phout^{j,\ell} + \sum_{i=1}^j\bH_i\phi^{j-i}_\ell\Bigr),
    \qquad \phin^{j,\ell}(0,\cdot) = -P_\ell\sum_{\ell'=1}^L\phout^{j,\ell'}(0,\cdot),
\end{align}
while $\phin^{p,\ell} :\equiv 0$. The initial condition in \eqref{eq:sm_phin} redistributes the
algebraic correctors of all sectors at $T = 0$, so that the approximate solution
\begin{align}\label{eq:sm_psi}
    \psi_\delta(t,x) := \sum_{\ell=1}^L e^{i(K\cdot x - E_\ell t)}\delta^{\frac d2}\phi_\ell(\delta t,\delta x),
    \qquad \phi_\ell := \sum_{j=0}^p\delta^j\phi^j_\ell,
\end{align}
matches the datum of \eqref{eq:sm_micro}. By Theorem \ref{thm:main_higher_order},
\begin{align}\label{eq:sm_estimate}
    \norm{(\psi_\delta - \varphi_\delta)(t,\cdot)}_{H^m} \le C(\delta^p + \delta^{p+\rate}t)\bigl(1+(\delta t)^{2m+2p}\bigr),
    \qquad t\ge0.
\end{align}

\paragraph{Band-adapted data.} For $\phi_0\in H^\infty(\mathbb{R}^d;\Ran P_\ell)$, the
band-adapted (or dressed) datum $\Pi^\delta_\ell\phi_0$ is $\phi_0$ plus the algebraic correctors
it generates at $T=0$ when the redistributed data in \eqref{eq:sm_phin} are set to zero
(Definition \ref{def:single}). By Lemma \ref{lem:Pi_S},
\begin{align}\label{eq:sm_Pi}
    \Pi^\delta_\ell = P_\ell + \sum_{j=1}^p\delta^jS^0_j, \qquad
    S^0_1 = -\ashl\bH_0P_\ell, \qquad
    S^0_2 = -\ashl\bigl(\bH_1P_\ell + \bH_0S^0_1 - S^0_1\bH_{0\ell}\bigr).
\end{align}
The band-adapted expansion of Theorem \ref{thm:equiv} launches each sector $\ell$ from
$P_\ell\mathcal T^{(q)}_\delta\phi_0$, where $\mathcal T^{(q)}_\delta$ is the truncated Neumann
inverse of the dressing map $\mathcal T_\delta := \sum_\ell\Pi^\delta_\ell P_\ell$, after which no
data are redistributed between sectors.

\paragraph{Effective Hamiltonians.} The effective Hamiltonian $\rH^{(\ell)}$ of the sector $\ell$,
defined in \eqref{eq:def_rH}, is the Hamiltonian induced on $\Ran P_\ell$. 
Its unbounded coefficients are truncated at frequencies
$|\zeta|\gtrsim\eps^{-1}$, with $\eta = \delta/\eps^2\ll1$, through
$\bH_j^\eps := \ow(\chi_\eps b_j\chi_\eps)$, $\chi_\eps(\zeta) = \chi(\eps\zeta)$ for a real
$\chi\in C^\infty_c$ equal to one near the origin, and $\bH^\eps_{0\ell} := P_\ell\bH^\eps_0P_\ell$.
By Remark \ref{rem:rH_expansion}, with $P = P_\ell$, $a^\sharp = \ashl$ and $\Re X := \frac12(X+X^*)$,
\begin{align}\label{eq:sm_rH}
\begin{split}
    \rH^{(\ell)} &\sim \sum_{j\ge0}\delta^j\rH_j, \qquad \rH_0 = P\bH_0P, \qquad
    \rH_1 = P\bH^\eps_1P - P\bH^\eps_0a^\sharp\bH^\eps_0P,\\
    \rH_2 &= P\bH^\eps_2P - 2\Re(P\bH^\eps_1a^\sharp\bH^\eps_0P) + P\bH^\eps_0a^\sharp\bH^\eps_0a^\sharp\bH^\eps_0P
    - \Re(P\bH^\eps_0P\bH^\eps_0(a^\sharp)^2\bH^\eps_0P),
\end{split}
\end{align}
and frequency cutoffs applied to bounded operators may be removed. By Theorem \ref{prop:rH} and
Corollary \ref{cor:rH_eps}, the dynamics generated by $\sum_{j<p}\delta^j\rH_j$ approximate the
tight-binding dynamics of band-adapted data with an error $O(\delta^{p+\rate}(1+t))$ in $L^2$,
for all $t\ge0$, on sectors of class $\cL_1$, and with an additional factor polynomial in
$\delta t$ on sectors of class $\cL_0$. The inter-sector couplings and the self-energies are
\begin{align}\label{eq:sm_MSigma}
    \mathbf M_{\ell\ell'} := P_\ell\bH_0P_{\ell'}\ (\ell\ne\ell'), \qquad
    \Sigma_\ell := -P_\ell\bH_0\ashl\bH_0P_\ell = \sum_{\ell'\ne\ell}\frac{\mathbf M_{\ell\ell'}\mathbf M_{\ell'\ell}}{E_\ell - E_{\ell'}},
\end{align}
so that $\rH_1 = P_\ell\bH_1P_\ell + \Sigma_\ell$ up to frequency truncation.

\section{Derivations for coupled one-dimensional chains}\label{sec:applications}

This section contains the derivations for the two models summarized in Section \ref{sec:summary_chains} of the main paper. These models illustrate the mechanisms of Section \ref{sec:tb} in closed form.
In each case we first expand the tight-binding symbol and verify that it satisfies Assumption \ref{assumption:a_higher_order_all}. We then present
the macroscopic operators of Section \ref{sec:sm_setting}, followed by the expansion of the effective solution $\phi$ from \eqref{eq:sm_psi}-\eqref{eq:sm_phin}.
We conclude with the band-adapted expansion of Theorem \ref{thm:equiv}.

\subsection{Two coupled SSH chains}\label{sec:ssh_two}

\subsubsection{Tight-binding symbol and expansion}\label{sec:ssh_two_symbol}

Suppose we have a system of two one-dimensional lattices, each with lattice constant $v > 0$. We assume each lattice is described by an SSH-type model, so that the symbol corresponding to lattice $j$ is
\begin{align}\label{eq:a_jj}
    a_{jj} (X, \xi; \delta) = (-1)^{j-1} E I_2 + \alpha_j (X) \begin{bmatrix}
        0 & e^{i v \xi/2} - e^{-iv \xi/2}\\
        e^{-i v \xi/2} - e^{iv \xi/2} & 0
    \end{bmatrix} + \delta M_j (X), \qquad (X, \xi) \in \mathbb{R}^2.
\end{align}
Here, $E > 0$ is a fixed constant, and $\alpha_j$ and $M_j$ are assumed to be $C^\infty_b$ (scalar and Hermitian-valued, respectively) functions, with $\alpha_j (X) \ge c > 0$.
The coupling between the lattices is given by
\begin{align*}
    a_{12} (X; \delta) = \delta \beta (X),
\end{align*}
for some $\beta \in C^\infty_b (\mathbb{R}; \mathbb{C}^{2 \times 2})$. The full tight-binding Hamiltonian is then $H_\delta = \ow a_\delta$, where
\begin{align}\label{eq:a_coupled_SSH}
    a_\delta (x, \xi) = a (\delta x, \xi; \delta), \qquad a (X, \xi; \delta) = \begin{bmatrix}
        a_{11} (X, \xi; \delta) & a_{12} (X; \delta)\\
        a_{12}^* (X; \delta) & a_{22} (X, \xi; \delta)
    \end{bmatrix}.
\end{align}
We now show that $a_\delta$ satisfies Assumption \ref{assumption:a_higher_order_all} with $K = 0$.
It is clear that $a \in S(1)$ uniformly in $\delta$. To obtain the decomposition \eqref{eq:sm_decomp}, we simply Taylor expand $a$ about $(\xi; \delta) = (0; 0)$. The symbols $a^0$ and $b_j$ are in the appropriate classes \eqref{eq:sm_decomp} and given by
\begin{align*}
    a^0 = \begin{bmatrix}
        E I_2 & 0 \\
        0 & -E I_2
    \end{bmatrix}, \qquad b_0 (X, \zeta) = \begin{bmatrix}
        -v \alpha_1 (X) \zeta \sigma_2 + M_1(X) & \beta (X)\\
        \beta^* (X) & -v \alpha_2 (X) \zeta \sigma_2 + M_2(X)
    \end{bmatrix}.
\end{align*}
We thus have 
\begin{align*}
E_1 = E, \qquad E_2 = -E, \qquad 
    P_1 = \begin{bmatrix}
        I_2 & 0\\
        0 & 0
    \end{bmatrix}, \qquad P_2 =
    \begin{bmatrix}
        0 & 0\\
        0 & I_2
    \end{bmatrix}.
\end{align*}
Since each $\alpha_\ell$ is bounded away from zero, we have $b_{0\ell} = P_\ell b_0 P_\ell \in S^1$ with
\begin{align*}
    \det \bigl( b_{0\ell} + (I_4 - P_\ell) \aver{\zeta} \bigr) = v^2\alpha_\ell^2 \zeta^2 \aver{\zeta}^2+ O(\aver{\zeta}^3),
\end{align*}
so that the ellipticity condition \eqref{eq:sm_ellipticity} holds with $N = 4$. Hence both sectors
belong to $\cL_1$ (cf. \eqref{eq:sm_ellipticity}). Therefore, Theorem \ref{thm:main_higher_order} applies, allowing for
approximations of the tight-binding model to arbitrary orders of accuracy.

\subsubsection{Macroscopic operators}\label{sec:ssh_two_ops}

The objects entering the expansions are the following. The resolvents on the complementary
subspaces are the constant matrices
\begin{align}\label{eq:ssh_ash}
    \inva_1 = (a^0 - E_1 I_4)^{-1}\big\vert_{\Ran (I_4-P_1)} = -\frac{1}{2E}P_2,
    \qquad
    \inva_2 = \frac{1}{2E}P_1 .
\end{align}
The sector Hamiltonians $\bH_{0\ell}$ are defined on $\Ran P_\ell$ by
\begin{align}\label{eq:ssh_b0l}
\bH_{0\ell} = \ow b_{0\ell}, \qquad 
    b_{0\ell}(X,\zeta) := -v\alpha_\ell (X)\zeta \sigma_2 + M_\ell (X) \in S^1,
\end{align}
each a one-dimensional Dirac operator with Fermi velocity $v\alpha_\ell$ and mass $M_\ell$.
The inter-sector couplings $\mathbf M_{\ell\ell'} := P_\ell\bH_0P_{\ell'}$, $\ell\ne\ell'$, are given by
\begin{align}\label{eq:ssh_M}
    \mathbf M_{12} = P_1 \bH_0 P_2 = \ow \beta, \qquad \mathbf M_{21} = \ow \beta^*,
\end{align}
and the corresponding self-energies
$\Sigma_\ell := -P_\ell\bH_0\ashl\bH_0P_\ell = \sum_{\ell'\ne\ell}\mathbf M_{\ell\ell'}\mathbf M_{\ell'\ell}/(E_\ell-E_{\ell'})$ are
\begin{align}\label{eq:ssh_Sigma}
    \Sigma_1 = \frac{1}{2E}\,\ow \beta\, \ow\beta^*, \qquad
    \Sigma_2 = -\frac{1}{2E}\,\ow\beta^*\,\ow\beta .
\end{align}
Note that the operators in \eqref{eq:ssh_M} and \eqref{eq:ssh_Sigma} are point-wise multiplications.
Since $b_1 = 0$, \eqref{eq:sm_rH} gives the first-order coefficients $\rH_1 = \Sigma_\ell$ of the
effective Hamiltonians \eqref{eq:def_rH}, and no frequency cutoff is needed at this order since $\beta$ is bounded.

The second-order coefficients of \eqref{eq:sm_rH} are $\rH_2 = P_\ell\bH^\eps_2P_\ell + \hat Q_\ell$, where
\begin{align*}
    \hat Q_1 &= \frac{1}{4E^2} \left(\ow(\beta) \bH^\eps_{02} \ow(\beta^*) - \frac12\bigl(\ow (\beta \beta^*) \bH^\eps_{01} + \bH^\eps_{01}\ow (\beta \beta^*)\bigr) \right), \\
    \hat Q_2 &= \frac{1}{4E^2} \left(\ow(\beta^*) \bH^\eps_{01} \ow(\beta) - \frac12\bigl(\ow (\beta^* \beta) \bH^\eps_{02} + \bH^\eps_{02}\ow (\beta^* \beta)\bigr) \right),
\end{align*}
with $\bH^\eps_{0\ell} = P_\ell\bH^\eps_0P_\ell$ the frequency-truncated sector Hamiltonians. Both are symmetric.

\subsubsection{Expansion with prescribed data}\label{sec:ssh_two_exp}

For $p=1$, the function $\psi_\delta$ in
\eqref{eq:sm_psi} is given by
\begin{align*}
    \psi_\delta (t,x) = \delta^{d/2} \left( e^{-iEt} \phi_1 (\delta t, \delta x; \delta) + e^{iEt} \phi_2 (\delta t, \delta x; \delta) \right),
\end{align*}
where the $\phi_\ell = \phi^0_\ell + \delta \phi^1_\ell$ satisfy 
\begin{align}\label{eq:phi_p_1}
    \phi^0_1 = \begin{bmatrix}
        \tilde\phi^0_1 \\ 0
    \end{bmatrix}, \qquad 
    \phi^0_2 = \begin{bmatrix}
        0 \\ \tilde\phi^0_2 
    \end{bmatrix}, \qquad \phi^1_1 = \frac{1}{2E} \begin{bmatrix}
        0 \\ \beta^* \tilde \phi^0_1
    \end{bmatrix}, \qquad \phi^1_2 = -\frac{1}{2E}\begin{bmatrix}
         \beta \tilde \phi^0_2 \\ 0
    \end{bmatrix},
\end{align}
with $\tilde \phi^0_\ell$ the solution to the initial value problem
\begin{align}\label{eq:Schrodinger_p_1}
    (D_T + \bH_{0\ell}) \tilde \phi^0_\ell = 0, \qquad \tilde \phi^0_\ell \vert_{T=0} = \tilde \phi_{0\ell}.
\end{align}
Here, $(\tilde \phi_{01}, \tilde \phi_{02})= \phi_0 \in H^\infty$ is the initial condition from \eqref{eq:sm_micro}.

If $p=2$, we instead have $\phi_\ell = \phi^0_\ell + \delta \phi^1_\ell + \delta^2 \phi^2_\ell$, where $\phi^0_\ell$ and $\tilde \phi_\ell^0$ are given by \eqref{eq:phi_p_1}-\eqref{eq:Schrodinger_p_1}, and
\begin{align*}
    \phi^1_1 = \frac{1}{2E} \begin{bmatrix}
        \tilde \phi^1_1 \\ \beta^* \tilde \phi^0_1
    \end{bmatrix}, \qquad \phi^1_2 = -\frac{1}{2E}\begin{bmatrix}
         \beta \tilde \phi^0_2 \\ \tilde \phi^1_2 \end{bmatrix}, \qquad
         \phi_1^2 = \frac{1}{4E^2} \begin{bmatrix}
             0 \\ \tilde \phi_1^2
         \end{bmatrix}, \qquad
         \phi_2^2 = \frac{1}{4E^2} \begin{bmatrix}
             \tilde \phi_2^2 \\ 0
         \end{bmatrix}.
\end{align*}
Here, 
\begin{align*}
    \tilde \phi_1^2 = (-\ow (\beta^*) \bH_{01} + \bH_{02} \ow (\beta^*)) \tilde \phi^0_1 + \beta^* \tilde \phi^1_1, \qquad 
    \tilde \phi_2^2 = (-\ow (\beta) \bH_{02} + \bH_{01} \ow (\beta)) \tilde \phi^0_2 + \beta \tilde \phi^1_2,
\end{align*}
with $\tilde \phi_1^1$ and $\tilde \phi_2^1$ respectively the solutions to the initial value problems
\begin{align}\label{eq:Schrodinger_p_2}
\begin{split}
    (D_T + \bH_{01}) \tilde \phi^{1}_1 = -(\beta \beta^* + 2E \bH_{11})\tilde \phi^0_1, \qquad \tilde \phi^1_1(0,\cdot) &= \beta \tilde \phi_{02},\\
    (D_T + \bH_{02}) \tilde \phi^{1}_2 = (-\beta^* \beta + 2E \bH_{12})\tilde \phi^0_2, \qquad \tilde \phi^1_2(0,\cdot) &= \beta^* \tilde \phi_{01},
\end{split}
\end{align}
where 
$\diag (\bH_{11}, \bH_{12}) : =\bH_1$. Note that the off-diagonal blocks of $\bH_j$ for $j \ge 1$ all vanish since the off-diagonal blocks of $a$ in \eqref{eq:a_coupled_SSH} are linear in $\delta$ and independent of $\xi$.

\subsubsection{Band-adapted expansion}\label{sec:ssh_two_ad}
As discussed in Section \ref{sec:dressed}, one could modify the construction so that only the leading-order Schr\"odinger equation has nontrivial initial data. That is, with $\phi_\ell = \phi_\ell^0 +\delta \phi_\ell^1 + \delta^2 \phi_\ell^2$ as before but now 
\begin{align*}
    \phi^0_1 = \begin{bmatrix}
        \mathring\phi^0_1 \\ 0
    \end{bmatrix}, \qquad 
    \phi^0_2 = \begin{bmatrix}
        0 \\ \mathring\phi^0_2 
    \end{bmatrix},
\end{align*}
the initial value problem \eqref{eq:Schrodinger_p_1} could be replaced by
\begin{align*}
    (D_T + \bH_{01}) \mathring \phi^0_1 &= 0, \qquad \mathring{\phi}^0_1 \vert_{T=0} = \tilde \phi_{01} + \frac{\delta}{2E} \beta \tilde \phi_{02},\\
    (D_T + \bH_{02}) \mathring \phi^0_2 &= 0, \qquad \mathring\phi^0_2 \vert_{T=0} = \tilde \phi_{02} - \frac{\delta}{2E} \beta^* \tilde \phi_{01}.
\end{align*}
In this case, we would define
\begin{align*}
    \phi^1_1 = \frac{1}{2E} \begin{bmatrix}
        \mathring \phi^1_1 \\ \beta^* \mathring \phi^0_1
    \end{bmatrix}, \qquad \phi^1_2 = -\frac{1}{2E}\begin{bmatrix}
         \beta \mathring \phi^0_2 \\ \mathring \phi^1_2 \end{bmatrix}, \qquad
         \phi_1^2 = \frac{1}{4E^2} \begin{bmatrix}
             0 \\ \mathring \phi_1^2
         \end{bmatrix}, \qquad
         \phi_2^2 = \frac{1}{4E^2} \begin{bmatrix}
             \mathring \phi_2^2 \\ 0
         \end{bmatrix},
\end{align*}
where
\begin{align*}
    \mathring \phi_1^2 = (-\ow (\beta^*) \bH_{01} + \bH_{02} \ow (\beta^*)) \mathring \phi^0_1 + \beta^* \mathring\phi^1_1, \qquad 
    \mathring \phi_2^2 = (-\ow (\beta) \bH_{02} + \bH_{01} \ow (\beta)) \mathring \phi^0_2 + \beta \mathring \phi^1_2,
\end{align*}
with $\mathring \phi_1^1$ and $\mathring \phi_2^1$ respectively the solutions to the initial value problems
\begin{align*}
    (D_T + \bH_{01}) \mathring \phi^{1}_1 = -(\beta \beta^* + 2E \bH_{11})\mathring \phi^0_1, \qquad \mathring \phi^1_1(0,\cdot) &= 0,\\
    (D_T + \bH_{02}) \mathring \phi^{1}_2 = (-\beta^* \beta + 2E \bH_{12})\mathring \phi^0_2, \qquad \mathring \phi^1_2(0,\cdot) &= 0.
\end{align*}
By Theorem \ref{thm:equiv}, this construction produces an $O(\delta^2)$ approximation of the tight-binding dynamics for times $t \lesssim \delta^{-1}$.

\subsection{An SSH chain coupled to a flat band}\label{sec:ssh_flat}

\subsubsection{Symbol and spectral data}\label{sec:ssh_flat_symbol}
Let us now replace the second chain in Section \ref{sec:ssh_two} by a flat band.
More specifically, take $a_{11}$ as in \eqref{eq:a_jj} and let $\gamma \in C^\infty_b (\mathbb{R}; \mathbb{C}^2)$. Suppose $H_\delta = \ow a_\delta$, where
\begin{align*}
    a_\delta (x, \xi) = a (\delta x, \xi; \delta), \qquad a (X, \xi; \delta) = \begin{bmatrix}
        a_{11} (X, \xi; \delta) & \delta \gamma (X)\\
        \delta \gamma^* (X) & 0
    \end{bmatrix} \in \mathbb{C}^{3 \times 3}.
\end{align*}
We then have
\begin{align*}
    a^0 = \begin{bmatrix}
        E I_2 & 0 \\
        0 & 0
    \end{bmatrix}, \qquad b_0 (X, \zeta) = \begin{bmatrix}
        -v \alpha_1 (X) \zeta \sigma_2 + M_1(X) & \gamma (X)\\
        \gamma^* (X) & 0
    \end{bmatrix},
\end{align*}
and
\begin{align*}
    E_1= E, \qquad E_2 = 0, \qquad P_1 = \begin{bmatrix}
        I_2 & 0\\
        0 & 0
    \end{bmatrix}, \qquad P_2 = \begin{bmatrix}
        0 & 0\\
        0 & 1
    \end{bmatrix}.
\end{align*}
The symbol $b_{01} = P_1 b_0 P_1 \in S^1$ satisfies the ellipticity condition
\eqref{eq:sm_ellipticity} as before, so that $1 \in \cL_1$. On the other hand, $b_{02} = P_2 b_0 P_2 = 0$,
so that $2 \in \cL_0$. 
Sectors of this kind fall
outside the scope of the effective mass theorems of \cite{allaire2005homogenization}, which
require a nondegenerate critical point of the band function, and outside that of the Peierls
substitution of \cite{nenciu1991dynamics, panati2003effective}, which requires an isolated
band with nonvanishing dispersion. The class $\cL_0$ of Assumption
\ref{assumption:a_higher_order_all} is introduced precisely to accommodate them.

\subsubsection{Macroscopic operators}\label{sec:ssh_flat_ops}

Here $E_1 = E$ and $E_2 = 0$, so that
\begin{align}\label{eq:flat_ash}
    \inva_1 = -\frac1E P_2, \qquad \inva_2 = \frac1E P_1.
\end{align}
The sector Hamiltonians are
$\bH_{01} = \ow b_{01}$ with $b_{01}$ defined by \eqref{eq:ssh_b0l}, and $\bH_{02} = 0$. The
couplings $\mathbf M_{\ell\ell'} = P_\ell\bH_0P_{\ell'}$ (cf. \eqref{eq:sm_MSigma}) are $\mathbf M_{12} = \ow \gamma$ and $\mathbf M_{21} = \ow \gamma^*$.
Here $b_1 = 0$ and $P_2b_2P_2 = 0$, and every term of \eqref{eq:sm_rH} containing $\bH_{02}$ vanishes.
The first coefficients $\rH_0$, $\rH_1$, $\rH_2$ of the effective Hamiltonian \eqref{eq:def_rH} of the flat
sector are therefore
\begin{align}\label{eq:flat_SigmaQ}
    \bH_{02} = 0, \qquad
    \Sigma_2 = -\frac1E \ow(|\gamma|^2),
    \qquad
    \rH^{(2)}_2 = \frac{1}{E^2}\ow(\gamma^*)\,\bH^\eps_{01}\,\ow(\gamma),
\end{align}
where $\bH^\eps_{01} = P_1\bH_0^\eps P_1$ is the frequency-truncated SSH Hamiltonian; the cutoff is
not needed in $\Sigma_2$, since $\gamma$ is bounded. Thus the flat band
acquires, at order $\delta$, the real frequency shift $-|\gamma|^2/E$, and at order
$\delta^2$ the dispersion inherited from the SSH band through
$\rH^{(2)}_2$. These effects will be evident when applying the construction
\eqref{eq:sm_phout}--\eqref{eq:sm_phin} below.

\subsubsection{Expansion with prescribed data}\label{sec:ssh_flat_exp}
Abusing notation, for $j \ge 0$ we write
\begin{align*}
    \phi^j_1 = \begin{bmatrix}
        \phin^{j,1} \\ \phout^{j,1}
    \end{bmatrix}, \qquad
    \phi^j_2 = \begin{bmatrix}
        \phout^{j,2} \\ \phin^{j,2}
    \end{bmatrix},
\end{align*}
where $\phin^{j,1}$ and $\phout^{j,2}$ are $\mathbb{C}^2$-valued functions, while $\phout^{j,1}$ and $\phin^{j,2}$ are scalars.
We consider the initial data
\begin{align*}
    \phi_0 = \begin{bmatrix}
        0 \\ \phi_{00}
    \end{bmatrix},
\end{align*}
for some $\phi_{00} \in H^\infty (\mathbb{R}; \mathbb{C})$.
Since $P_1 \phi_0 = 0$, it immediately follows that $\phi^0_1 \equiv 0$.
On the other hand, we have $\bH_{02} = \ow b_{02}= 0$, which implies $D_T \phi^0_2 \equiv 0$, and thus $\phi^0_2 \equiv \phi_0$.
Using that
\begin{align*}
    \inva_2 = \frac{1}{E}\begin{bmatrix}
        I_2 & 0\\
        0 & 0
    \end{bmatrix},
\end{align*}
we find from \eqref{eq:sm_phout} that
\begin{align*}
    \phout^{1,2} = -\frac{1}{E} (D_T \phout^{0,2} + \bH_{01} \phout^{0,2} + \gamma \phin^{0,2}) = -\frac{1}{E} \gamma \phi_{00}.
\end{align*}
We now obtain higher-order terms.
Using that
\begin{align*}
    P_2\bH_0 \begin{bmatrix}
        \phout^{1,2}\\0
    \end{bmatrix} = -\frac{1}{E} |\gamma|^2 \phi_{00},
\end{align*}
we conclude by \eqref{eq:sm_phin} that
$
    D_T \phin^{1,2} = \frac{1}{E} |\gamma|^2 \phi_{00}
$ and
$\phin^{1,2} (0,\cdot) \equiv 0$,
and so $$\phin^{1,2} (T,X) = \frac{iT}{E}|\gamma(X)|^2 \phi_{00}(X).$$
The fact that $\phi^0_1 \equiv 0$ immediately implies that $\phout^{1,1} \equiv 0$. Since $\phout^{1,2}$ is independent of $T$, it follows that $\phin^{1,1}$ is the solution to the initial value problem
\begin{align*}
    (D_T + \bH_{01}) \phin^{1,1} = 0, \qquad \phin^{1,1}\vert_{T=0} = -\phout^{1,2}.
\end{align*}
Set $\phi^0 := \phi^0_1 + \phi^0_2 = \phi^0_2$. Using that $P_1 D_T \phi^1_2 \equiv 0$ 
and $P_1 \bH_1 \phi^0 \equiv \bH_1 P_1 \phi^0 \equiv 0$, and that $\inva_2 = \frac1E \diag (I_2, 0)$, we find that
\begin{align*}
    \phout^{2,2} = -\frac{1}{E} (D_T \phout^{1,2} + \bH_{01} \phout^{1,2} + \ow(\gamma) \phin^{1,2} ) = -\frac{1}{E} (\bH_{01} \phout^{1,2} + \ow(\gamma) \phin^{1,2} ).
\end{align*}
It follows that
\begin{align*}
    \phout^{2,2} = \frac{1}{E^2} \left(
        \bH_{01} [\phi_{00} \gamma] - iT |\gamma|^2  \phi_{00}\gamma \right).
\end{align*}
Now, the fact that $\phout^{1,1} \equiv 0$ implies that $P_2 D_T \phi^1_1 \equiv 0$. 
Applying \eqref{eq:sm_phout} with $\phi^0_1 \equiv 0$, and recalling that
$$\inva_1 = (a^0 - E_1 I_3)^{-1}\big\vert_{\Ran (I_3 - P_1)} = -\frac1E P_2,$$
we conclude that
\begin{align*}
    \phout^{2,1} = \frac{1}{E} (D_T \phout^{1,1} + \ow (\gamma^*) \phin^{1,1}+ \bH_{02} \phout^{1,1} ) = \frac{1}{E} \ow (\gamma^*) \phin^{1,1}= \frac{1}{E^2} 
        \gamma^* e^{-iT \bH_{01}} [\phi_{00}\gamma ].
\end{align*}
We have thus shown that a second-order approximation to the tight-binding dynamics is given by
\begin{align}\label{eq:ssh_flat_psi}
    \psi_\delta (t,x) = e^{-iEt} \delta^{1/2} \phi_1 (\delta t, \delta x) + \delta^{1/2} \phi_2 (\delta t, \delta x),
\end{align}
where $\phi_\ell (T,X) = \phi_\ell = \phi^0_\ell + \delta \phi^1_\ell + \delta^2 \phi^2_\ell$, with
\begin{align*}
    \phi^0_1 \equiv 0, \qquad \phi^1_1 = \frac{1}{E} \begin{bmatrix}
        e^{-iT \bH_{01}} [\phi_{00} \gamma]\\
        0
    \end{bmatrix}, \qquad \phi^2_1 = \frac{1}{E^2} \begin{bmatrix}
        0\\
        \gamma^* e^{-iT \bH_{01}} [\phi_{00} \gamma]
    \end{bmatrix},
\end{align*}
and
\begin{align*}
    \phi^0_2 = \begin{bmatrix}
        0 \\ \phi_{00}
    \end{bmatrix}, \qquad
    \phi^1_2 = \frac{iT}{E} |\gamma|^2 \begin{bmatrix}
        0 \\ \phi_{00}
    \end{bmatrix} - \frac{1}{E} \begin{bmatrix}
        \phi_{00} \gamma \\ 0
    \end{bmatrix}, \qquad
    \phi^2_2 = \frac{1}{E^2} \begin{bmatrix}
        \bH_{01} [\phi_{00} \gamma] - iT |\gamma|^2 \phi_{00} \gamma \\
        0
    \end{bmatrix}.
\end{align*}
Observe that
\begin{align*}
\phi_2 =: \begin{bmatrix}
    \phi_2^{\rm out} \\ \phi_2^{\rm in}
\end{bmatrix}, \qquad
    \phi_2^{\rm in} = \phi_{00} + \delta \frac{iT}{E} |\gamma|^2 \phi_{00} = e^{i \delta T |\gamma|^2 /E} \phi_{00} + O(\delta^2)= e^{-i \delta T \Sigma_2} \phi_{00} + O(\delta^2),
\end{align*}
with $\Sigma_2$ as in \eqref{eq:flat_SigmaQ}. Thus, as predicted by \eqref{eq:sm_rH}, the truncated effective Hamiltonian $\rH_0 + \delta\rH_1 = \bH_{02} + \delta \Sigma_2 = \delta \Sigma_2$ of the flat sector describes the Schr\"odinger dynamics of $\phi_2^{\rm in}$ to leading order.

\subsubsection{Band-adapted expansion}\label{sec:ssh_flat_ad}
We have $\phout^{1,2}(0,\cdot) = -E^{-1}\gamma\phi_{00}$, so that the truncated inverse $\mathcal T^{(2)}_\delta$ of the dressing map
of Section \ref{sec:sm_setting} reads
\begin{align}\label{eq:flat_T}
    \mathcal T^{(2)}_\delta \phi_0 = \phi_0 + \frac{\delta}{E}\begin{bmatrix} \gamma \phi_{00}\\ 0\end{bmatrix} =
    \begin{bmatrix}
        \delta \gamma \phi_{00}/E \\ \phi_{00}
    \end{bmatrix}.
\end{align}
Two band-adapted descriptions are available here, and they answer different questions.

First, the construction of Definition \ref{def:single} may be applied to the datum $\phi_0$
itself, which lies in $\Ran P_2$. By Proposition \ref{prop:dressed} (item 1), it gives
$\check\phi_1 \equiv 0$ at every order: the SSH sector is not excited at all. Its initial
value, however, is the band-adapted datum
\begin{align*}
    \phi^\delta_0 = \Pi^\delta_2\phi_0 = \phi_0 + \delta\phout^{1,2}(0,\cdot)
    = \begin{bmatrix} -\delta\gamma\phi_{00}/E \\ \phi_{00}\end{bmatrix}
\end{align*}
of \eqref{eq:sm_Pi}, so that this expansion approximates the solution of
\eqref{eq:sm_micro} with $\phi^\delta_0$ in place of $\phi_0$, by Proposition
\ref{prop:dressed} (item 4). The $O(\delta)$ SSH component of $\phi^\delta_0$ is exactly the algebraic
corrector $\delta\phout^{1,2}(0,\cdot)$ attached to the flat sector, so that no data need to be
redistributed and no SSH packet is emitted.

Second, Theorem \ref{thm:equiv} keeps the prescribed datum $\phi_0$ and launches each sector
from $P_\ell\mathcal T^{(2)}_\delta\phi_0$, with no redistribution of data at any order. By
\eqref{eq:flat_T}, the flat sector starts from $\phi_{00}$ and the SSH sector from
$\delta\gamma\phi_{00}/E$, so that the latter is excited, with leading term
$\frac{\delta}{E}e^{-iT\bH_{01}}[\gamma\phi_{00}]$. This is the term $\delta\phi^1_1$ of
Section \ref{sec:ssh_flat_exp}: the $O(\delta)$ SSH packet is still present, but it is carried
by the datum of its own sector rather than transferred at $T=0$ by the mechanism (M2) of
Remark \ref{rem:mechanisms}.

\section{Derivations for graphene with Kekul\'e-O modulation}\label{sec:kekule}
%
%
%
%

This section contains the derivations of the results summarized in Section \ref{sec:summary_kekule} of the main paper. We derive effective macroscopic Hamiltonians for all three band sectors of the Kekul\'e-O tight-binding model on a honeycomb lattice using the framework from Section \ref{sec:tb}. 

We refer to the review paper \cite{andrade2025topical} for derivations and motivations for this $6\times 6$ tight-binding symbol modeling the folding of $K$ and $K'$ valleys as well as the $\Gamma$ sector after tripling of the graphene unit cell. The zero-energy sector yields a $4$-component Dirac operator with
Kekul\'e mass at leading order.
The two high-energy sectors are asymptotically flat to leading order and acquire quadratic dispersions of opposite signs, together with a coupling to the Dirac sector mediated by the Kekul\'e order parameter $\Delta(X)$. This coupling enters the wave function at order $\delta^2$, through an algebraic corrector and, for expansions with $p \ge 3$, through the initial conditions of the $\Gamma$-point sectors, and it enters the effective Hamiltonians at order $\delta^3$ through the self-energies; see Sections \ref{sec:kekule_ops} and \ref{sec:kekule_system_sub}. The leading term of the effective Hamiltonian obtained below is a massive two-dimensional Dirac operator. We refer to \cite{fefferman2012honeycomb, fefferman2014wave} for rigorous derivations of such Dirac operators in the absence of remote bands.
For applications of this model to topological lasing cavities, see \cite{yang2022topological}.

\subsection{Setup and tight-binding symbol}\label{subsec:TB_Kekule}

\paragraph{Lattice and notation.}

We take the lattice constant equal to one (it is denoted by $a$ in \cite{andrade2025topical}; here $a$ denotes the symbol). Define the three nearest-neighbor vectors
\begin{equation*}
  a_1 = (0,1)^T,\quad
  a_2 = R_{2\pi/3}\,a_1,\quad
  a_3 = R_{-2\pi/3}\,a_1,
\end{equation*}
and the Kekul\'e wave vector $G = K'$, of length $|G| = \frac{4\pi}{3\sqrt{3}}$ and
orthogonal to $a_1$, that is
\begin{equation}\label{eq:Kekule_G}
  G = \frac{4\pi}{3\sqrt{3}}(1,0)^T ,
\end{equation}
so that $G \cdot a_1 = 0$ and $G\cdot a_{2,3} = \mp \frac{2\pi}{3}$ and hence
$\kdisp (\pm G) = 1 + e^{2\pi i/3} + e^{-2\pi i /3} = 0$. (Any $G$ parallel to $a_1$
would give $|\kdisp (G)| = |e^{-ic} + 2 e^{ic/2}| \ge 1$ for every real
$c = G \cdot a_1$, and could not fold a Dirac point onto $\xi = 0$.)
The three folded dispersion functions are
\begin{equation*}
  \kdisp(\xi) = \sum_{j=1}^3 e^{i\xi\cdot a_j},\qquad
  \kdisp_n(\xi) = \kdisp(\xi + nG),\quad n\in\{-1,0,1\}.
\end{equation*}
In the basis $(\fA_1,\fA_2,\fA_3,\fB_1,\fB_2,\fB_3)$ of sublattice $\times$
folded-valley degrees of freedom,
the symbol of the Kekul\'e-O tight-binding Hamiltonian is the $6\times 6$
Hermitian matrix
\begin{equation}\label{eq:fullsymbol}
  a_\delta(x,\xi) = a(\delta x,\xi)
  = t_1\begin{pmatrix} 0 & A(\delta x,\xi) \\ A^*(\delta x,\xi) & 0 \end{pmatrix},
  \qquad
  A (X, \xi) = A^0(\xi) + \delta A^1(X,\xi),
\end{equation}
where $t_1 \in \mathbb{R}$ and with $\De\in
C^\infty_b(\mathbb{R}^2;\C)$ the complex Kekul\'e order parameter,
\begin{equation}
  A^0
  = \begin{pmatrix} \kdisp_0 & 0 & 0 \\ 0 & \kdisp_{-1} & 0 \\ 0 & 0 & \kdisp_1 \end{pmatrix},
  \qquad
  A^1(X,\xi)
  = \begin{pmatrix}
      0 & \De (X) \kdisp_1 (\xi) & \De^* (X) \kdisp_{-1} (\xi)\\
      \De^*(X) \kdisp_1(\xi) & 0 & \De (X) \kdisp_0 (\xi)\\
      \De(X) \kdisp_{-1}(\xi) & \De^*(X) \kdisp_0 (\xi) & 0
    \end{pmatrix}.
\end{equation}

\paragraph{Taylor expansion in $\delta$.}

We henceforth take $t_1 = 1$ for simplicity.
Setting $\xi=\delta\ze$ (so that we expand about the folded Dirac point $\xi=0$), with
$\ze=\ze_1+i\ze_2\in\C$ we find
\begin{align}\label{eq:eps_expansions}
\begin{split}
  \kdisp_0(\delta\ze) &= 3 - \tfrac{3}{4}\delta^2|\ze|^2
                       + \tfrac{1}{16}\delta^3 (\ze^3 - \bar\ze^3) + O(\delta^4),
  \\
  \kdisp_1(\delta\ze) &= -\tfrac{3}{2}\delta\bar\ze
                       + \tfrac{3}{8}\delta^2\ze^2
                       + \tfrac{3}{16}\delta^3\bar\ze|\ze|^2 + O(\delta^4),
                       \\
  \kdisp_{-1}(\delta\ze) &= \tfrac{3}{2}\delta\ze
                          + \tfrac{3}{8}\delta^2\bar\ze^2
                          - \tfrac{3}{16}\delta^3\ze|\ze|^2 + O(\delta^4).
\end{split}
\end{align}
These follow from $\kdisp_n (\xi) = \sum_j e^{inG\cdot a_j} \sum_{k \ge 0} \frac{(i \xi \cdot a_j)^k}{k!}$, together with the elementary sums
\begin{align}\label{eq:kekule_sums}
\begin{split}
  \sum_j (\ze\cdot a_j)^k &= 0, \ \tfrac32 |\ze|^2, \ \tfrac34 \ze_2 (\ze_2^2 - 3 \ze_1^2), \qquad k = 1,2,3,\\
  \sum_j e^{iG\cdot a_j}(\ze\cdot a_j)^k &= \tfrac{3i}{2}\bar\ze, \ -\tfrac34 \ze^{2}, \ \tfrac{9i}{8}\bar\ze\,|\ze|^2, \qquad k=1,2,3,\\
  \sum_j e^{-iG\cdot a_j}(\ze\cdot a_j)^k &= -\tfrac{3i}{2}\ze, \ -\tfrac34 \bar\ze^{2}, \ -\tfrac{9i}{8}\ze\,|\ze|^2, \qquad k=1,2,3.
\end{split}
\end{align}
Writing $A(X, \delta \zeta) = \sum_{j\ge 0}\delta^j A_j (X,\zeta)$ and
$b_j = \begin{pmatrix}0 & A_{j+1} \\ A_{j+1}^* & 0\end{pmatrix}$, so that
\begin{equation}
  a(X,\delta\ze) = a^0 + \delta b_0(X,\ze) + \delta^2 b_1(X,\ze)
                 + \delta^3 b_2(X,\ze) + O(\delta^4),
\end{equation}
the first four coefficient blocks are
\begin{equation}\label{eq:A0}
  A_0 = \begin{pmatrix} 3&0&0\\0&0&0\\0&0&0 \end{pmatrix},
\end{equation}
\begin{equation}\label{eq:A1}
  A_1 (X,\zeta) = \begin{pmatrix}
    0 & 0 & 0 \\
    0 & \tfrac{3}{2}\ze & 3\De(X) \\
    0 & 3\De^*(X) & -\tfrac{3}{2}\bar\ze
  \end{pmatrix},
\end{equation}
\begin{equation}\label{eq:A2}
  A_2(X,\zeta) = \begin{pmatrix}
    -\tfrac{3}{4}|\ze|^2 & -\tfrac{3}{2}\De(X)\bar\ze & \tfrac{3}{2}\De^*(X)\ze \\
    -\tfrac{3}{2}\De^*(X)\bar\ze & \tfrac{3}{8}\bar\ze^2 & 0 \\
    \tfrac{3}{2}\De(X)\ze & 0 & \tfrac{3}{8}\ze^2
  \end{pmatrix},
\end{equation}
\begin{equation}\label{eq:A3}
  A_3(X,\zeta) = \begin{pmatrix}
    \tfrac{1}{16}(\ze^3 - \bar\ze^3) & \tfrac{3}{8}\De(X)\ze^2 & \tfrac{3}{8}\De^*(X)\bar\ze^2 \\
    \tfrac{3}{8}\De^*(X)\ze^2 & -\tfrac{3}{16}\ze|\ze|^2 & -\tfrac{3}{4}\De(X)|\ze|^2 \\
    \tfrac{3}{8}\De(X)\bar\ze^2 & -\tfrac{3}{4}\De^*(X)|\ze|^2 & \tfrac{3}{16}\bar\ze|\ze|^2
  \end{pmatrix}.
\end{equation}
Note that $b_j\in S(\langle\ze\rangle^{j+1})$ for each $j\ge 0$, as required by
Assumption \ref{assumption:a_higher_order_all}.

\subsection{Eigenstructure of $a^0$ and projections}\label{sec:kekule_eigen}

The constant matrix $a^0 = \begin{pmatrix}0&A_0\\A_0&0\end{pmatrix}$ has
eigenvalues
\begin{equation}
  E_+ = +3 \;(\text{mult.\ }1),\quad
  E_0 = 0 \;(\text{mult.\ }4),\quad
  E_- = -3 \;(\text{mult.\ }1).
\end{equation}
Let $e_1=(1,0,0)^T\in\C^3$ denote the first standard basis vector. The rank-one projections onto $E_\pm$ are
\begin{equation*}
  P_\pm = \frac{1}{2}
  \begin{pmatrix} e_1e_1^T & \pm e_1e_1^T \\ \pm e_1e_1^T & e_1e_1^T \end{pmatrix},
\end{equation*}
and the rank-four projection onto $E_0$ is
\begin{equation*}
  P_0 = I_6 - P_+ - P_- = \diag (0,1,1,0,1,1).
\end{equation*}

\paragraph{Resolvents.}

The Green's functions on the complementary subspaces are
\begin{align*}
  \asharp_0 &= (a^0)^{-1}\big|_{\Ran(I-P_0)} = \tfrac{1}{3}P_+ - \tfrac{1}{3}P_-,\\
  \ashift_+ &:= (a^0-3I_6)^{-1}\big|_{\Ran(I-P_+)} = -\tfrac{1}{3}P_0 - \tfrac{1}{6}P_-,\\
  \ashift_- &:= (a^0+3I_6)^{-1}\big|_{\Ran(I-P_-)} = \tfrac{1}{3}P_0 + \tfrac{1}{6}P_+.
\end{align*}

\paragraph{Vanishing of $b_0$ on $E_\pm$.}

We observe that $A_1$ has a zero first row and first column. Consequently
\begin{equation}\label{eq:b0_vanishing}
  P_\pm b_0 P_\pm = 0,\qquad P_\pm b_0 P_0 = 0, \qquad P_0 b_0 P_\pm = 0, \qquad P_\pm b_0 P_\mp = 0.
\end{equation}
Thus $b_{0,\pm}:=P_\pm b_0 P_\pm = 0\in S^0$, placing the $E_\pm$ sectors in
the class $\cL_0$ of \eqref{eq:sm_ellipticity}. The leading dynamics of the
$E_\pm$ sectors is therefore trivial at order $\delta$, and nontrivial structure
first appears at order $\delta^2$. Hence the $E_\pm$ sectors correspond to approximately flat bands.

\subsection{Macroscopic operators}\label{sec:kekule_ops}

We now compute the operators entering the expansions of Section
\ref{sec:sm_setting}. This includes the projected symbols $P_\ell b_j P_\ell$ for each of the three
sectors, the resulting effective Hamiltonians, and the inter-sector couplings.

\subsubsection{$E_0 = 0$ sector: Dirac operator with Kekul\'e mass}\label{sec:kekule_E0}
For each sector $\ell\in\{0,+,-\}$, we denote by $\rH^{(\ell)}_{j}$ the coefficient $\rH_j$ of \eqref{eq:sm_rH}
for that sector, written without frequency cutoff, so that the effective Hamiltonian
\eqref{eq:def_rH} satisfies
\begin{align*}
    \rH^{(\ell)} \sim \sum_{j \ge 0} \delta^j \rH^{(\ell)}_{j},
\end{align*}
up to the frequency truncation \eqref{eq:sm_rH} of the unbounded coefficients with $j\ge1$.
By \eqref{eq:b0_vanishing}, $(I_6-P_\ell)b_0P_\ell = 0$ for every $\ell$, so that $S_1 = 0$ in
\eqref{eq:sm_Pi} for every sector and all terms of \eqref{eq:sm_rH} containing
$\ashl\bH_0P_\ell$ vanish. We explicitly calculate $\rH^{(0)}_{j}$ for $j=0,1,2$.
\paragraph{Leading order ($O(1)$).}

The projected symbol $b_{0,0}:=P_0 b_0 P_0$, acting on $\Ran(P_0)\cong\C^4$
in the basis $(\fA_2,\fA_3,\fB_2,\fB_3)$, is
\begin{equation}\label{eq:b00}
  b_{0,0}
  = \begin{pmatrix} 0 & A_1^{(0)} \\ (A_1^{(0)})^* & 0 \end{pmatrix},
  \qquad
  A_1^{(0)} (X,\ze) = \begin{pmatrix} \tfrac{3}{2}\ze & 3\De(X) \\ 3\De^*(X) & -\tfrac{3}{2}\bar\ze \end{pmatrix}.
\end{equation}
In terms of the Pauli matrices (with $\sig_0=I_2$), this is
\begin{equation}\label{eq:b_0_0}
  b_{0,0}(X,\ze)
  = \tfrac{3}{2}\bigl(\ze_1\,\sig_1\otimes\sig_3 - \ze_2\,\sig_2\otimes\sig_0\bigr)
  + 3\bigl(\Re\De(X)\,\sig_1\otimes\sig_1 - \Im\De(X)\,\sig_1\otimes\sig_2\bigr).
\end{equation}
We refer to $\ow (b_{0,0})$ as a ``Dirac operator with a Kekul\'e mass''.
We then verify that
\begin{equation*}
  b_{0,0}^2(X,\ze) = \Bigl(\tfrac{9}{4}|\ze|^2 + 9|\De(X)|^2\Bigr)I_4,
\end{equation*}
whence (with $P_0 b_0 P_0 \in \mathbb{C}^{6 \times 6}$) 
\begin{equation*}
  \det \bigl( P_0 b_0 P_0 + (I_6-P_0)\aver{\ze}\bigr)
  = \aver{\ze}^2 \det b_{0,0}
  = \tfrac{81}{16}\aver{\ze}^2 \bigl( |\ze|^2 + 4|\De(X)|^2\bigr)^2 ,
\end{equation*}
so that \eqref{eq:sm_ellipticity} holds with $N=6$. We have thus shown that $b_{0,0}\in S^1$ is elliptic, so that $0 \in \cL_1$
and Assumption \ref{assumption:a_higher_order_all} is satisfied for this sector.
The leading-order operator in the effective Hamiltonian is
\begin{align*}
    \rH^{(0)}_{0} = \ow b_{0,0}.
\end{align*}

\paragraph{First order ($O(\delta)$).}

The direct term $b_{1,0} := P_0 b_1 P_0$, restricted to the zero-mode block, gives
\begin{equation*}
  b_{1,0}
  = \begin{pmatrix} 0 & A_2^{(0)} \\ (A_2^{(0)})^* & 0 \end{pmatrix},
  \qquad
  A_2^{(0)} (\ze) = \begin{pmatrix} \tfrac{3}{8}\bar\ze^2 & 0 \\ 0 & \tfrac{3}{8}\ze^2 \end{pmatrix}.
\end{equation*}
This can be rewritten as
\begin{align}\label{eq:trigonal}
    b_{1,0} (\ze) = \tfrac{3}{8}\bigl(
      \Re (\ze^2)\, \sig_1\!\otimes\!\sig_0 + \Im (\ze^2)\, \sig_2\!\otimes\!\sig_3
    \bigr),
\end{align}
which belongs to $S^2 \subset S(\aver{\ze}^2)$.
The self-energy correction $P_0 b_0\,\ashift_0\,b_0\,P_0$ vanishes because
$b_0 P_\pm = 0$. Therefore we conclude that
\begin{align*}
    \rH^{(0)}_{1} = \ow b_{1,0}.
\end{align*}

\paragraph{Second order ($O(\delta^2)$).}
We find that
\begin{align}\label{eq:b2_zero_block}
\begin{split}
  b_{2,0} &:= P_0 b_2 P_0
  = \begin{pmatrix} 0 & A_3^{(0)} \\ (A_3^{(0)})^* & 0 \end{pmatrix},
  \\
  A_3^{(0)} (X, \ze)
  &= -\frac{3}{16}|\ze|^2 \begin{pmatrix} \ze & 4 \De(X) \\ 4\De^*(X) & -\bar\ze\end{pmatrix}
  = -\frac{|\ze|^2}{8} A^{(0)}_1 - \frac38 |\ze|^2 \begin{pmatrix} 0 & \De(X)\\ \De^*(X) & 0\end{pmatrix},
  \end{split}
\end{align}
which belongs to $S^3 \subset S(\aver{\ze}^3)$. Note that no other terms appear at this order, since $b_0 \inva_0 = \inva_0 b_0 = 0$. Hence
\begin{align*}
    \rH^{(0)}_{2} = \ow b_{2,0}.
\end{align*}

\paragraph{Full effective Hamiltonian for $E_0$.}
Collecting the terms above, we have shown that
\begin{equation}\label{eq:H0eff}
\rH^{(0)} = \sum_{j=0}^2 \delta^j \rH^{(0)}_{j} + O(\delta^3), \qquad  \rH^{(0)}_{j} = \ow b_{j,0},
\end{equation}
with the $b_{j,0}$ given by \eqref{eq:b_0_0}, \eqref{eq:trigonal}, and \eqref{eq:b2_zero_block}.
We emphasize that none of the operators $\rH^{(0)}_{j}$ in \eqref{eq:H0eff} couple the $E_0$ sector to the $E_\pm$ sectors. Indeed, since $\bH_0 \inva_0 = \inva_0 \bH_0 =0$, we see from \eqref{eq:sm_rH} and the computation of $\rH^{(\ell)}_{3}$ in Section \ref{sec:kekule_Gamma} that coupling between these sectors does not occur until $O(\delta^3)$.

\subsubsection{$E_\pm = \pm3$ sectors: flat bands with quadratic dispersion}\label{sec:kekule_Epm}

We showed that  $b_{0,+}=0$. For the rank-one sectors $E_\pm$ the projected
symbols are scalars, given for each $j$ by
\begin{equation}\label{eq:kekule_upm_rule}
  P_\pm b_j P_\pm = \uu_\pm^* b_j \uu_\pm = \pm \Re (A_{j+1})_{11},
  \qquad \uu_\pm = \tfrac{1}{\sqrt2}(e_1, \pm e_1)^T.
\end{equation}
Since $(A_2)_{11} = -\tfrac{3}{4}|\ze|^2$, we get
\begin{equation}\label{eq:b_1_pm}
  b_{1,+} := P_+ b_1 P_+ = -\tfrac{3}{4}|\ze|^2, \qquad b_{1,-} := P_- b_1 P_- = +\tfrac{3}{4}|\ze|^2
  \qquad (\text{scalars on }\Ran(P_\pm)).
\end{equation}

The self-energy term $P_+ b_0\,\ashift_+\,b_0\,P_+$ vanishes because $b_0 P_+ = 0$.
Higher-order Schur complement terms $P_+ b_j\,\ashift_+\,b_k\,P_+$ for $j+k=1$
similarly vanish. The direct second-order term also vanishes, in contrast with the $E_0$
sector. Indeed, \eqref{eq:kekule_upm_rule} and \eqref{eq:A3} imply
\begin{equation*}
  P_\pm b_2 P_\pm = \pm \Re (A_3)_{11} = 0.
\end{equation*}
With the $b_{1,\pm}$ as in \eqref{eq:b_1_pm}, we thus have
\begin{align}\label{eq:Hpeff}
    \rH^{(\pm)} = \delta \rH^{(\pm)}_{1} + O(\delta^3), \qquad \rH^{(\pm)}_{1} = \ow b_{1,\pm},
\end{align}
acting as a scalar operator on the one-dimensional space $\Ran(P_\pm)$. For the $+$ sector, this is a
negative quadratic dispersion (downward-curving flat band), with effective
mass $m^*_+ = -2/3$. The $-$ sector produces a positive quadratic dispersion with effective mass $m^*_- = 2/3$. The $O(\delta^3)$ term contains the
self-energy generated by the coupling to the Dirac sector, which is computed in
\eqref{eq:Sigma0} below.

\subsubsection{Inter-sector coupling}\label{sec:kekule_Gamma}

\paragraph{Coupling covectors $\Gamma_\pm$.}
Take $\uu_\pm$ as in \eqref{eq:kekule_upm_rule} so that
$P_\pm=\uu_\pm\uu_\pm^*$. Let
$v=(v_2,v_3,w_2,w_3)\in\C^4\cong\Ran P_0$ in the basis
$(\fA_2,\fA_3,\fB_2,\fB_3)$.
Since $b_1=\begin{psmallmatrix}0&A_2\\A_2^*&0\end{psmallmatrix}$, the coupling
covectors $\Gamma_\pm := \uu_\pm^* b_1 P_0$ are
\begin{align}\label{eq:Gamma}
\begin{split}
\Gamma_\pm(X,\zeta)
&=\frac{1}{\sqrt2}\Bigl(\pm e_1^TA_2^*(X,\zeta) e_2,\ \pm e_1^TA_2^*(X,\zeta)e_3,\
 e_1^TA_2(X,\zeta)e_2,\ e_1^TA_2(X,\zeta)e_3\Bigr)\\
&=\frac{3}{2\sqrt2}\Bigl(\mp\Delta(X)\zeta,\ \pm\Delta^*(X)\bar\zeta,\
 -\Delta(X)\bar\zeta,\ \Delta^*(X)\zeta\Bigr).
 \end{split}
\end{align}
Above, we used the identities $\uu_\pm^* b_1 = \frac{1}{\sqrt2} (\pm e_1^T A_2^*, e_1^T A_2)$ and (recalling \eqref{eq:A2})
\begin{equation}\label{eq:A2rows}
e_1^TA_2e_2=-\tfrac32\Delta\bar\zeta,\quad
e_1^TA_2e_3=\tfrac32\Delta^*\zeta,\quad
e_1^TA_2^*e_2=-\tfrac32\Delta\zeta,\quad
e_1^TA_2^*e_3=\tfrac32\Delta^*\bar\zeta .
\end{equation}
Both $\Gamma_\pm$ lie in $S(\langle\zeta\rangle)$ and satisfy
$|\Gamma_\pm(X,\zeta)|^2=\tfrac92|\Delta(X)|^2|\zeta|^2$. They are related by
\begin{equation}
\Gamma_-(X,\zeta)=\Gamma_+(X,\zeta)\,\mathrm{diag}(-1,-1,1,1).
\end{equation}
The operators 
$\ow (\Gamma_\pm)$ and their Hermitian conjugates first appear in the third-order coefficients
$\rH^{(\ell)}_{3}$. Indeed, since $S_1 = 0$, Definition \ref{def:S} gives
$S = \delta^2S_2 + O(\delta^3)$ with $S_2 = -\ashl\bH_1P_\ell$, and $\cG^{-1/2} = P_\ell + O(\delta^4)$. Expanding
\eqref{eq:def_rH} to third order as in \eqref{eq:sm_rH}, and using \eqref{eq:b0_vanishing}, we find
\begin{align*}
     \rH^{(\ell)}_{3} = P_\ell \bH_3 P_\ell + \Sigma^{(2)}_\ell, \qquad \Sigma^{(2)}_\ell := -P_\ell \bH_1 \ashl \bH_1 P_\ell,
\end{align*}
where a direct calculation reveals
\begin{align}\label{eq:Sigma0}
    \Sigma^{(2)}_0 = \tfrac13\Bigl(\Opw(\Ga_-)^*\Opw(\Ga_-)-\Opw(\Ga_+)^*\Opw(\Ga_+)\Bigr), \qquad \Sigma^{(2)}_\pm = \pm \frac13 \Opw (\Ga_\pm) \Opw (\Ga_\pm)^*.
\end{align}
Thus the inter-sector coupling effects are of $O(\delta^3)$ with respect to the effective Hamiltonians $\rH^{(\ell)}$. 
We will see in Section \ref{sec:kekule_system_sub} that the above inter-sector coupling operators appear at $O(\delta^2)$ in the construction \eqref{eq:sm_phout}-\eqref{eq:sm_phin} of the macroscopic solution $\phi$ from \eqref{eq:sm_psi}.

\subsection{Expansion with prescribed data in the Dirac sector}\label{sec:kekule_system_sub}

Throughout this subsection $K = 0$ and $p = 2$, so that
Assumption \ref{assumption:a_higher_order_all} holds with $\rate = 1$ and with
$b_0, b_1$ read off from \eqref{eq:A1}--\eqref{eq:A2}. Let
$\phi_0 = P_0 \phi_0 \in H^\infty (\mathbb{R}^2 ; \Ran P_0)$ in
\eqref{eq:sm_micro}, so that the initial condition is supported in the
Dirac sector. Recall from Section
\ref{sec:kekule_eigen} that $A_1$ has vanishing first row and column, which implies
\begin{equation}\label{eq:b0P}
P_\pm b_0 = b_0 P_\pm = 0,
\qquad (I_6-P_0)\,b_0\,P_0 = 0 .
\end{equation}
Set $\bH_j = \ow b_j$ and $\bH_{j0} = P_0 \bH_j P_0$, so that $\bH_{j0} = \ow b_{j,0}$ with the $b_{j,0}$ for $0 \le j \le 1$ given by \eqref{eq:b00} and \eqref{eq:trigonal}. 
We run \eqref{eq:sm_phout}--\eqref{eq:sm_phin} with $\asharp_0 = \tfrac13 (P_+ - P_-)$.

\paragraph{Order $\delta^0$.} We have $\phi^0_\pm\vert_{T=0} = P_\pm \phi_0 \equiv 0$, hence $\phi^0_\pm \equiv 0$. On the other hand,
$\phi^0_0$ solves the Dirac equation with Kekul\'e mass,
\begin{equation}\label{eq:phi_0_0_Kekule}
(D_T + \bH_{00}) \phi^0_0 = 0, \qquad \phi^0_0 (0, \cdot) = \phi_0.
\end{equation}

\paragraph{Order $\delta^1$.} By \eqref{eq:b0P},
$\phout^{1,0} = -\asharp_0 (I_6-P_0) \bH_0 \phi^0_0 \equiv 0$ and $\phi^1_\pm \equiv 0$. Hence $\phin^{1,0}$ solves
\begin{equation}\label{eq:warp}
(D_T + \bH_{00}) \phin^{1,0} = -\bH_{10} \phi^0_0,
\qquad \phin^{1,0}(0,\cdot) = 0.
\end{equation}

\paragraph{Order $\delta^2$.} Since $\phi^1_0 = \phin^{1,0} \in \Ran P_0$ and
$(I_6-P_0)\bH_0 P_0 = 0$, both $D_T \phi^1_0$ and $\bH_0 \phi^1_0$ are annihilated by
$\asharp_0$ in \eqref{eq:sm_phout}, so that
\begin{equation}\label{eq:out2}
\phout^{2,0} = -\asharp_0 \bH_1 \phi^0_0
= -\tfrac13\,\uu_+\,\Opw(\Ga_+)\phi^0_0
+\tfrac13\,\uu_-\,\Opw(\Ga_-)\phi^0_0 .
\end{equation}
Recall that $\phi_\pm^j \equiv 0$ for $j\in \{0,1\}$ and thus $\phout^{2,\pm} \equiv 0$ by \eqref{eq:sm_phout}. The redistribution rule \eqref{eq:sm_phin} at order $j = 2$ would thus assign to the $\Gamma$-point sectors the data
\begin{equation}\label{eq:seed}
\phin^{2,\pm}(0,\cdot) = -P_\pm \phout^{2,0}(0,\cdot)
= \pm\tfrac13\,\Opw(\Ga_\pm)\,\phi_0 .
\end{equation}
Since $p = 2$, however, the convention $\phin^{p,\ell} \equiv 0$ of Section \ref{subsec:macroscopic} sets $\phin^{2,\pm} \equiv 0$, so that the data \eqref{eq:seed} do not enter the approximation below. The $\Gamma$-point packets they generate are of order $\delta^2$, which is the size of the error at $p=2$, and they are retained by the construction as soon as $p \ge 3$. In that case, because $\bH_{0\pm} = \Opw b_{0,\pm} = 0$ and, by \eqref{eq:b0P},
$P_\pm \bH_0 \phout^{2,\pm} = 0$, the source in
\eqref{eq:sm_phin} vanishes and $\phin^{2,\pm} \equiv \phin^{2,\pm} (0, \cdot)$ is constant in $T$: the seeded packets are frozen at this order, and begin to disperse only through the term $\delta\,\ow b_{1,\pm}$ of \eqref{eq:Hpeff}.
The identities \eqref{eq:b0P} also imply that
$P_0 \bH_0 \phout^{2,0} = \Opw (P_0 b_0 P_\pm)(\cdots) = 0$, so that $\Ga_\pm$ does
not contribute to the source term in \eqref{eq:sm_phin} for $(j,\ell) = (2,0)$.

\paragraph{Full wave packet ansatz.}
We have thus shown that the function
\begin{align*}
    \psi_\delta (t,x) = \delta^{d/2} \left( \phi^0_0 (\delta t, \delta x) + \delta \phin^{1,0} (\delta t, \delta x) + \delta^2 \phout^{2,0} (\delta t, \delta x)\right)
\end{align*}
given by \eqref{eq:phi_0_0_Kekule}, \eqref{eq:warp}, and \eqref{eq:out2} satisfies the estimate \eqref{eq:sm_estimate} of Theorem \ref{thm:main_higher_order} with $\rate = 1$ and $p=2$. That is, $\psi_\delta$ provides an $O(\delta^2)$ approximation of the tight-binding dynamics for all times $t \lesssim \delta^{-1}$. It contains no $\Gamma$-point packet, since $\phin^{2,\pm} \equiv 0$ for $p = 2$. The $\Gamma$-point sectors enter only through the algebraic corrector $\delta^2\phout^{2,0}$, which lies in $\Ran(P_+ + P_-)$ but travels with the Dirac packet and carries its phase.

\subsection{Band-adapted expansion, and comparison}\label{sec:kekule_ad}

Using that $(I_6-P_0)b_0P_0 = 0$, a direct calculation reveals
\begin{align}\label{eq:kekule_T}
    \mathcal T^{(3)}_\delta \phi_0 = \phi_0 - \delta^2 \phout^{2,0}(0,\cdot)
    = \phi_0 + \frac{\delta^2}{3}\bigl( \uu_+ \Opw (\Ga_+) - \uu_-\Opw(\Ga_-)\bigr)\phi_0,
\end{align}
whose $P_\pm$ components are $\pm\frac{\delta^2}{3}\Opw (\Ga_\pm)\phi_0$, that is, $\delta^2$ times the data \eqref{eq:seed}. In the
band-adapted expansion of Theorem \ref{thm:equiv}, each of the $\Gamma$ packets carries this datum, evolves under $\bH_{0\pm}$ alone, and receives
nothing from the Dirac sector at any order. These packets carry the fast phases $e^{\mp3it}$ and separate from the Dirac packet. In the expansion with prescribed data and $p=2$ of Section \ref{sec:kekule_system_sub}, the same packets are absent, since $\phin^{2,\pm} \equiv 0$. The two expansions therefore differ by $O(\delta^2)$, which is the size of the error of either one: with $p=2$ and $q=3$, Theorem \ref{thm:equiv} gives the bound $C(\delta^3 + \delta^3 t)(1+(\delta t)^{2m+4})$ in \eqref{eq:ad_error}, while Theorem \ref{thm:main_higher_order} gives $C(\delta^2 + \delta^3 t)(1+(\delta t)^{2m+4})$, so that both are $O(\delta^2)$ approximations of the same solution for $t \lesssim \delta^{-1}$. Resolving the $\Gamma$-point packets in the expansion with prescribed data, where they arise through \eqref{eq:seed} and cancel the transverse admixture $\delta^2\phout^{2,0}$ at $T=0$, requires $p \ge 3$ and hence the symbol $b_2$ of \eqref{eq:A3}.

\subsection{The Kekul\'e vortex}\label{sec:kekule_vortex}

The case of principal interest for applications is that of a vortex profile, say
$\De (X) = \De_0 (R) e^{i \vartheta}$ in polar coordinates $X = R(\cos\vartheta,\sin\vartheta)$,
with $\De_0 (0) = 0$ and $\De_0$ bounded away from zero at infinity; see
\cite{andrade2025topical, yang2022topological}. 
The condition $\De_0 (0) = 0$ ensures that $\Delta$ is continuous. If we further assume that $\De_0 \in C^\infty_b [0,\infty)$ with its derivatives satisfying $\De^{(2j)}_0 (0) = 0$ for all $j \in \mathbb{N}$, then $\De \in C^\infty_b (\mathbb{R}^2 ; \C)$. The winding number of $\De$ is
\begin{align*}
    \mathcal{W} [\De] := \frac{1}{2\pi i} \lim_{R \to \infty} \int_{0}^{2\pi} |\De (R, \vartheta)|^{-2}\bar \De (R, \vartheta) \partial_\vartheta \De (R,\vartheta) d\vartheta = 1.
\end{align*}
The leading term of \eqref{eq:H0eff} is then a
two-dimensional Dirac operator with a mass vortex of winding number one. Note that the Kekul\'e
mass couples the two valleys: in \eqref{eq:b_0_0} it multiplies $\sig_1\otimes\sig_1$ and
$\sig_1\otimes\sig_2$, which are off-diagonal in the valley index, whereas the kinetic terms
$\sig_1\otimes\sig_3$ and $\sig_2\otimes\sig_0$ are diagonal in it. The vortex is therefore
a single defect of the four-component operator rather than one vortex per valley. Its
spectrum contains a topologically protected zero mode localized near the vortex core $R = 0$, by the
index computation of Jackiw and Rossi \cite{jackiw1981zero}. The relevance of this mechanism
to Kekul\'e-modulated honeycomb structures was identified in \cite{hou2007electron}, and it
underlies the topological cavity designs of \cite{yang2022topological}.
A rigorous spectral comparison of tight-binding to their effective macroscopic Hamiltonians is left for future work. See also \cite[Section 5]{bal2026macroscopic} for a relevant discussion on bulk-difference invariants, which applies to the setting $L = 1$ (cf. Assumption \ref{assumption:a_higher_order_all}) of the present paper.

\section{Derivations for magnetized cold plasmas}\label{sec:plasma}

This section contains the derivations for the cold plasma model summarized in Section
\ref{sec:summary_plasma} of the main paper. The model is not of tight-binding type: its symbol is a
first-order differential symbol, unbounded in $\xi$, so that Assumption
\ref{assumption:a_higher_order_all} does not apply as stated, and we work instead
under Assumption \ref{assumption:pde} of Section \ref{sec:pde_class}.

\subsection{The model and its spectral data}\label{sec:plasma_model}

\paragraph{Physical model.}
A cold magnetized plasma in two spatial dimensions $(x,y)$ with a
background magnetic field $\mathbf{B}_{\rm ext} = B_{\rm ext}\hat{e}_z$ supports three
types of fields: the electron fluid velocity $\mathbf{v}=(v_x,v_y,v_z)^T$,
the electric field $\mathbf{E}=(E_x,E_y,E_z)^T$, and the magnetic field
$\mathbf{B}=(B_x,B_y,B_z)^T$.  The linearized cold plasma equations are
\begin{equation}\label{eq:plasma_eqs}
  \partial_t\mathbf{v} = \Omega\hat{e}_z\times\mathbf{v} - \omega_p(\delta x)\mathbf{E},
  \qquad
  \partial_t\mathbf{E} = \omega_p(\delta x)\mathbf{v} + \nabla\times\mathbf{B},
  \qquad
  \partial_t\mathbf{B} = -\nabla\times\mathbf{E},
\end{equation}
where $\Omega>0$ is the (constant) cyclotron frequency and $\omega_p(\delta x)>0$
is a spatially slowly-varying plasma frequency.  Setting $\varphi=(\mathbf{v},\mathbf{E},\mathbf{B})^T
\in\mathbb{C}^9$, these equations take the Hamiltonian form
$(D_t + H_\delta)\varphi = 0$ with
\begin{equation}\label{eq:plasma_H}
  H_\delta = \begin{pmatrix}
    i\Omega\hat{e}_z\times & -i\omega_p(\delta x) & 0 \\
    i\omega_p(\delta x) & 0 & i\nabla\times \\
    0 & -i\nabla\times & 0
  \end{pmatrix}.
\end{equation}
The system \eqref{eq:plasma_eqs} differs from the form in which the linearized cold plasma
equations are usually written, namely
\begin{equation}\label{eq:plasma_eqs_standard}
  \partial_t\mathbf{v} = -\Omega\hat{e}_z\times\mathbf{v} + \omega_p\mathbf{E},
  \qquad
  \partial_t\mathbf{E} = -\omega_p\mathbf{v} + \nabla\times\mathbf{B},
  \qquad
  \partial_t\mathbf{B} = -\nabla\times\mathbf{E}.
\end{equation}
One can obtain \eqref{eq:plasma_eqs_standard} from \eqref{eq:plasma_eqs} by the relabelling $(\Omega, \mathbf E, \mathbf B) \mapsto (-\Omega, -\mathbf E, -\mathbf B)$, which corresponds to a reversal of the background field together with the opposite sign convention for the electric and magnetic fields. This is a change of convention and not of model; we adopt \eqref{eq:plasma_eqs}
because it is the form for which \eqref{eq:plasma_H} and \eqref{eq:plasma_symbol_full} below
hold as written.
We refer to, e.g., \cite{parker2020topological,parker2020nontrivial,fu2021topological,fu2022dispersion,han2022anomalous,frazier2025topological,serra2025influence} for references on the derivation and topological properties of cold plasma models. In particular, for a particular value $\omega_-$ of the plasma frequency $\omega_p$,
two frequency bands meet and a topological phase transition occurs; see \eqref{eq:def_omega_minus} and \eqref{eq:plasma_spectrum}. Assuming that $\omega_p$ is an $O(\delta)$ perturbation of $\omega_-$, the effective model describing this topological transition is of Dirac type. The aim of this section is to show that the theory developed in the present paper allows one to retrieve this Dirac model as well as higher-order perturbations coming from inter-band mixing. 

We now make our further assumptions on the function $\omega_p$ precise. These assumptions will be explained in the following paragraphs; see in particular \eqref{eq:spec_tilde}--\eqref{eq:plasma_spectrum}. Fix $k_z \in \mathbb{R} \setminus \{0\}$. As the map $x \mapsto x^2 (1 - \frac{x^2}{x (x-\Omega)})$ is strictly monotonically increasing on the interval $[0,\Omega)$ with range equal to $[0,\infty)$, we can define $\omega_- \in (0,\Omega)$ as the unique solution to 
\begin{align}\label{eq:def_omega_minus}
    k_z^2 = \omega_-^2 \left( 1 - \frac{\omega_-^2}{\omega_- (\omega_- - \Omega)}\right).
\end{align}
Now, assume that $\omega_p (X) = \omega_- + \delta \tilde \omega_p (X)$ for some $\tilde \omega_p \in C^\infty_b (\mathbb{R}^2)$.
Then the Weyl symbol of $H_\delta$ is the $9\times 9$ Hermitian matrix $a_\delta (x, \xi) = a (\delta x, \xi)$, where
\begin{equation}\label{eq:plasma_symbol_full}
  a(X,\xi) = \begin{pmatrix}
    i\Omega\hat{e}_z\times & -i\omega_p(X) & 0 \\
    i\omega_p(X) & 0 & -\xi\times \\
    0 & \xi\times & 0
  \end{pmatrix},
  \qquad (\xi\times) v = \xi \times v, \quad\text{i.e.}\quad (\xi\times)_{jk} = \epsilon_{jlk}\xi_l,
\end{equation}
with $\xi=(\xi_1,\xi_2,k_z)^T$, and $k_z$ fixed.

\paragraph{Affine structure and crossing.}
Setting $K=(0,0,k_z)^T$ (zero in-plane momentum) and writing
$\xi = K+\delta\zeta$ with $\zeta=(\zeta_1,\zeta_2,0)^T$, we obtain
\begin{equation}\label{eq:plasma_affine}
  a_\delta(x,\xi) = a^0 + \delta\,b_0\!\bigl(\delta x,\tfrac{\xi-K}{\delta}\bigr),
  \qquad
  b_0(X,\zeta) = \begin{pmatrix}
    0 & -i\tilde\omega_p(X) & 0 \\
    i\tilde\omega_p(X) & 0 & -\zeta\times \\
    0 & \zeta\times & 0
  \end{pmatrix},
\end{equation}
where
\begin{align}\label{eq:a0_plasma}
    a^0  = \begin{pmatrix}
    i\Omega\hat{e}_z\times & -i\omega_- & 0 \\
    i\omega_- & 0 & -K\times \\
    0 & K\times & 0
\end{pmatrix}
\end{align}
is the constant matrix obtained by evaluating \eqref{eq:plasma_symbol_full}
at $\omega_p=\omega_-$ and $\xi=K$, and $\delta\tilde\omega_p(X)=\omega_p(X)-\omega_-$
is the slowly varying deviation from the crossing value $\omega_-$ defined by \eqref{eq:def_omega_minus}.
Observe that the symbol $a_\delta$ is exactly affine in $\delta$. That is, \eqref{eq:sm_decomp} holds for every $p$ with $b_j = 0$ for all $j\ge1$, and the coefficients of \eqref{eq:sm_rH} involve $\bH_0$ alone.

Let us now verify Assumption \ref{assumption:pde}. It is clear that the symbol $a$ lies in $S(\aver{\xi})$ uniformly in $\delta$. The condition \eqref{eq:slow_part_bounded} holds because $\nabla_X a$ is $O(\delta)$ in $S(1)$, by assumption on $\omega_p$.
Finally, $H_\delta$ is self-adjoint on $L^2 (\mathbb{R}^2;\C^9)$: its $\delta x$-independent part $a_{\rm free}(D)$, with $a_{\rm free}$ the value of \eqref{eq:plasma_symbol_full} at $\omega_p = \omega_-$, is a Fourier multiplier by a Hermitian matrix, hence self-adjoint on its maximal domain $\{u\in L^2:\ a_{\rm free}(D)u\in L^2\}$, and the remaining part, involving $\omega_p(\delta x)$, is a bounded symmetric perturbation (Kato--Rellich). This domain contains $H^1$ but is strictly larger, since $\xi\times$ annihilates longitudinal fields and the principal part is not elliptic; only $H^1\subset\mathcal D(H_\delta)$ is required in Assumption \ref{assumption:pde}. By Proposition \ref{prop:pde_class}, all results of Section \ref{sec:tb} therefore apply to this model for every $p \in \mathbb{N}$ and with $\rate = 1$; see also Remark \ref{rem:pde_class_gain}.

\paragraph{Plasma dispersion relations.}
Define the Hermitian matrix-valued function
\begin{align*}
    \tilde a^0 (\omega) := \begin{pmatrix}
    i\Omega\hat{e}_z\times & -i\omega & 0 \\
    i\omega & 0 & -K\times \\
    0 & K\times & 0
\end{pmatrix}, \qquad \omega \in \mathbb{R},
\end{align*}
so that $\tilde a^0 (\omega_-)= a^0$. We now identify the eigenvalues of $\tilde a ^0 (\omega)$. For $\omega \in \mathbb{R}$, define the functions
\begin{equation}\label{eq:RL}
  R_{\omega} (\lambda) = 1 - \frac{\omega^2}{\lambda(\lambda+\Omega)},
  \qquad
  L_{\omega} (\lambda) = 1 - \frac{\omega^2}{\lambda(\lambda-\Omega)}, \qquad \lambda \in (0,\infty) \setminus \{\Omega\}.
\end{equation}
Assume that $\omega > 0$.
One can verify that $k_z^2/\lambda^2 = R_\omega (\lambda)$ has exactly one positive solution $\lambda > 0$, which we denote by $\omega_R$. On the other hand, $k_z^2/\lambda^2 = L_\omega (\lambda)$ has two positive solutions $\omega_{L\pm}$, and they satisfy $\omega_{L-} < \Omega < \omega_{L+}$. The solutions $\omega_R$ and $\omega_{L\pm}$ are all eigenvalues of $\tilde a^0 (\omega)$. The fourth positive eigenvalue of $\tilde a^0 (\omega)$ is $\omega$. Since $0$ is also an eigenvalue of $\tilde a^0 (\omega)$ and the spectrum of $\tilde a^0 (\omega)$ is symmetric about $0$, we conclude that the $9$ eigenvalues of $\tilde a^0 (\omega)$ are
\begin{align}\label{eq:spec_tilde}
    \sigma (\tilde a^0 (\omega)) = \{\pm \omega_R, \pm \omega_{L+}, \pm \omega_{L-}, \pm \omega, 0\}.
\end{align}

\paragraph{Spectral structure at the crossing.}
Recalling \eqref{eq:def_omega_minus}, the frequency $\omega_-$ is the value of $\omega$ in \eqref{eq:spec_tilde} for which $\omega = \omega_{L-}$. Thus the constant matrix $a^0$ in \eqref{eq:a0_plasma} has degenerate eigenvalues $\pm \omega_-$, each of multiplicity $2$.
This gives $L=7$ distinct sectors,
\begin{equation}\label{eq:plasma_spectrum}
  \sigma(a^0) = \bigl\{
    \underbrace{\pm\omega_-}_{\mathrm{mult.\,2}},\;
    \pm\omega_R,\;\pm\omega_{L+},\;0
  \bigr\},
  \qquad
  0 < \omega_- < \omega_R < \omega_{L+}.
\end{equation}
Recall that $\omega_{L+}$ is the unique value of $\lambda > \Omega$ for which $k_z^2/\lambda^2 = L_{\omega_-} (\lambda)$, while $\omega_R$ is the unique solution $\lambda > 0$ to $k_z^2/\lambda^2 = R_{\omega_-} (\lambda)$, with $L_{\omega_-}$ and $R_{\omega_-}$ given by \eqref{eq:RL}.

The eigenvectors $\Psi_1$ and $\Psi_2$ corresponding to the eigenvalue $\omega_-$ are given by
\begin{equation}\label{eq:evecs}
  \Psi_1 = \frac{1}{\sqrt{2}}\begin{pmatrix}\hat{e}_z\\i\hat{e}_z\\0\end{pmatrix},
  \qquad
  \Psi_2 = \frac{1}{\sqrt{N_-}}\begin{pmatrix}
    i\frac{\omega_-}{\Omega-\omega_-}\hat{e}_+\\\hat{e}_+\\-i\frac{k_z}{\omega_-}\hat{e}_+
  \end{pmatrix},
  \qquad
  \hat{e}_\pm = \frac{1}{\sqrt{2}}\begin{pmatrix}1\\\pm i\\0\end{pmatrix},
\end{equation}
where $N_j = \omega_-^2/(\Omega\pm\omega_j)^2+1+k_z^2/\omega_j^2$, with the sign $+$ for $j = R$ and $-$ for $j \in \{-,L+\}$ (and $\omega_j$ the corresponding frequency), so that each eigenvector has unit norm.
The simple non-negative eigenvalues $\omega_R, \omega_{L+}, 0$ have eigenvectors
\begin{equation}\label{eq:evecs_simple}
  \Psi_R = \frac{1}{\sqrt{N_R}}\begin{pmatrix}
    -i\frac{\omega_-}{\Omega+\omega_R}\hat{e}_-\\\hat{e}_-\\i\frac{k_z}{\omega_R}\hat{e}_-
  \end{pmatrix},
  \quad
  \Psi_{L+} = \frac{1}{\sqrt{N_{L+}}}\begin{pmatrix}
    i\frac{\omega_-}{\Omega-\omega_{L+}}\hat{e}_+\\\hat{e}_+\\-i\frac{k_z}{\omega_{L+}}\hat{e}_+
  \end{pmatrix},
  \quad
  \Psi_0 = \begin{pmatrix}0\\0\\\hat{e}_z\end{pmatrix}.
\end{equation}
Using the identity
\begin{equation}\label{eq:parity_identity}
  \Gamma_P \,\overline{a^0}\, \Gamma_P = -a^0, \qquad \Gamma_P := \diag (I_3, I_3, -I_3),
\end{equation}
we find that the eigenvectors associated with the negative eigenvalues are given by
\begin{equation}\label{eq:parity}
  \Psi_{-\ell} = \Gamma_P \overline{\Psi_\ell}, \qquad \ell \in \{1,2,R,L+\}.
\end{equation}
That is, $a^0 \Psi_{-j} = -\omega_- \Psi_{-j}$ for $j \in \{1,2\}$, and $a^0 \Psi_{-\ell} = -\omega_\ell \Psi_{-\ell}$ for $\ell \in \{R,L+\}$.

\subsection{Macroscopic operators}\label{sec:plasma_ops}

\paragraph{Classification of sectors.}
The projected symbols $b_{0,\ell}=P_\ell b_0 P_\ell$ are computed from
the eigenvectors in \eqref{eq:evecs}--\eqref{eq:parity}.  For the degenerate $\omega_-$ sector, we find that $b_{0,\omega_-}= P_{\omega_-}b_0P_{\omega_-} $ in the basis $(\Psi_1, \Psi_2)$ is given by
\begin{equation}\label{eq:D_minus}
  b_{0,\omega_-}(X,\zeta)
  = \begin{pmatrix}
      \tilde\omega_p & \dfrac{i k_z \zeta}{2\omega_-\sqrt{N_-}} \\[2ex]
      -\dfrac{i k_z \bar\zeta}{2\omega_-\sqrt{N_-}} & -\dfrac{2\omega_-\,\tilde\omega_p}{(\Omega-\omega_-)N_-}
    \end{pmatrix}
  = \bigl(\tfrac{1}{2}-\beta\bigr)\tilde\omega_p\,I_2
  - \tfrac{\zeta_2}{\alpha}\sigma_1 - \tfrac{\zeta_1}{\alpha}\sigma_2
  + \bigl(\tfrac{1}{2}+\beta\bigr)\tilde\omega_p\,\sigma_3.
\end{equation}
Here, $\zeta = \zeta_1 + i \zeta_2$, $\sigma_{1,2,3}$ are the Pauli matrices, and
\begin{equation}\label{eq:alpha_beta}
  \alpha = \frac{2\omega_-\sqrt{N_-}}{k_z}, \qquad
  \beta = \frac{\omega_-}{(\Omega-\omega_-)N_-}.
\end{equation}
It follows that $b_{0,\omega_-} \in S^1$.
The operator $\ow b_{0,\omega_-}$ is a Dirac operator with Fermi velocity $v_F = \alpha^{-1}$, topological mass
$(\frac12+\beta)\tilde\omega_p(X)$ and potential
$(\frac12-\beta)\tilde\omega_p$. Since
$\det (P_{\omega_-} b_{0} P_{\omega_-} + (I_9-P_{\omega_-})\aver{\zeta}) = \aver{\zeta}^7 \det b_{0,\omega_-}$ (with $P_{\omega_-} b_0 P_{\omega_-} \in \mathbb{C}^{9 \times 9}$ here) and
$\det b_{0,\omega_-} = -\alpha^{-2}|\zeta|^2 + O(1)$, the condition \eqref{eq:sm_ellipticity} holds with
$N = 9$. See \cite{frazier2025topological,fu2022dispersion} for a detailed topological and
quantitative analysis of this elliptic operator.

For the (non-negative) simple sectors, we find that
\begin{equation}\label{eq:flat_symbols}
  b_{0,R} = f_R\tilde\omega_p,\quad
  b_{0,L+} = f_{L+}\tilde\omega_p,\quad
  b_{0,0} = 0,
\end{equation}
where
\begin{equation}\label{eq:fR_fLp}
  f_R = \frac{2\omega_-}{N_R(\Omega+\omega_R)} > 0, \qquad
  f_{L+} = -\frac{2\omega_-}{N_{L+}(\Omega-\omega_{L+})} > 0.
\end{equation}
Thus $b_{0,\ell}\in S^0$ is of class $\cL_0$ for $\ell \in \{\omega_R, \omega_{L+}, 0\}$.

With $\Gamma_P$ defined in \eqref{eq:parity_identity}, the identities
\begin{align*}
    \Gamma_P \,\overline{b_0}\, \Gamma_P = -b_0
\end{align*}
and \eqref{eq:parity} imply that the effective operators on the negative sectors are given by
\begin{align*}
    b_{0,-\ell} = -\overline{b_{0,\ell}}.
\end{align*}

\paragraph{Inter-sector coupling operators.}
The coupling operators $\mathbf{M}_{\ell\ell'}=P_\ell \bH_0 P_{\ell'}$ (cf. \eqref{eq:sm_MSigma}) are computed using the above eigenvectors. Set $m_{\ell, \ell'} := P_\ell b_0 P_{\ell'}$ so that $\ow (m_{\ell \ell'}) = \mathbf{M}_{\ell \ell'}$. Writing $\zeta = \zeta_1 + i \zeta_2$, 
we find
\begin{align}
\begin{split}
  m_{\omega_-,\,-\omega_-}
  &= \frac{ik_z\bar\zeta}{2\omega_-\sqrt{N_-}}\begin{pmatrix} 0 & -1\\ 1 & 0\end{pmatrix}, \qquad
  m_{\omega_-,\,0}
  = \begin{pmatrix} 0 \\ -\dfrac{i\bar\zeta}{\sqrt{2N_-}}\end{pmatrix},\\[4pt]
  m_{\omega_-,\,\pm R}
  &= \begin{pmatrix} \dfrac{i k_z (\pm \zeta_1 - i \zeta_2)}{2\omega_R \sqrt{N_R}} \\[2ex]
     - \dfrac{\omega_- \tilde\omega_p (2\Omega - \omega_- + \omega_R)}{\sqrt{N_- N_R}\,(\Omega-\omega_-)(\Omega+\omega_R)}\,\mathbbm{1}_\mp \end{pmatrix},\\
  m_{\omega_-,\,\pm L+}
  &= \begin{pmatrix} \dfrac{i k_z (\pm \zeta_1 + i \zeta_2)}{2\omega_{L+}\sqrt{N_{L+}}}\\[2ex]
     -\dfrac{\omega_-\tilde\omega_p (2\Omega-\omega_--\omega_{L+})}{\sqrt{N_-N_{L+}}\,(\Omega-\omega_-)(\Omega-\omega_{L+})}\,\mathbbm{1}_\pm\end{pmatrix},
  \label{eq:M_pm}
  \end{split}
\end{align}
where $\mathbbm{1}_\pm := \frac{1}{2} \pm \frac{1}{2}$ is $1$ for the upper sign and $0$ for the lower one. Note that the definition of $m_{\ell, \ell'}$ implies
\begin{align}\label{eq:m_conjugation}
    m_{\ell, \omega_-} = (m_{\omega_-, \ell})^*
\end{align}

\paragraph{Self-energies.}
We denote the eigenvalues in \eqref{eq:plasma_spectrum} by
\begin{align}\label{eq:evals}
    E_{\pm \omega_-} := \pm \omega_-, \qquad E_{\pm \omega_R} := \pm \omega_R, \qquad E_{\pm \omega_{L+}} := \pm \omega_{L+}, \qquad E_0 := 0;
\end{align}
that is, $E_\ell := \ell$.
Since $b_j=0$ for $j\geq 1$, \eqref{eq:sm_rH} shows that the effective Hamiltonian \eqref{eq:def_rH} of the topological sector satisfies
\begin{align*}
    \rH^{(\omega_-)} = \bH_{0,\omega_-}+ \delta \Sigma_{\omega_-} + O(\delta^2),
\end{align*}
with $\bH_{0,\omega_-} = \ow b_{0,\omega_-}$ as in \eqref{eq:D_minus}, and
\begin{align*}
    \Sigma_{\omega_-} = -P_{\omega_-}\bH_0\ashift_{\omega_-}\bH_0P_{\omega_-} = \sum_{\ell\neq\omega_-}\frac{\mathbf{M}_{\omega_-,\ell}\mathbf{M}_{\ell,\omega_-}}{E_{\omega_-}-E_{\ell}} =: \sum_{\ell \ne \omega_-} \ow (\sigma_{\omega_-, \ell})
\end{align*}
obtained using \eqref{eq:M_pm}--\eqref{eq:m_conjugation}. Since the couplings are unbounded, $\Sigma_{\omega_-}$ enters $\rH^{(\omega_-)}$ with the frequency cutoff of Section \ref{sec:sm_setting}, that is, with $\bH_0$ replaced by $\bH_0^\eps$.
In particular, we have
\begin{align}\label{eq:symbol_self_energies}
    \sigma_{\omega_-, -\omega_-}= \frac{m_{\omega_-, -\omega_-}m_{-\omega_-,\omega_-}}{2\omega_-} = \frac{k_z^2 |\zeta|^2}{8N_-\omega_-^3}\, I_2,
  \qquad
  \sigma_{\omega_-, 0} = \frac{m_{\omega_-, 0}m_{0,\omega_-}}{\omega_-}
  = \frac{|\zeta|^2}{2N_-\omega_-}\begin{pmatrix}0&0\\0&1\end{pmatrix}.
\end{align}
Note that the formulas in \eqref{eq:symbol_self_energies} follow from the composition calculus and the fact that the involved symbols are independent of $X$.

\subsection{Expansion with prescribed data in the topological sector}\label{sec:plasma_exp}

Since $b_j = 0$ for $j \ge 1$, we have $\bH_j = 0$ for $j \ge 1$, and the construction
\eqref{eq:sm_phout}--\eqref{eq:sm_phin} reduces to
\begin{equation}\label{eq:plasma_rec}
  \phout^{j,\ell} = -\ashl (D_T + \bH_0)\phi^{j-1}_\ell,
  \qquad
  (D_T + \bH_{0\ell})\phin^{j,\ell} = -P_\ell \bH_0 \phout^{j,\ell},
\end{equation}
with the initial conditions of \eqref{eq:sm_phin}.  Take $\phi_0 = P_{\omega_-}\phi_0$, so that $\phi^0_{\omega_-}$ solves
the Dirac equation 
\begin{align}\label{eq:plasma_Dirac_omega_-}
    (D_T + \bH_{0,\omega_-}) \phi^0_{\omega_-} = 0, \qquad \phi^0_{\omega_-} \vert_{T=0} = \phi_0,
\end{align}
with $\bH_{0,\omega_-} =  \ow (b_{0,\omega_-})$ of \eqref{eq:D_minus}, and $\phi^0_\ell \equiv 0$ for
$\ell \neq \omega_-$.  At order $\delta$, we find
\begin{align}\label{eq:plasma_out}
\begin{split}
    \phout^{1,\omega_-} = 
    \frac{\Psi_{0}}{\omega_-} \mathbf{M}_{0, \omega_-} \phi^0_{\omega_-} +
    \sum_{\ell \in \{R, L+\}}\frac{\Psi_{\ell}}{\omega_- -\omega_\ell} \mathbf{M}_{\ell, \omega_-} \phi^0_{\omega_-}
    &+\sum_{\ell \in \{-R, -L+\}}\frac{\Psi_{\ell}}{\omega_- +\omega_\ell} \mathbf{M}_{\ell, \omega_-} \phi^0_{\omega_-}\\
    &+ \frac{\begin{bmatrix}
        \Psi_{-1} & \Psi_{-2}
    \end{bmatrix}}{2\omega_-} \mathbf{M}_{-\omega_-, \omega_-} \phi^0_{\omega_-},
\end{split}
\end{align}
where the $\Psi_\ell$ are given by \eqref{eq:evecs}, \eqref{eq:evecs_simple}, and \eqref{eq:parity}, and the $\mathbf{M}_{\ell, \omega_-} = \ow (m_{\ell, \omega_-})$ by \eqref{eq:M_pm}--\eqref{eq:m_conjugation}.
Thus we have shown that the function
\begin{align*}
    \psi_\delta (t,x):= \delta^{d/2} e^{-i \omega_- t} (\phi^0_{\omega_-} (\delta t, \delta x) + \delta \phout^{1,\omega_-} (\delta t, \delta x))
\end{align*}
given by \eqref{eq:plasma_Dirac_omega_-}--\eqref{eq:plasma_out} approximates the tight-binding dynamics to $O(\delta)$ over times $t \lesssim \delta^{-1}$. That is, with $H_\delta$ defined by \eqref{eq:plasma_H} and $\varphi_\delta$ the solution to
\begin{align*}
    (D_t + H_\delta) \varphi_\delta = 0, \qquad \varphi_\delta (0, x) = \delta^{d/2} \phi_0 (\delta x),
\end{align*}
Theorem \ref{thm:main_higher_order} implies
\begin{align*}
    \|(\psi_\delta-\varphi_\delta)(t, \cdot)\|_{H^m(\Rm^d)} \leq C_m(\delta + \delta^{2} t) \left(1 + (\delta t)^{2m+2}\right), \qquad t \ge 0, \quad 0 < \delta \le 1, \quad m \ge 0.
\end{align*}
The expansion could be continued to arbitrary orders of accuracy, but we do not pursue this here.


\section{Numerical experiments on the effective Hamiltonian}\label{sec:sm_numerics}

The simulations of Section \ref{sec:numerics} cover moderate times and effective Hamiltonians of order $p \le 2$. In this
section, we illustrate the results of Section \ref{sec:effective} beyond that range: the effective
Hamiltonian of order $p = 3$ (Remark \ref{rem:rH_expansion}), long times on dispersive and on flat
sectors (Theorem \ref{prop:rH} and Corollary \ref{cor:rH_eps}), the frequency cutoff, and prescribed
data treated sector by sector, as in Remark \ref{rem:all_sectors}. Unless stated
otherwise, the models, coefficients and data are those of Section \ref{sec:numerics}.

\subsection{Methods}\label{sec:sm_num_methods}
The tight-binding and macroscopic discretizations are those of Section
\ref{sec:numerics_methods}. The approximation of Theorem \ref{prop:rH} is evaluated with $\rH$
replaced by its truncation $\tilde\rH = \sum_{j<p}\delta^j\rH_j$, with the coefficients of Remark
\ref{rem:rH_expansion}, with $S = \sum_{j\le p}\delta^jS_j$, and with $\cG^{\pm1/2}$ expanded to
order $\delta^p$. Since the data are smooth, no frequency cutoff is used, except in Section
\ref{sec:sm_num_cutoff}. Where $\tilde\rH - \bH_{0\ell}$ is bounded, this is covered by Remark \ref{rem:rH_expansion} and the last statements of Corollary \ref{cor:rH_eps}. Where $\tilde\rH$ contains unbounded corrections (e.g.\ $\rH_1$ in the Kekul\'e and plasma models, or $\rH_2$ for $p=3$), the pseudo-spectral discretization itself truncates frequencies at the grid scale, which then plays the role of $\eps^{-1}$ in Theorem \ref{prop:rH}. The approximation is compared with the tight-binding solution started from
$(P+S^0)\phi_0$, as in the last statement of Corollary \ref{cor:rH_eps}; by Lemma \ref{lem:Pi_S},
this is the band-adapted datum of Definition \ref{def:single} of the same order. The operators
$\tilde\rH$ are Hermitian, and the effective dynamics are computed by Krylov subspace methods. For
the one-dimensional models, they are small enough to be assembled as matrices on the Fourier grid
and diagonalized, which gives the effective solution at any $T$ directly. In the same way, the
generator of the expansions of order $p = 2$ is block lower triangular, with diagonal blocks
$\bH_{0\ell}$, so that its exponential follows from the eigendecomposition of $\bH_{0\ell}$ and the
divided differences of $\lambda\mapsto e^{-iT\lambda}$. Over long times, the tight-binding evolutions
are computed by Krylov approximations in substeps.

\subsection{Rates of convergence up to third order}\label{sec:sm_num_rates}
\begin{figure}[tb]
  \centering
  \includegraphics[width=5.125in]{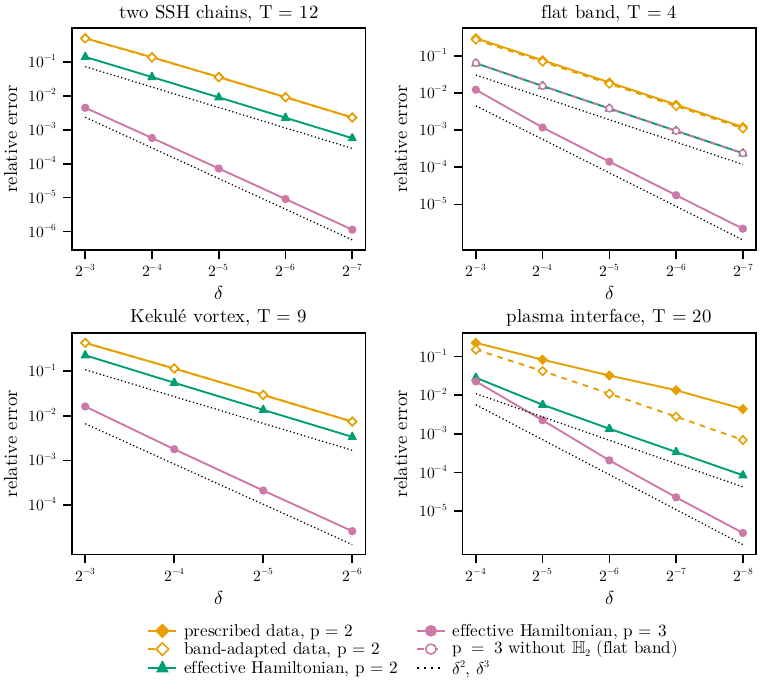}
  \caption{Relative error as a function of $\delta$ for the four models of Section
  \ref{sec:numerics}: two SSH chains with a barrier at $T = 12$ (top left), SSH chain coupled to a
  flat band with a flat-band datum at $T = 4$ (top right), scattering off the Kekul\'e vortex at
  $T = 9$ (bottom left) and cold plasma interface at $T = 20$ (bottom right). Filled diamonds:
  prescribed datum, $p = 2$; open diamonds: band-adapted datum, $p = 2$ (except for the plasma, the two
  nearly coincide); triangles and circles: effective Hamiltonian of Theorem \ref{prop:rH} with $p = 2$
  and $p = 3$; open circles (top right): $p = 3$ without $\rH_2$, which coincides with $p = 2$. Dotted
  lines: $\delta^2$ and $\delta^3$.}
  \label{fig:sm_rates}
\end{figure}

Figure~\ref{fig:sm_rates} shows, for the four models of Section \ref{sec:numerics}, the error of the
effective Hamiltonian with $p = 2$ and $p = 3$ at a time after the interaction with the
inhomogeneity, together with that of the expansions of order $p = 2$. For $p = 3$, the effective
Hamiltonian is $\rH_0 + \delta\rH_1 + \delta^2\rH_2$. For the two SSH chains, $\rH_2 = P_\ell\bH_2P_\ell +
\hat Q_\ell$ as in Section \ref{sec:ssh_two_ops}, where $b_2 = \frac{v^3}{24}\diag\bigl(\alpha_1\zeta^3\sigma_2,
\alpha_2\zeta^3\sigma_2\bigr)$ is the next term of the Taylor expansion of \eqref{eq:sum_ssh_ajj}. In the
Dirac sector of the Kekul\'e-O model, $S_1 = 0$, so that $\rH_2 = P_0\bH_2P_0$. For the cold plasma,
$b_j = 0$ for $j\ge1$, and $\rH_2$ consists of the terms of third order in $\bH_0$. In all four models,
the effective Hamiltonian converges at the rate $\delta^2$ for $p = 2$ and $\delta^3$ for $p = 3$, as
predicted by Corollary \ref{cor:rH_eps} (and by Remark \ref{rem:pde_class_gain} for the plasma), and its
error is markedly smaller than that of the expansions. On the flat band, $\rH_2 =
\frac{1}{E^2}\gamma^*\bH_{01}\gamma$, the dispersion of \eqref{eq:sum_flat_Heff}, is exactly the term
$P\bH_0a^\sharp\bH_0a^\sharp\bH_0P$; without this term, the construction with $p = 3$ is no more
accurate than the one with $p = 2$.

\subsection{Range of validity}\label{sec:sm_num_longtime}
\begin{figure}[tb]
  \centering
  \includegraphics[width=5.125in]{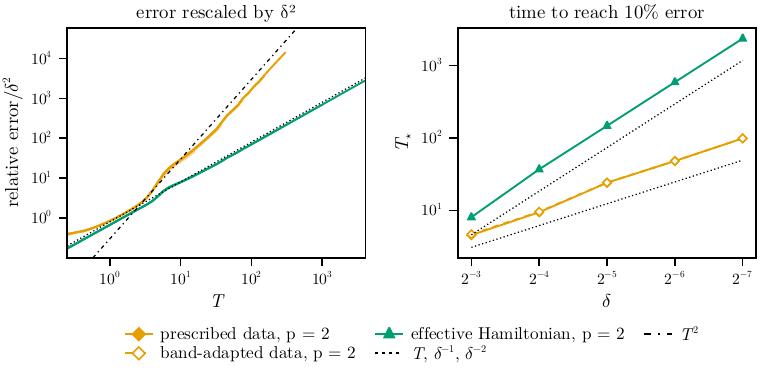}
  \caption{Two coupled SSH chains with the barrier of Section \ref{sec:numerics_chains}, over long
  times. Left: relative $\ell^2$ error divided by $\delta^2$, as a function of $T$, for
  $\delta = 2^{-3},\dots,2^{-7}$ (darker for smaller $\delta$), for the expansion with prescribed
  data and $p = 2$ and for the effective Hamiltonian $\rH_0 + \delta\rH_1$ of Theorem
  \ref{prop:rH}; dotted and dash-dotted lines: $T$ and $T^2$. Right: slow time $T_\star$ at which the
  relative error reaches $10\%$, as a function of $\delta$; dotted lines: $\delta^{-1}$ and
  $\delta^{-2}$.}
  \label{fig:sm_longtime}
\end{figure}

For the two coupled SSH chains with the barrier, we follow the solution up to $T = \delta^{-2}/4$,
that is, $t = \delta^{-3}/4$, during which the wave packet crosses the periodic domain and the
barrier many times. Divided by $\delta^2$, the errors collapse onto a single curve for each
construction (Figure~\ref{fig:sm_longtime}, left). The error of the expansions grows quadratically in
$T$: they contain the self-energy $\delta\Sigma_1$ of \eqref{eq:sum_ssh_Heff} only through a Duhamel
source, that is, through the phase it generates expanded in powers of $\delta T$. The error of the
effective Hamiltonian grows only linearly in $T$, over the whole range, far beyond
$t\sim\delta^{-1}$, as Corollary \ref{cor:rH_eps} predicts for a sector of class $\cL_1$.
Accordingly, the time at which the error reaches a given level grows like $\delta^{-2}$ for the
effective Hamiltonian, and like $\delta^{-1}$ for the expansions (right).

\subsection{Flat and dispersive sectors}\label{sec:sm_num_flat}
\begin{figure}[tb]
  \centering
  \includegraphics[width=5.125in]{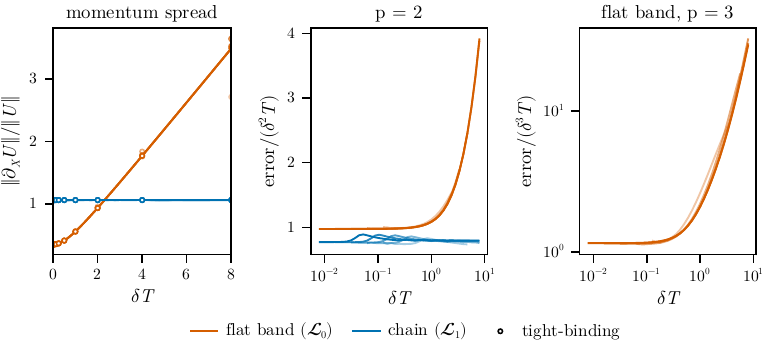}
  \caption{SSH chain coupled to a flat band, with data on the flat band (sector of class $\cL_0$)
  and on the chain (class $\cL_1$), for $\delta = 2^{-3},\dots,2^{-7}$ (darker for smaller
  $\delta$), as functions of $\delta T$. Left: $\norm{\partial_X U}/\norm{U}$ for the component $U$ of
  the solution in the sector of the data, for the tight-binding solution (circles) and the effective
  solution (lines). Center: relative $\ell^2$ error of the effective Hamiltonian with $p = 2$, divided
  by $\delta^2 T$. Right: flat band, relative error with $p = 3$, divided by $\delta^3 T$.}
  \label{fig:sm_flat}
\end{figure}

For the SSH chain coupled to a flat band, we compare the flat-band datum of Section
\ref{sec:numerics_chains}, in the sector $2\in\cL_0$, with a wave packet on the upper branch of the
chain, in the sector $1\in\cL_1$, and follow both up to $\delta T = 8$. For $p = 2$, the effective
Hamiltonians are $\bH_{01} + \frac{\delta}{E}\gamma\gamma^*$ and $-\frac{\delta}{E}|\gamma|^2$. On the
flat band, the gradient of the self-energy acts as a force: the momentum of the wave packet drifts
like $\delta T\,\nabla|\gamma|^2/E$, and its Sobolev norms grow linearly in $\delta T$, in the
tight-binding solution and in the effective one alike, while on the chain they remain bounded
(Figure~\ref{fig:sm_flat}, left). This growth is genuine, and it is the one that limits the range
of Theorem \ref{prop:rH} on sectors of class $\cL_0$ (Lemma \ref{lemma:Sobolev_rH}). Accordingly, the
error divided by $\delta^2 T$ remains constant on the chain, while on the flat band it is a function
of $\delta T$ which increases once $\delta T$ is of order one (center). The same holds for $p = 3$,
with an error of order $\delta^3T$ that increases faster with $\delta T$ (right), since the neglected
terms involve more derivatives of the drifting wave packet.

\subsection{The frequency cutoff}\label{sec:sm_num_cutoff}
\begin{figure}[tb]
  \centering
  \includegraphics[width=5.125in]{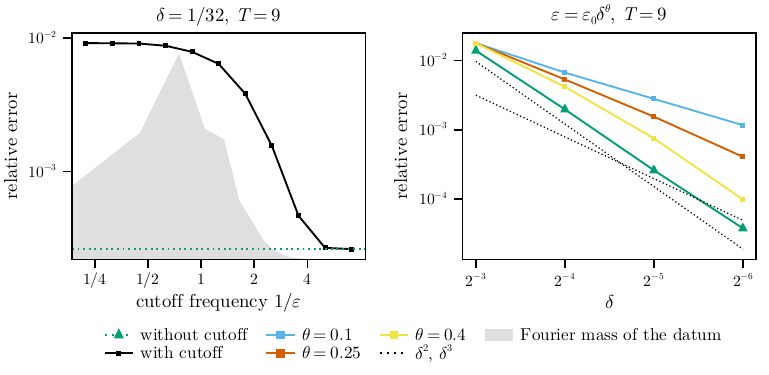}
  \caption{Frequency cutoff, Kekul\'e vortex with the zero mode of $\ow b_{0,0}$ as datum, effective
  Hamiltonian with $p = 2$, relative $\ell^2$ error at $T = 9$. Left: as a function of the cutoff
  frequency $1/\eps$, for $\delta = 1/32$; dotted line: without cutoff; shaded: radial distribution
  of the Fourier mass of the datum. Right: as a function of $\delta$, without cutoff (triangles) and
  with $\eps = \eps_0\delta^\theta$, $\theta = 0.1, 0.25, 0.4$ (squares), where $\eps_0$ is chosen so
  that, for $\delta = 1/8$, $1/\eps$ is the radius containing $90\%$ of the Fourier mass of the datum;
  dotted lines: $\delta^2$ and $\delta^3$.}
  \label{fig:sm_cutoff}
\end{figure}

In the Dirac sector of the Kekul\'e-O model, the operators $\rH_1 = P_0\bH_1P_0 = \ow b_{1,0}$ and $S_2 = -\ashift_0\bH_1P_0$ are unbounded, so that
Theorem \ref{prop:rH} involves the frequency cutoff of Section \ref{sec:effective}. We take the zero mode
of Section \ref{sec:numerics_kekule} as datum, and a smooth radial cutoff $\chi$ equal to one on
$[0,1]$ and to zero on $[2,\infty)$. The construction of Theorem \ref{prop:rH} with $p = 2$, in which
$\rH_1$ and $S_2$ carry the cutoff, is compared with the tight-binding solution started from the
datum $(P+S^0)\phi_0$ without cutoff, as in the last statement of Corollary \ref{cor:rH_eps}. As a
function of the cutoff frequency $1/\eps$, at fixed $\delta$, the error decreases as the cutoff moves
past the Fourier content of the datum, and coincides with that of the construction without cutoff
once $1/\eps$ exceeds it (Figure~\ref{fig:sm_cutoff}, left): the cutoff then plays no role. With
$\eps = \eps_0\delta^\theta$, the cutoff frequency increases as $\delta$ decreases, and the error
approaches that of the construction without cutoff, the faster the larger $\theta$ (right). For
small $\theta$, the frequency-truncation error, the term $\delta\eps^n$ of Theorem \ref{prop:rH},
dominates over the whole range of $\delta$: Corollary \ref{cor:rH_eps} then requires a large $n$, hence
a large constant. Since the data are smooth, the construction without cutoff, used everywhere
else in this section and in Section \ref{sec:numerics}, is the more accurate one.

\subsection{Data in all sectors}\label{sec:sm_num_allsectors}
For prescribed data, Remark \ref{rem:all_sectors} builds the approximation sector by
sector: with $u = (I+\sum_\ell S^\ell)^{-1}\phi_0$, each sector $\ell$ is evolved by its own
effective Hamiltonian from $\cG_\ell^{1/2}P_\ell u$, and the contributions
$(P_\ell + S^\ell)\cG_\ell^{-1/2}w_\ell$ are summed. Without frequency cutoff,
$I+\sum_\ell S^\ell P_\ell$ is the dressing map $\mathcal T_\delta$ of Section \ref{sec:sm_setting}, and
$P_\ell u$ is the datum from which the band-adapted expansion of Theorem \ref{thm:equiv} launches
the sector $\ell$. For a datum in the sector $\ell_0$, the components
$P_\ell u$, $\ell\ne\ell_0$, are small, of order $\delta$ in general: they are the wave packets emitted
by the mechanism (M2) of Remark \ref{rem:mechanisms}. In this construction, each of them is evolved by the effective
Hamiltonian of its own sector, whereas the expansions of Theorem \ref{thm:main_higher_order} propagate
it with $\bH_{0\ell}$ only. We compute $u$ by the Neumann series, to the order of the construction.

\begin{figure}[tb]
  \centering
  \includegraphics[width=5.125in]{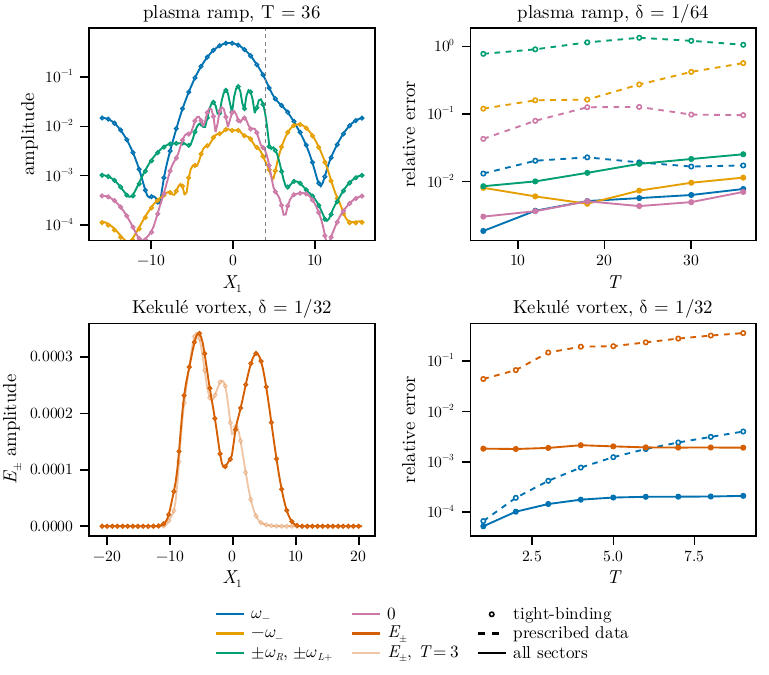}
  \caption{Prescribed data treated sector by sector. Top: cold plasma with a ramp, $p = 3$. Left:
  amplitudes of the components of the solution in the sectors $\omega_-$, $-\omega_-$,
  $\{\pm\omega_R, \pm\omega_{L+}\}$ and $0$, normed in $X_2$, as functions of $X_1$, at $T = 36$ and
  $\delta = 1/64$: tight-binding solution (circles) and sector-by-sector construction (lines); dashed
  line: position of the ramp. Right: relative error in each group of sectors as a function of $T$,
  for $\delta = 1/64$: expansion with prescribed data and $p = 2$ (dashed, open circles) and
  sector-by-sector construction (solid). Bottom: Kekul\'e vortex, $p = 3$, $\delta = 1/32$. Left:
  amplitude of the component in the sectors $E_\pm$, normed in $X_2$, at $T = 3$ and $T = 9$. Right:
  relative error in the Dirac sector and in the sectors $E_\pm$, for the expansion with prescribed
  data and $p = 3$ (dashed) and the sector-by-sector construction (solid).}
  \label{fig:sm_allsectors}
\end{figure}

\paragraph{Cold plasma.}
We take $k_z = 0.4$, $\Omega = 1$, and a ramp of the plasma frequency,
\begin{align*}
    \wt(X) = \omega_{\max}\,r(X_1 - 4)\,r\bigl(2(12 - X_1)\bigr), \qquad r(x) = \tfrac12(1+\tanh x),
    \qquad \omega_{\max} = 1.2,
\end{align*}
on the torus $[-16, 16)^2$. The initial condition is a Gaussian wave packet on the upper branch of
$\ow b_{0,\omega_-}$, of width $2$ and momentum $\zeta_0 = (2, 0)$, centered at $X_0 = (-3, 0)$. The main
wave packet is reflected by the ramp, while its partner in the sector $-\omega_-$ is transmitted
(Figure~\ref{fig:sm_allsectors}, top left): since $b_{0,-\omega_-} = -\overline{b_{0,\omega_-}}$, the two
sectors see opposite potentials and masses. The sectors $\pm\omega_R$, $\pm\omega_{L+}$ and $0$ are of class
$\cL_0$, with $b_{0,\ell} = f_\ell\wt$ by \eqref{eq:flat_symbols}: at leading order, the wave packets
seeded there do not move, and their dynamics is that of the effective Hamiltonian at order $\delta$.
The expansion with prescribed data, which propagates them with $\bH_{0\ell}$ only, gets them wrong at
leading order, and its error in these sectors is of order one relative to their size; in the sector
$-\omega_-$, it grows with $T$. With the sector-by-sector construction, the error is small in every
sector (top right). We use $p = 3$ here: for the ramp, the coefficient $\rH_2$ of the main sector is
large, the gap between $\omega_-$ and $\omega_R$ being small, and with $p = 2$ its omission dominates
the error of the main sector over these times. The expansion with prescribed data is shown with $p=2$; at any order it propagates the seeded packets with $\bH_{0\ell}$, and contains their self-energies only through Duhamel sources, that is, expanded in powers of $\delta T$.

\paragraph{Kekul\'e vortex.}
For the scattering off the Kekul\'e vortex of Section \ref{sec:numerics_kekule}, the sectors $E_\pm$ at
the $\Gamma$ point are of class $\cL_0$, with the effective Hamiltonians \eqref{eq:sum_kekule_Hpm}, of
order $\delta$. The components $P_\pm u$ are of order $\delta^2$, since
$S_1 = 0$ in the Dirac sector. In the tight-binding solution, the $\Gamma$-sector content consists of
these seeds, which stay near the launch site, and of the algebraic corrector carried along with the
Dirac wave packet (bottom left). With $p = 3$, the sector-by-sector construction reproduces both. Its
error is small in both groups of sectors and nearly constant in $T$, while that of the expansion
with prescribed data grows, to a sizeable fraction of the $\Gamma$-sector content (bottom right).

\section{List of notation}\label{sec:sm_notation}

The following list collects the notation of the main paper and of this supplement. Equation and
section numbers without the prefix SM refer to the main paper. Symbols used only inside a single
proof are listed at the end of the second group. A few letters are used for unrelated objects in
well-separated contexts; these are marked with $\dagger$ and listed together after the last group.

{\small
\renewcommand{\arraystretch}{1.15}
\begin{longtable}{@{}p{0.2\textwidth}p{0.58\textwidth}p{0.16\textwidth}@{}}
\toprule
Symbol & Meaning & Reference\\
\midrule
\endfirsthead
\toprule
Symbol & Meaning & Reference\\
\midrule
\endhead
\bottomrule
\endfoot
\multicolumn{3}{@{}l}{\emph{Scales, variables and parameters}}\\[2pt]
$\delta$, $\delta_0$ & scale-separation parameter, $0<\delta\le\delta_0$ & \eqref{eq:a_delta}\\
$t$, $x$; $T$, $X$ & microscopic time and space; slow variables $T=\delta t$, $X=\delta x$ & Sect.~\ref{sec:intro}\\
$d$ & space dimension ($d=1$ for the chains, $d=2$ otherwise) & \eqref{eq:a_delta}\\
$\xi$; $\zeta$ & momentum; rescaled momentum $\zeta=(\xi-K)/\delta$, identified with $\zeta_1+i\zeta_2\in\C$ when $d=2$ & \eqref{eq:decomposition_higher_order_all}\\
$K$ & reference quasi-momentum ($K=0$ for the chains and the Kekul\'e model; in-plane $K=0$ and fixed $k_z$ for the plasma) & \eqref{eq:microWP_higher_order}\\
$p$ & number of terms (order) of the expansion & Assumpt.~\ref{assumption:a_higher_order_all}\\
$\rate$ & exponent of the remainder $\delta^{p+\rate}b_p$, $0<\rate\le1$ & \eqref{eq:decomposition_higher_order_all}\\
$q$ & order of the truncated Neumann series $\mathcal T^{(q)}_\delta$ & Def.~\ref{def:dressing}\\
$\kappa$ & time exponent of Corollary~\ref{cor:timescale} & Cor.~\ref{cor:timescale}\\
$\eps$ & frequency-cutoff parameter (cutoff at $|\zeta|\gtrsim\eps^{-1}$) & Sect.~\ref{sec:effective}\\
$\eta$, $\eta_0$ & $\eta=\delta/\eps^2$ and an upper bound for it & \eqref{eq:def_eta}\\
$c_0$ & smallness constant in the hypothesis $0<\delta\le c_0\eps$ & below \eqref{eq:norm_S}\\
$n$\textsuperscript{$\dagger$} & order of the truncation error $\delta\eps^n$ & Thm.~\ref{prop:rH}\\
$\theta$, $\delta_1$ & exponent in $\eps=\delta^\theta$, $0<\theta<\frac12$, and threshold for $\delta$ & Cor.~\ref{cor:rH_eps}\\
$\eps_0$ & prefactor in $\eps=\eps_0\delta^\theta$ in the experiments of Sect.~\ref{sec:sm_num_cutoff} & Sect.~\ref{sec:sm_num_cutoff}\\
$m$, $s$, $k$ & Sobolev indices; $k$ also the order of $D_T^k$ in Lemma~\ref{lemma:Sobolev_higher_order_all} & ---\\
$C$, $c$ & generic constants (independent of $\delta$, $t$) & Sect.~\ref{sec:intro}\\[4pt]
\multicolumn{3}{@{}l}{\emph{Function spaces, symbols and operators}}\\[2pt]
$\aver{Y}$ & $(1+|Y|^2)^{1/2}$ & Sect.~\ref{sec:intro}\\
$D_z$ & $\frac1i\partial_z$ & Sect.~\ref{sec:intro}\\
$H^m$, $H^\infty$, $C^\infty_b$ & Sobolev spaces, $H^\infty=\cap_mH^m$; smooth functions bounded with all derivatives & Sect.~\ref{sec:intro}\\
$\ow$ & Weyl quantization & \eqref{eq:weylquanth}\\
$\fm$; $S(\fm)$, $S^m$\textsuperscript{$\dagger$} & order function; symbol classes & Defs.~\ref{def:symbol_class}, \ref{def:hormander_class}\\
$\ncv$ & number of derivatives in the Calder\'on--Vaillancourt bound & Lemma~\ref{lemma:CV}\\
$N$\textsuperscript{$\dagger$} & size of the matrix-valued symbols, $a\in\C^{N\times N}$ & Assumpt.~\ref{assumption:a_higher_order_all}\\
$a$, $a_\delta$ & symbol; $a_\delta(x,\xi)=a(\delta x,\xi;\delta)$ & \eqref{eq:a_delta}\\
$H_\delta$ & tight-binding (microscopic) Hamiltonian $\ow a_\delta$ & \eqref{eq:H_delta}\\
$a^0$ & constant leading matrix $a^0=\sum_\ell E_\ell P_\ell$ & Assumpt.~\ref{assumption:a_higher_order_all}\\
$E_\ell$, $P_\ell$, $L$ & distinct eigenvalues of $a^0$, orthogonal projections, their number & Assumpt.~\ref{assumption:a_higher_order_all}\\
$\ell$, $\ell'$ & sector indices; $\Ran P_\ell$ is the \emph{sector} $\ell$ & Sect.~\ref{sec:intro}\\
$b_j$, $b_{0\ell}$ & coefficients of the expansion of $a$; $b_{0\ell}=P_\ell b_0P_\ell$ & \eqref{eq:b_j_symbol_class}, \eqref{eq:decomposition_higher_order_all}\\
$\cL_0$, $\cL_1$ & sectors with $b_{0\ell}\in S(1)$ (bounded), resp.\ elliptic $b_{0\ell}\in S^1$ & \eqref{eq:cL_def}, \eqref{eq:ellipticity}\\
$\bH_j$, $\bH_{0\ell}$ & $\ow b_j$, $\ow b_{0\ell}$ (boldface: operators in the slow variables) & Sect.~\ref{sec:tb}\\
$\ashl$ & reduced resolvent $(a^0-E_\ell)^{-1}$ on $\Ran(I_N-P_\ell)$ & \eqref{eq:inva_higher_order}\\
$\mathbf M_{\ell\ell'}$ & inter-sector couplings $P_\ell\bH_0P_{\ell'}$, $\ell\ne\ell'$ & \eqref{eq:sm_MSigma}\\
$\Sigma_\ell$, $\Sigma^{(2)}_\ell$ & self-energies $-P_\ell\bH_0\ashl\bH_0P_\ell$ and $-P_\ell\bH_1\ashl\bH_1P_\ell$ & \eqref{eq:sm_MSigma}, \eqref{eq:sum_kekule_Sigma}\\
$\Re X$ & $\frac12(X+X^*)$ for an operator $X$ & Rem.~\ref{rem:rH_expansion}\\
$\tilde H_\delta$ & slow-variable form of $H_\delta$ & \eqref{eq:tHdelta}\\
\multicolumn{3}{@{}p{0.95\textwidth}}{\emph{Symbols local to a proof:} $\mathcal Q_\delta$, $\mathcal K_\delta$, $\tilde\chi$, $\zeta_*$, $M$ (Lemma~\ref{lemma:sa_H0l}); $\lambda$, $Q=I-P$, $A$, $u$, $v$, $h$, $R$, $R_{\rm new}$ (Lemma~\ref{lem:Pi_S}); $\Lambda$, $\ds$ (Lemmas~\ref{lemma:elliptic_reg}, \ref{lemma:bdd_reg}, \ref{lemma:Sobolev_rH}); $R_\delta$, $M$, $r_\delta$ (Lemma~\ref{lemma:u_delta_reg}); $\uni_\delta$, $r^k_\ell$ (Proposition~\ref{lemma:res_higher_order_all}); $\rho$, $\rho_0$, $\rho_1$, $\mathcal E$, $B_0$ (Proposition~\ref{prop:res_rH}); $g_k$ (Theorem~\ref{thm:equiv}).}\\[4pt]
\multicolumn{3}{@{}l}{\emph{Solutions and the expansion (Sections~\ref{subsec:macroscopic}--\ref{sec:dressed})}}\\[2pt]
$\phi_0$ & macroscopic initial profile & \eqref{eq:microWP_higher_order}\\
$\varphi_\delta$ & exact solution of the microscopic problem & \eqref{eq:microWP_higher_order}\\
$\phi^0_\ell$ & leading amplitude in the sector $\ell$ & \eqref{eq:def_phi_0_ell}\\
$\phout^{j,\ell}$, $\phin^{j,\ell}$, $\phi^j_\ell$ & algebraic correctors, in-sector correctors, $\phi^j_\ell=\phin^{j,\ell}+\phout^{j,\ell}$ & \eqref{eq:def_phout}, \eqref{eq:def_phin}\\
$\phi_\ell$, $\phi$ & $\sum_j\delta^j\phi^j_\ell$, $\sum_\ell\phi_\ell$ & \eqref{eq:phi_expansion_higher_order}\\
$\psi_\delta$ & approximate solution & \eqref{eq:psi_def_higher_order_all}\\
(M1)--(M3) & mechanisms of inter-band coupling & Rem.~\ref{rem:mechanisms}\\
$\cphin^{j,\ell}$, $\cphout^{j,\ell}$, $\check\phi$, $\check\psi_\delta$ & single-sector construction & Def.~\ref{def:single}\\
$\Pi^\delta_\ell$, $\phi^\delta_0$ & band-adapted (dressed) datum $\phi^\delta_0=\Pi^\delta_{\ell_0}\phi_0$ & \eqref{eq:dressed}\\
$\Psi_\delta[\ell,\phi_0]$\textsuperscript{$\dagger$} & band-adapted wave packet & \eqref{eq:Psi_single}\\
$\mathcal T_\delta$, $\mathcal A_\delta$, $\mathcal T^{(q)}_\delta$ & dressing map $\mathcal T_\delta=I_N+\delta\mathcal A_\delta$, truncated Neumann inverse & \eqref{eq:Tdelta}\\
$\psi^{\rm ad}_\delta$ & band-adapted expansion of arbitrary accuracy & \eqref{eq:psi_ad}\\[4pt]
\multicolumn{3}{@{}l}{\emph{Effective Hamiltonians (Section~\ref{sec:effective})}}\\[2pt]
$P$, $E$, $a^\sharp$ & $P_\ell$, $E_\ell$, $\ashl$ for a fixed sector $\ell$, whose index is dropped & Sect.~\ref{sec:effective}\\
$\chi$, $\chi_\eps$ & real cutoff function equal to one near the origin; $\chi_\eps(\zeta)=\chi(\eps\zeta)$ & Sect.~\ref{sec:effective}\\
$b^\eps_j$, $\bH^\eps_j$, $\bH^\eps_{0\ell}$ & $\chi_\eps b_j\chi_\eps$, $\ow b^\eps_j$, $P\bH^\eps_0P$ & \eqref{eq:bH_bd}\\
$\bHe$ & cut-off sum $\sum_{j<p}\delta^j\bH^\eps_j$ & \eqref{eq:bH}\\
$S$, $S_j$; $S^0$, $S^0_j$ & coupling operator $\Ran P\to\Ran(I-P)$ and its coefficients; the same without cutoff & \eqref{eq:norm_S}, Def.~\ref{def:S}\\
$S^\ell$ & the operator $S$ of the sector $\ell$ & Rem.~\ref{rem:all_sectors}\\
$\cG$ & $P+S^*S$ & Sect.~\ref{sec:effective}\\
$B$ & operator defining $S$ through $(I-P)B(P+S)$ & \eqref{eq:def_B}, \eqref{eq:riccati_S}\\
$\rH$, $\rH^{(\ell)}$ & effective Hamiltonian (of the sector $\ell$) & \eqref{eq:def_rH}\\
$\rH_j$, $\rH^{(\ell)}_j$, $\tilde\rH$ & coefficients of $\rH\sim\sum_j\delta^j\rH_j$ (of the sector $\ell$); $\tilde\rH=\sum_{j<p}\delta^j\rH_j$ & Rem.~\ref{rem:rH_expansion}\\
$\hat Q_\ell$ & second-order coefficient for the coupled SSH chains & Sect.~\ref{sec:ssh_two_ops}\\
$\psi$, $\varphi$, $\varphi^0$ & effective approximation; exact solutions with data $(P+S)\phi_0$, $(P+S^0)\phi_0$ & \eqref{eq:psi_def_higher_order_rH}, \eqref{eq:microWP_higher_order_rH}\\
$w$ & effective amplitude ($w$ in \eqref{eq:num_reduced}, in the proof of Cor.~\ref{cor:rH_eps}, $w_\ell$ in Sect.~\ref{sec:sm_num_allsectors}) & ---\\[4pt]
\multicolumn{3}{@{}l}{\emph{Coupled chains (Sections~\ref{sec:summary_chains}, \ref{sec:applications})}}\\[2pt]
$v$\textsuperscript{$\dagger$} & lattice constant of the chains & \eqref{eq:sum_ssh_ajj}\\
$E$\textsuperscript{$\dagger$} & on-site energy offset, $E_1=E$, $E_2=-E$ (resp.\ $0$ for the flat band) & \eqref{eq:sum_ssh_ajj}\\
$\alpha_j$\textsuperscript{$\dagger$}, $M_j$ & hopping amplitudes and mass matrices of chain $j$ & \eqref{eq:sum_ssh_ajj}\\
$\beta$\textsuperscript{$\dagger$}, $\gamma$ & inter-chain coupling; coupling to the flat band & \eqref{eq:sum_ssh_a}, \eqref{eq:sum_flat_a}\\
$\sigma_0,\dots,\sigma_3$\textsuperscript{$\dagger$} & identity and Pauli matrices & \eqref{eq:sum_ssh_ajj}\\
$\tilde\phi_{0\ell}$, $\tilde\phi^0_\ell$ & components of the datum; their free evolutions & \eqref{eq:sum_ssh_psi}\\[4pt]
\multicolumn{3}{@{}l}{\emph{Graphene with Kekul\'e-O modulation (Sections~\ref{sec:summary_kekule}, \ref{sec:kekule})}}\\[2pt]
$a_1,a_2,a_3$; $R_{\pm2\pi/3}$ & nearest-neighbor vectors; rotations & Sect.~\ref{sec:summary_kekule}\\
$K$, $K'$; $G$ & graphene valleys; Kekul\'e wave vector ($G=K'$) & \eqref{eq:Kekule_G}\\
$\kdisp$, $\kdisp_n$ & dispersion function and folded dispersions, $n\in\{-1,0,1\}$ & \eqref{eq:sum_kekule_symbol}\\
$\De$, $\De_0$ & complex Kekul\'e order parameter; its radial profile for a vortex & \eqref{eq:sum_kekule_symbol}\\
$t_1$ & hopping amplitude (set to one) & \eqref{eq:fullsymbol}\\
$A$, $A_j$ & off-diagonal block of the symbol and its expansion coefficients & \eqref{eq:sum_kekule_symbol}, \eqref{eq:sum_kekule_A01}\\
$\fA_j$, $\fB_j$ & sublattice $\times$ folded-valley basis & Sect.~\ref{sec:summary_kekule}\\
$E_0$, $E_\pm$; $P_0$, $P_\pm$; $\uu_\pm$ & Dirac and $\Gamma$-point sectors; projections; unit vectors with $P_\pm=\uu_\pm\uu_\pm^*$ & \eqref{eq:sum_kekule_spec}\\
$b_{j,0}$, $b_{j,\pm}$ & $P_0b_jP_0$, $P_\pm b_jP_\pm$ & \eqref{eq:sum_kekule_b}\\
$\Gamma_\pm$\textsuperscript{$\dagger$} & covectors $\uu_\pm^*b_1P_0$ coupling the Dirac and $\Gamma$-point sectors & \eqref{eq:sum_kekule_Gamma}\\
$m^*_\pm$ & effective masses of the $\Gamma$-point sectors & \eqref{eq:sum_kekule_Hpm}\\
$(R,\vartheta)$; $\mathcal W[\De]$ & polar coordinates; winding number & Sect.~\ref{sec:kekule_vortex}\\[4pt]
\multicolumn{3}{@{}l}{\emph{Magnetized cold plasma (Sections~\ref{sec:summary_plasma}, \ref{sec:plasma})}}\\[2pt]
$\mathbf v$, $\mathbf E$\textsuperscript{$\dagger$}, $\mathbf B$ & electron fluid velocity, electric and magnetic fields; $\varphi=(\mathbf v,\mathbf E,\mathbf B)^T$ & \eqref{eq:sum_plasma_eqs}\\
$\mathbf B_{\rm ext}$ & background magnetic field & Sect.~\ref{sec:plasma_model}\\
$\Omega$; $\omega_p$, $\wt$ & cyclotron frequency; plasma frequency, $\omega_p=\omega_-+\delta\wt$ & \eqref{eq:sum_plasma_eqs}\\
$k_z$; $\xi\times$, $\epsilon_{jlk}$ & out-of-plane wavenumber; cross-product matrix, Levi-Civita symbol & \eqref{eq:sum_plasma_symbol}\\
$\omega_-$, $\omega_R$, $\omega_{L\pm}$ & crossing frequency; right-hand and left-hand polarized frequencies & \eqref{eq:sum_def_omega_minus}, \eqref{eq:RL}\\
$R_\omega$, $L_\omega$ & dispersion functions of the polarized modes & \eqref{eq:RL}\\
$\Psi_1,\Psi_2,\Psi_\ell$\textsuperscript{$\dagger$}; $\hat e_z$, $\hat e_\pm$ & eigenvectors of $a^0$; unit and circular polarization vectors & \eqref{eq:sum_plasma_evecs}, \eqref{eq:evecs}\\
$N_j$\textsuperscript{$\dagger$}; $f_R$, $f_{L+}$ & normalization constants; coefficients of the flat sectors & \eqref{eq:sum_plasma_flat}\\
$\alpha$, $\beta$\textsuperscript{$\dagger$} & constants of the Dirac operator $b_{0,\omega_-}$ & \eqref{eq:sum_plasma_Dirac}, \eqref{eq:alpha_beta}\\
$\Gamma_P$\textsuperscript{$\dagger$} & antiunitary symmetry, $\Gamma_P\overline{a^0}\Gamma_P=-a^0$ & \eqref{eq:sum_plasma_flat}\\
$m_{\ell\ell'}$; $\sigma_{\omega_-,\ell}$ & symbols of $\mathbf M_{\ell\ell'}$; symbols of the contributions to $\Sigma_{\omega_-}$ & \eqref{eq:M_pm}, \eqref{eq:symbol_self_energies}\\
$\mathbbm 1_\pm$ & $1$ for the upper sign, $0$ for the lower one & \eqref{eq:M_pm}\\[4pt]
\multicolumn{3}{@{}l}{\emph{Numerical simulations (Sections~\ref{sec:numerics}, \ref{sec:sm_numerics})}}\\[2pt]
$\bar m$, $V_0$, $\beta_0$, $\gamma_0$ & mass, barrier height, and coupling strengths of the chain experiments & Sect.~\ref{sec:numerics_chains}\\
$X_0$, $\zeta_0$ & center and momentum of the Gaussian data & Sect.~\ref{sec:numerics}\\
$\omega_{\max}$, $W$, $X_{\max}$ & amplitude and width of the interface profile; half-period of the domain & Sect.~\ref{sec:numerics_plasma}\\
$T_\star$ & slow time at which the relative error reaches $10\%$ & Sect.~\ref{sec:sm_num_longtime}\\
$u$; $r$ & datum $(I+\sum_\ell S^\ell P_\ell)^{-1}\phi_0$; ramp profile $r(x)=\frac12(1+\tanh x)$ & Sect.~\ref{sec:sm_num_allsectors}\\
$\ell^2$ & relative $\ell^2$ error on the lattice (not a sector index) & Sect.~\ref{sec:numerics_methods}\\[4pt]
\multicolumn{3}{@{}p{0.95\textwidth}}{$^\dagger$\emph{Letters with unrelated meanings in separate contexts:} $S$ (coupling operator; symbol classes $S(\fm)$, $S^m$); $N$ (matrix size; normalizations $N_j$); $n$ (truncation order; index of $\kdisp_n$); $E$ (eigenvalues $E_\ell$, energy offset of the chains; electric field $\mathbf E$); $v$ (lattice constant; fluid velocity $\mathbf v$); $\alpha$, $\beta$ (chains; plasma constants; multi-indices in Appendix~\ref{sec:pseudo}); $\sigma$ (Pauli matrices; spectrum $\sigma(\cdot)$; symbols $\sigma_{\omega_-,\ell}$); $\Gamma$ (the $\Gamma$ point; covectors $\Gamma_\pm$; symmetry $\Gamma_P$); $\Psi$ (wave packet $\Psi_\delta[\ell,\phi_0]$; plasma eigenvectors $\Psi_j$); $M$ (exponent in Lemma~\ref{lemma:u_delta_reg}; matrix in Lemma~\ref{lemma:sa_H0l}; mass matrices $M_j$; couplings $\mathbf M_{\ell\ell'}$); $R$ (radius; rotations $R_{\pm2\pi/3}$; frequency label $\omega_R$; functions $R_\omega$; prefactor $R_\delta$).}\\
\end{longtable}
}

\end{document}